\documentclass[useAMS,usenatbib,usegraphicx]{mn2e}
\usepackage{graphicx}
\usepackage{times}
\usepackage[british]{babel}
\usepackage{amstext}
\usepackage{psfig,epsfig}
\usepackage{morefloats}

\newcounter{subfigure}
\title[Line-strengths of late-type spirals observed with {\tt SAURON}]
{Absorption line-strengths of 18 late-type spiral galaxies observed with {\tt SAURON}}
\author[Ganda et al.]{Katia Ganda$^1$\thanks{E-mail: katia@astro.rug.nl}, Reynier F.\ Peletier$^1$, Richard M.\ McDermid$^2$, Jes\'us Falc\'on-Barroso$^{3,2}$,
\newauthor
 P.T. de Zeeuw$^{2}$, Roland Bacon$^{4}$, Michele Cappellari$^{5}$, Roger L. Davies$^{5}$,   
\newauthor 
Eric Emsellem$^{4}$, Davor Krajnovi\'c$^{5}$, Harald Kuntschner$^{6}$, Marc Sarzi$^{7,5}$, 
\newauthor Glenn van de Ven$^{8,2}$\thanks{Hubble Fellow}\\
$^1$Kapteyn Astronomical Institute, University of Groningen, P.O. Box 800, 9700 AV Groningen,
The Netherlands\\
$^2$Sterrewacht Leiden, University of Leiden, Niels Bohrweg~2, 2333~CA Leiden, The Netherlands\\
$^3$European Space Agency / ESTEC, Keplerlaan 1, 2200 AG Noordwijk, The Netherlands\\
$^4$Universit\'e de Lyon 1, CRAL, Observatoire de Lyon, 9 av. Charles Andr\'e,
F-69230 Saint-Genis Laval; CNRS, UMR 5574 ; ENS de Lyon, France\\
$^5$Sub-Department of Astrophysics, University of Oxford, Denys Wilkinson
Building, Keble Road, Oxford OX1 3RH, United Kingdom\\
$^6$Space Telescope European Coordinating Facility, European Southern
Observatory, Karl-Schwarzschild-Str~2, 85748 Garching, Germany\\
$^7$Centre for Astrophysics Research, University of Hertfordshire, Hatfield, UK\\
$^8$Institute for Advanced Study, Einstein Drive, Princeton, NJ 08540, USA}

\date{Released 2007 Xxxxx XX}

\pagerange{\pageref{firstpage}--\pageref{lastpage}} \pubyear{2007}

\def\LaTeX{L\kern-.36em\raise.3ex\hbox{a}\kern-.15em
    T\kern-.1667em\lower.7ex\hbox{E}\kern-.125emX}

\begin{document}

\label{firstpage}

\maketitle

\begin{abstract}
 We present absorption line-strength maps for a sample of 18 Sb-Sd galaxies observed using the integral-field spectrograph {\tt SAURON} operating at the William Herschel 
 Telescope on La Palma, as part of a project devoted to the investigation of the kinematics and stellar 
 populations of late-type spirals, a relatively unexplored field. The {\tt SAURON} spectral range allows the measurement of the Lick/IDS indices  
 H$\beta$, Fe5015 and Mg{\textit{b}}, which can be used to estimate the stellar population parameters. 
 We present here the two-dimensional line-strength maps for each galaxy. From the maps, we learn that late-type spiral galaxies tend to have 
 high H$\beta$ and low Fe5015 and Mg{\textit{b}} values, and that the H$\beta$ index has often a positive gradient over the field, while the metal indices peak in 
 the central region.\\
  We investigate the relations between the central line-strength indices and their correlations with 
 morphological type and central velocity dispersion, and compare the observed behaviour with that for ellipticals, lenticulars and early-type 
 spirals from the {\tt SAURON} survey. We find that our galaxies lie
 below the Mg - $\sigma$ relation determined for elliptical galaxies and that the indices show a clear trend with
 morphological type.\\
From the line-strength maps we calculate age, metallicity and abundance ratio maps via a comparison
 with model predictions; we discuss the results from a one-SSP (Single Stellar Population)  approach and from a 
 two-SSP approach, considering the galaxy as a superposition of an old ($\approx$ 13 Gyr) and a 
 younger (age $\leq$ 5 Gyr) population. We confirm that late-type galaxies are generally younger and more metal poor than ellipticals and have 
 abundance ratios closer to solar values. We also explore a continuous star formation scenario, and try to recover the star formation
 history using the evolutionary models of Bruzual \& Charlot
 (2003), assuming constant or exponentially declining star formation rate (SFR). In this last case, fixing the galaxy age to 10 Gyr, we find a correlation between the $e$-folding
 time-scale $\tau$ of the starburst and the central velocity dispersion, in the sense that more massive galaxies tend to have shorter $\tau$, 
 suggesting that the star
 formation happened long ago and has now basically ended, while for smaller objects with larger values of $\tau$ it is still active now.
 \end{abstract}


\clearpage

\section{Introduction}
Over the last few decades, stellar population synthesis has become one of the most popular and powerful techniques to study 
the formation and evolution histories of galaxies too distant to be resolved into individual stars. In particular, the 
measurement of absorption line-strengths combined with stellar population models has been used to 
investigate the luminosity-weighted age, metallicity and abundance ratios in integrated stellar populations. Extensive work has been
carried out on stellar populations in early-type galaxies, mainly using long-slit spectra (see for example \citealt{danziger}, Davies,
Sadler \& Peletier 1993, Fisher, Franx \& Illingworth 1995, Fisher, Franx \& Illingworth 1996, \citealt{longhetti}, \citealt{jorg99}, \citealt{harald2000}, \citealt{patricia}) and recently 
also integral-field spectroscopy (Kuntschner et al. 2006, hereafter Paper VI, McDermid et al. 2006, hereafter Paper VIII). The main advantage brought by the integral-field observations 
is that one can obtain two-dimensional maps and easily identify extended structures. In particular, the two-dimensional coverage 
gives the possibility to straightforwardly connect the stellar populations with the kinematical structures and therefore represents a powerful 
tool towards a comprehensive understanding of the structure of the galaxy.\\
\indent Some literature is also available on absorption line-strengths in early-type spiral galaxies, mainly S0 and Sa galaxies; 
(see for example Jablonka, Martin \& Arimoto 1996, \citealt{idiart}, \citealt{proctor}, \citealt{silchenko}, and Peletier et al. 2007, hereafter Paper XI). 
In particular, \citet{reynier} presents two-dimensional absorption line-strength maps of 24 spiral galaxies, mostly of type Sa, observed 
with the integral-field spectrograph {\tt SAURON} (Bacon et al. 2001, hereafter Paper I); these reveal young 
populations in Sa galaxies, possibly formed in mini-starbursts in structures such as nuclear discs, circular star forming rings and bars. These mini-starbursts generate a 
large scatter in index - index diagrams, larger than observed for elliptical galaxies. As a consequence, there is a large range in luminosity-weighted ages. Different star formation modes -starbursts or quiescent- are also reflected in the wide range observed in the
abundance ratios. All the galaxies of Paper XI lie on or below the Mg{\textit{b}} - $\sigma$  relation for ellipticals determined by 
J\o rgensen, Franx \& Kj\ae rgaard (1996). The authors argue that, if one considers that relation as valid for old galaxies, then the early-type spirals 
present a scatter in age, with the biggest scatter for the smallest velocity dispersions.\\
\indent Spiral galaxies towards the end of the Hubble sequence are, on the other hand, still poorly-studied objects, due to the difficulty of 
obtaining reliable measurements in these low-surface brightness objects, full of dust and star forming regions, whose spectra can 
be dramatically contaminated by emission-lines. In particular, the study of their stellar populations is extremely complicated. 
An approach based on broad-band colours is basically invalidated by the huge amount of dust, which is very difficult to take into account, especially when 
working with data from ground-based telescopes. A spectroscopic approach, relying 
for example on stellar absorption-strength indices, has instead to deal with the almost ubiquitous presence of gas 
with consequent emission, which makes it necessary to accurately remove the emission-lines from the spectra before performing any 
population analysis. Only in the last decade several HST-based imaging surveys have targeted these late-type galaxies. 
Particularly relevant are the papers by \citet{marcella97}, Carollo \& Stiavelli 1998, Carollo, Stiavelli \& Mack 1998, 
\citet{marcella99}, \citet{marcella02}, \citet{boker}, \citet{laine}, where the authors reveal the presence of a variety of
structures in the inner regions, such as bulges, nuclear star clusters, stellar discs, small bars, double bars and star forming rings 
whose formation and evolutionary pattern are not properly understood yet. \citet{marcella06} also performed a population analysis on nine late-type spiral
galaxies (types between Sa and Sc) on the basis of HST ACS and NICMOS optical and near-infrared colours. The high spatial 
resolution of their space-based data allowed them to mask dust features and measure the properties of the stellar populations from the colours; 
this method, together with the high spatial resolution provided by HST, produced results that do not suffer much from the effects of dust; only a smooth dust
distribution would not be detected. From their analysis, they found 
a large range in the colour properties of bulges in late-type spirals, reflecting a large range in their stellar population properties. From a 
comparison with population models, they concluded that 
in about half of their bulges the bulk of the stellar mass formed at early epochs (more than 50\% of their mass was formed more than 9 Gyr ago), 
and that in the other half a non-negligible fraction of the stellar mass (up to 25\%) was formed recently, in the 
last $\approx$ 3 Gyr.\\
\indent Until now, only little spectroscopic counterpart for the quoted imaging was available, due to the mentioned observational
difficulties. The published work (\citealt{gallag}, Zaritsky, Kennicutt \& Huchra 1994, \citealt{boker01},
\citealt{boker03}, \citealt{walcher}) mainly refers to the emission-line properties, to the characteristics of the 
HII regions or to the nature of the innermost component. In particular, 
no study has addressed yet in a systematic way the stellar populations of late-type spiral galaxies with a two-dimensional coverage. Existing studies on abundances 
concern topics such as elemental abundances in HII regions and the distribution of oxygen within discs \citep{zaritsky}, not 
properties inferred from the integrated stellar absorption spectrum. As for the absorption-lines properties, there is ongoing work 
by MacArthur, Gonzalez \& Courteau (2007), who analyse long-slit spectra (Gemini - {\tt GMOS}) of eight spiral galaxies of type between Sa and Scd on a wide spectral range, measuring
many absorption indices out to 1-2 disc scale-lengths. Also, \citet{moorthy} presented line-strengths in the bulges and inner discs of 38 galaxies of morphological type
between S0 and Sc for which they acquired long-slit spectra on a broad spectral range including Balmer lines, Mg and Fe features. They concluded that the
central regions of bulges span a wide range both in SSP metallicity and age, confirming the results of \citet{marcella06}. They also 
found that luminosity-weighted metallicities and abundance ratios are sensitive to the value of the central velocity dispersion and of the maximum disc rotational
velocity, and that red bulges (defined as those with B$-$K $>$ 4) of all types are similar to luminous ellipticals and obey the same scaling relations; 
for the blue bulges (B$-$K $<$ 4) they observed instead 
some differences. They also addressed the radial variations of the population parameters; in most cases, they measured negative metallicity gradients 
with increasing radius, in the bulge-dominated region; 
positive gradients in age were found only in barred galaxies.\\
\indent As mentioned, there is still a lack of work on populations in late-type spirals on a full
two-dimensional field: we therefore started a project on a sample of 18 late-type spiral galaxies, with the purpose of investigating the nature of their inner
regions, addressing the bulge and disc formation and evolution, the interconnection between stellar and gaseous components and 
the star formation history. Given the high complexity in the inner regions of these objects, integral-field 
spectroscopy has to be preferred with respect to long-slit, providing a full picture of the kinematics and populations on a two-dimensional field of view. 
We observed our 18 late-type spiral galaxies using {\tt SAURON}, an integral-field spectrograph that was built for a representative census of ellipticals, lenticulars and early-type spiral galaxies, 
to which in the rest of this Paper we will refer as `the {\tt SAURON} 
survey'. Our project can be regarded as an extension of the {\tt SAURON} survey towards later-type objects. First results based on our data are
presented by \citet{ganda} and focus on the stellar and gas kinematical maps. The main findings of that Paper are that in many cases the stellar kinematics suggests the 
presence of a cold inner region, as visible from a central drop in the velocity dispersion, detected in about one third of the sample; that the ionized gas is almost 
ubiquitous and often presents more irregularities in its kinematics than the stellar component; and that the line ratio [OIII]/H$\beta$ assumes often low values 
over most of the field, possible indication 
for wide-spread star formation, in contrast to early-type galaxies.\\
\indent As a following step in our project, we subtracted the emission-lines from our {\tt SAURON} spectra, measured as explained in \citet{ganda}, and on the emission-cleaned 
spectra derived absorption line-strength indices in the Lick/IDS system (\citealt{lick1}, \citealt{lick2}, \citealt{lick3}), which allows an easier 
comparison with existing data and with other classes of objects. Here we present and analyse the absorption line-strength maps. The data and maps presented in this
Paper will be made
available via the {\tt SAURON} WEB page {\tt http://www.strw.leidenuniv.nl/sauron/}.\\
\indent The Paper is structured as follows. 
Section 2 briefly describes the sample selection, observations and data reduction. Section 3 reviews the analysis methods. In Section 4 we present the 
line-strength maps for all the galaxies in our sample. In Section 5 we investigate the correlations between different indices and between indices and other galactic parameters, 
such as the morphological type and the central velocity dispersion, mainly focussing on the central values. In Section 6 we present our estimates for the ages 
and metallicities, obtained by comparison of our observed line-strength indices with model predictions; 
we also explore the star formation history and give estimates for the time-scale $\tau$, in an
exponentially declining star formation rate scenario. Section 7 summarises the main results. In Appendix A we deal with the radial variations of the indices 
and the stellar population parameters,
presenting azimuthally averaged radial profiles. Appendix B shortly describes the individual galaxies and their
peculiarities, giving a brief summary of relevant literature information. For more literature reviews on these objects, we 
suggest the reader to check Section 6.4 in \citet{ganda}.

\section{Sample selection, observations and data reduction}\label{samplesec}
The galaxies were optically selected ($B_{T}$ $<$ 12.5, according to the values given in \citealt{RC3},
hereafter RC3) with
HST imaging available from WFPC2 and/or NICMOS. Their morphological type ranges
between Sb and Sd, following the classification given in NED\footnote{http://nedwww.ipac.caltech.edu} (from the RC3). Galaxies in close
interaction and Seyferts were discarded. The resulting sample contains 18 nearby galaxies, whose 
main properties are listed and illustrated respectively in Table 1 and Fig. 1 in \citet{ganda}. The NGC\, numbers and some characteristics are reported also
in the tables in the present Paper.
Observations of the 18 sample galaxies were carried out during 6 nights in January 2004, using the integral-field spectrograph {\tt SAURON} attached to the 4.2-m WHT. The
exposure details are listed in Table 2 in \citet{ganda}.\\
\indent We used the low spatial resolution mode of {\tt SAURON}, giving a field-of view
(FoV) of 33\arcsec\/ $\times$ 41\arcsec\/. The spatial sampling of individual exposures is
determined by an array of 0\farcs94 $\times$ 0\farcs94 square lenses. This produces 1431
spectra per pointing over the {\tt SAURON} FoV; another 146 lenses sample a region 1\farcm9 away
from the main field in order to measure simultaneously the sky background. 
{\tt SAURON} delivers a spectral resolution of 4.2 \AA\, FWHM, and its maximum spectral coverage is 
4800-5380 \AA; the wavelength interval common to all lenses is the narrow spectral range 4825-5275 \AA\, (1.1 \AA\, per pixel). This
wavelength range includes a number of important stellar absorption lines (e.g.
H$\beta$, Fe5015, Mg{\textit{b}}), and potential emission-lines as well 
(H$\beta$, [OIII], [NI]). For a more exhaustive description of the instrument, see \citet{paper1}.\\
\indent Data reduction was carried out using the dedicated software {\tt XSAURON} developed at Centre de Recherche Astronomique de Lyon (CRAL). 
For details on the data reduction we refer
the reader to \citet{paper1} and \citet{ganda}. For an explanation of the spectrophotometric calibration we recommend the 
reader to study Paper VI, in particular its Section 3.1.

\section{Analysis and methods}
\begin{figure*}
{\includegraphics[width=0.99 \linewidth]{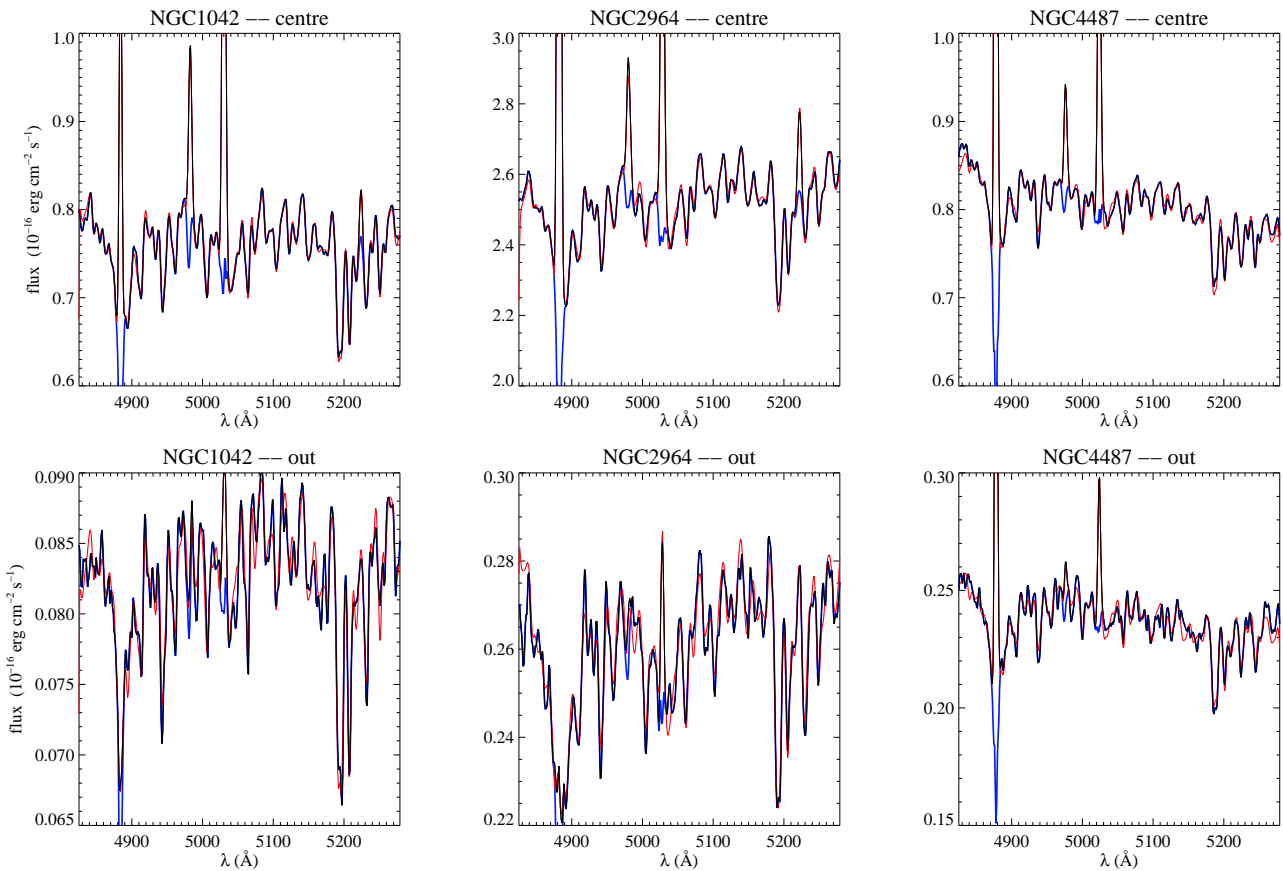}}
\caption{Top row: central aperture spectra of some representative galaxies in our sample: from left to right, we plot NGC\,1042, 2964
and 4487. Along the vertical axis, the flux is in units of 10$^{-16}$ erg cm$^{-2}$ s$^{-1}$. The black lines -hardly distinguishable from the
other ones-
represent the observed spectra; the red ones the best-fit and 
the blue ones the spectra after emission removal. Bottom row: as in the top row, but for bins located $\approx$ 10\arcsec\/ away from 
the galaxy's centre.}
\label{spec_864}
\end{figure*}
The fully reduced and calibrated spectra were first spatially binned using a Voronoi tessellation \citep{voronoi} in order to reach a minimum signal-to-noise ratio 
of $\approx$ 60 per spectral resolution element per bin. 
The kinematics of stars and gas were then extracted from the binned spectra via the pixel-fitting methods extensively described in \citet{ppxf}, Emsellem 
et al. (2004, hereafter Paper III), and Sarzi et al. (2006, hereafter Paper V), that we already 
applied to calculate the kinematical maps shown in \citet{ganda}. The spectra were fitted using the SSP models from \citet{vazdekis}. We selected a library of 
48 models evenly sampling a wide range in both age and
metallicity (0.50 $\leq$ Age $\leq$ 17.78 Gyr, $-$1.68 $\leq$ [Fe/H] $\leq$ +0.20). We used this SSP library and additional Gaussian templates reproducing the 
emission-lines to fit the observed spectra, separating the emission-lines from the underlying absorption line spectrum (see \citealt{paper5} for details on the 
procedure). We notice that for our objects this separation is crucial if we want to measure absorption strengths, since in spiral galaxies the gas 
is almost
ubiquitous and emission in several cases dominates the spectrum, filling in completely the absorption features, as shown for example in the cases of some
central aperture spectra in the top row of Fig. \ref{spec_864} (black lines); in 
the same Figure we overplot the best-fit to the spectra obtained with our method
(red lines) and the emission-removed spectra (blue lines), illustrating the reliability of the applied gas-cleaning procedure. The bottom row of 
Fig. \ref{spec_864} presents similar plots for bins located $\approx$ 10\arcsec\/ from the galaxy's centre, showing that our fits and emission
correction are reliable also outside the central region, given the fact that a minimum signal-to-noise of 60 is garanteed by the spatial binning. The reader can find more examples 
of this in Fig. 1 in \citet{paper5}, Fig. 3 in \citet{ganda}, and Fig. 3 in Falc\'on-Barroso et al. (2006), hereafter Paper
VII.\\
\indent We also notice that in two cases (namely, NGC\,2964 and NGC\,4102) we detected regions of activity (see \citealt{ganda} 
for a description), where the spectra are characterised by double-peaked emission-lines due to the presence of multiple velocity
components in the gas, because of the presence of a jet. In those 
regions our method of fitting the lines with single Gaussians fails to reproduce accurately the line profile, but since this is limited to only a few spectra, we 
did not modify the method. In any case, this does not have any impact on our following analysis, since the double-peaked emission-lines do not 
contaminate the spectra defining the central aperture (see Section \ref{correlationsec}) on which most of this study is focused on.
\subsection{Line-strengths measurement and Fe5015 correction}\label{correctionsec}
\begin{figure*}
{\includegraphics[width=0.99 \linewidth]{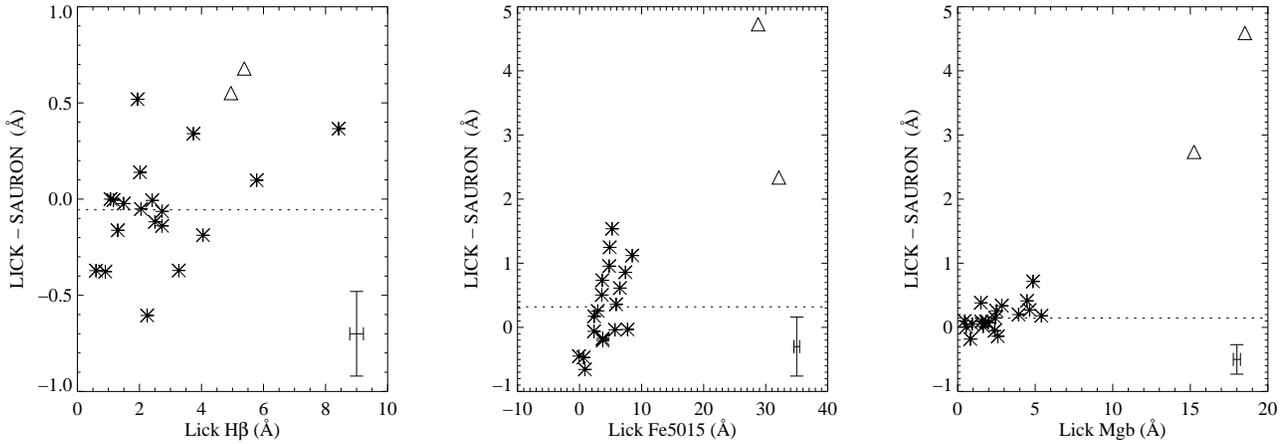}}
\caption{Differences between the Lick/IDS and our measurements for the 21 stars in common, for the three indices. For two stars (HD18191 and HD114961, spectral 
type respectively M6III and M7III) our measurements of the metal indices differ dramatically from the Lick/IDS measurements; therefore we excluded them from the determination 
of the offsets and plotted them with a different symbol (open triangles). The overplotted dashed line shows the adopted offset, 
derived by a bi-weight estimator. In the bottom right corner we place a typical errorbar.}
\label{lickoffsets}
\end{figure*}

\begin{figure*}
{\includegraphics[width=0.74 \linewidth]{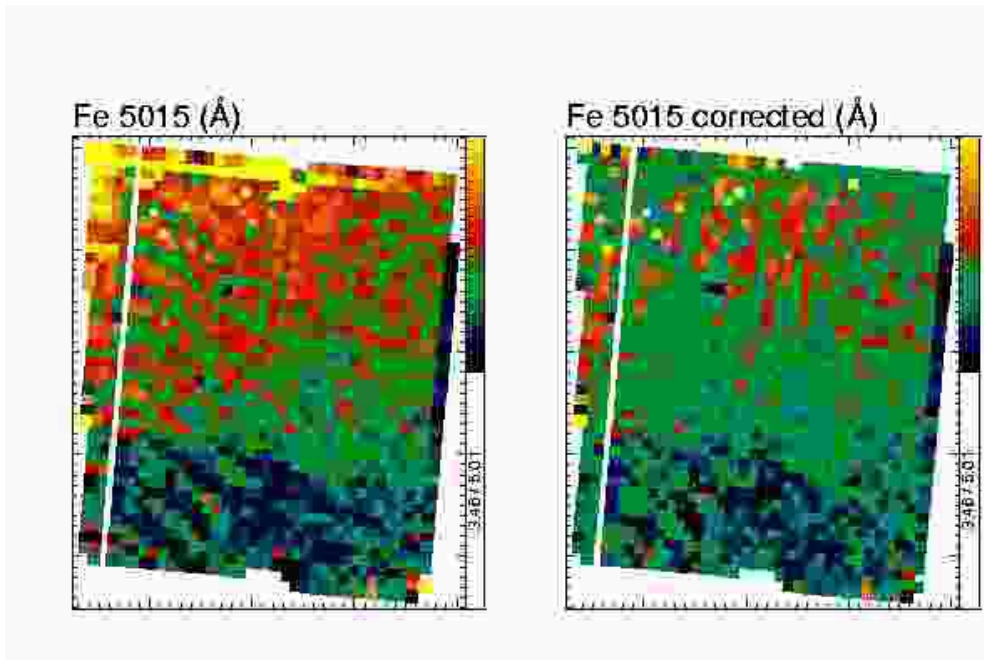}}
\caption{Maps of the Fe5015 index for the twilight frame, before and after correction, in \AA. The two maps are plotted with the same cuts, indicated on their side, 
together with the colour bar. In the corrected frame, the spatial gradient is largely removed.}
\label{twilightcorrection}
\end{figure*}

\begin{figure*}
{\includegraphics[width=0.74 \linewidth]{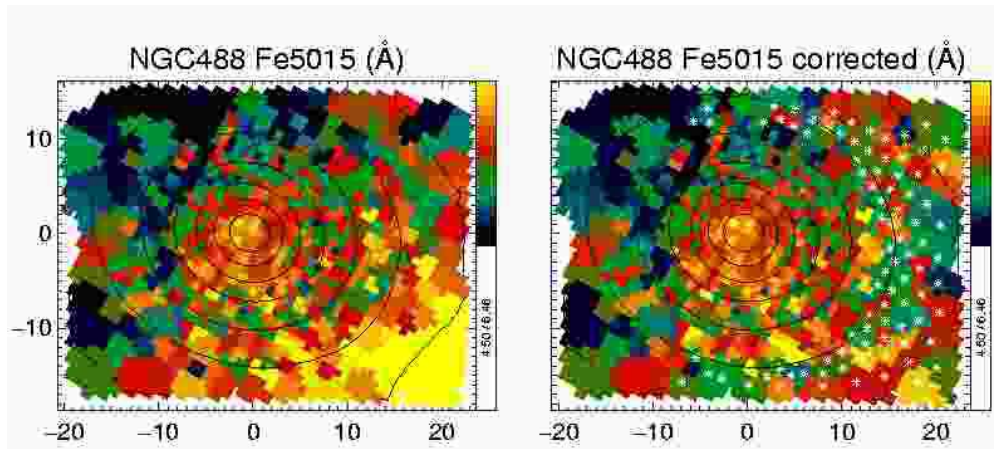}}
\caption{Maps of the Fe5015 index for NGC\,488, before and after correction, in \AA. The white asterisks in the right-hand map mark the position of the centroids of the bins
to which we have applied the correction. Overplotted on the maps are the isophotal contours. The two maps are plotted with the same cuts, indicated on their side, 
together with the colour bar; they are oriented with the horizontal and vertical axis aligned, respectively, to the long- and to the short- axis of the {\tt SAURON} field.}
\label{correctionexample}
\end{figure*}

\begin{figure*}
{\includegraphics[width=0.99 \linewidth]{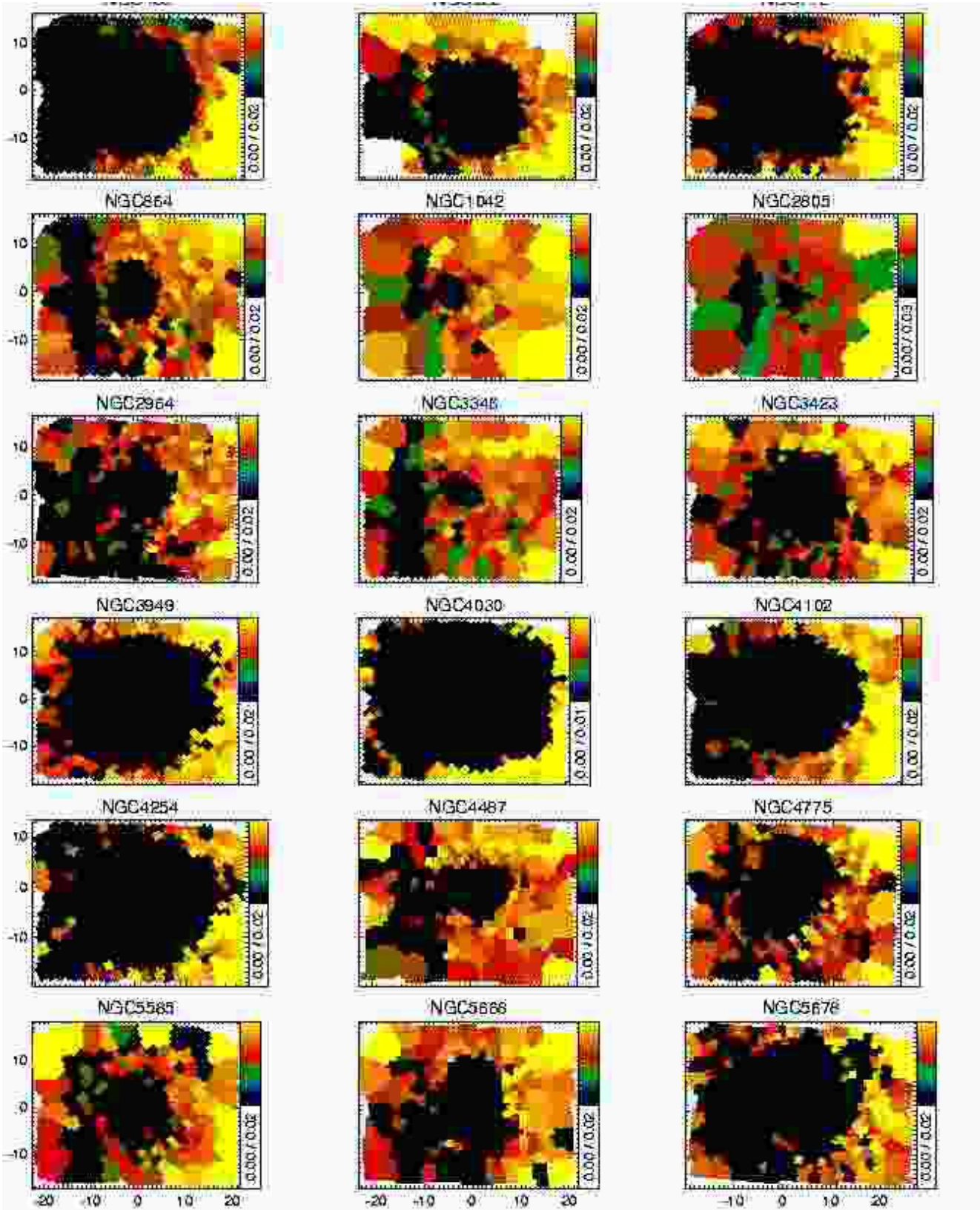}}
\caption{Galaxy by galaxy maps of the RMS deviations of the fitted polynomial continuum after removal of a linear slope, in the correction procedure for the continuum problem affecting the Fe5015 measurement; 
in the bins in the black areas the correction will not be applied. All the maps are plotted with the same spatial
scale and oriented with the 
horizontal and vertical axis aligned, respectively, to the long- and to the short- axis of the {\tt SAURON} field.}
\label{correctedregions}
\end{figure*}

On the emission-removed spectra we calculated line-strength indices on the Lick/IDS system. For details on the actual computation of the line-strengths, we refer to
\citet{paper6}. The calibration to the Lick/IDS system 
takes into account several effects: the difference in spectral resolution between the Lick/IDS system and the {\tt SAURON} instrumental setup, the internal velocity broadening of the 
observed galaxies and small systematic offsets rising from differences in the shape of the continuum, due to the fact that the original Lick/IDS 
spectra have not been flux calibrated. In our observing run, we obtained spectra for 21 different stars in common with the Lick/IDS stellar library
\citep{lick3}, for a total of 39 repeated observations. Fig. \ref{lickoffsets} shows the difference between the Lick/IDS measurements and ours (for stars with 
repeated observations, we considered the average of the different values available). 
We calculated the mean offsets to the Lick/IDS system and their dispersion using bi-square weighting 
in order to minimize the influence of the outliers, after removing from the sample two stars of spectral type M6III and M7III, whose 
measured metal indices differ more than 2 \AA\ from the Lick measurements. The adopted offsets that we applied to all our measurements are: $-$0.06 \AA\ ($\pm$ 
0.06 \AA) for H$\beta$, +0.32 \AA\, ($\pm$ 0.15 \AA) for Fe5015 and +0.15 \AA\ ($\pm$ 0.05 \AA) for Mg{\textit{b}}.\\
\indent From a visual inspection of our maps of the Fe5015 index, we saw that our measurements tend to assume high values 
in the outer parts, especially in the bottom-right corner, revealing a systematic effect that makes the maps look asymmetric. The most striking 
manifestation of the problem is this `high-iron corner', but there could be as well regions where the index is too low. A detailed 
inspection of the spectra showed that the cause is a wrong shape of the continuum in the Fe5015 spectral region, as proved by the fact 
that we are unable to fit our twilight spectra using our SSP template library: the Fe5015 region is not matched by the templates, and the Fe5015 map 
calculated on the twilight spectra is not flat over the field, showing a range of about 1.6 \AA\ from minimum to maximum value. This continuum problem 
could be due to instrumental instabilities, to the high sensitivity of the grism to the incidence angle of the light, 
to the imperfect extraction of the spectra, to the misalignment of the spectra with respect to the columns of the CCD (see \citealt{ganda}). 
We can exclude, instead, relations with problems in the gas removal, since the Fe5015 measurement is affected also in the twilight frame, 
where the spectra are gas-free. In addition to this, continuum problems should not influence the gas-cleaning procedure: in the fitting, a polynomial is included in order to account 
for differences in the flux calibration between the observed spectra and the model ones.\\ 
\indent Unfortunately, we could not take this effect into account properly at the data reduction stage, and decided to apply a correction to the affected spectra before 
measuring the Fe5015 index. A problem related to the shape of the continuum was experienced also for the early-type spirals of the 
{\tt SAURON} survey and affected the H$\beta$ measurements, due to the closeness of the spectral feature to the edge of the 
spectral range; despite the fact that the problem is different, we worked out a correction similar to the one that was there applied in the case of H$\beta$ and 
described in \citet{paper6} (Section 3.1.2).\\
\indent For each galaxy, we averaged the spectra within an aperture of 10\arcsec\/ radius, obtaining a `global spectrum' for that galaxy. 
We then fitted this global spectrum using a linear combination of SSP templates, masking the spectral regions possibly affected by emission lines. In this 
way we obtained for each galaxy a `global optimal template' that approximates very well the general spectral features of that galaxy. Then we considered the two-dimensional 
galaxy's datacube, spatially binned and cleaned from emission as explained in the previous Sections; for each bin, we fitted the cleaned spectrum 
using the `global optimal template' together with an 11-th order multiplicative polynomial, over the whole spectral range. The fit determined the best-fitting polynomial continuum, 
from which we then removed the linear component (which does not change the line indices); we considered the 
residuals and calculated their RMS variation over the wavelength range of Fe5015. For bins covering four or more single-lens spectra and with an RMS greater than 0.01, we used the fitted 
polynomial as a correction to the spectral shape by dividing it through the emission-cleaned spectrum. For these bins, we finally measured the 
Fe5015 index on the corrected spectrum.\\
\indent This correction depends critically on the assumption that the optimal
template is an accurate representation of the true galaxy spectrum. Whilst
this is generally true, in the central, sometimes bulge-dominated, regions
of our galaxies, non-solar abundance ratios can create a fit residual
similar in size to the systematic effect we wish to remove. By correcting
only bins containing four or more lenses, we avoid these central parts, since
the data there are unbinned. Correcting large bins is also important
because of their larger contribution to the spatial appearance of the
maps.\\
\indent We verified the Fe5015 correction using exposures of the twilight sky. As we mentioned above, in
the twilight maps of Fe5015, a spatial gradient was also visible, as is
apparent in the galaxy data. In the uncorrected twilight, the field centre
gave values of Fe5015 close $\approx$ 4 \AA, consistent with a typical G2V
- G4V sun-like spectrum, showing that the field centre is not
significantly affected by the problem. After applying our correction
technique\footnote{applied in this case to all single-lens spectra with an RMS greater than 0.01: given the high signal-to-noise of the twilight frame, no spatial 
binning is necessary}, the gradient was reduced, and the global level remained around
4 \AA. We show this in Fig. \ref{twilightcorrection}, where we display the Fe5015 map calculated on the twilight frame, before and after applying the correction. 
Similarly, in Fig. \ref{correctionexample} we show the Fe5015 map for NGC\,488 before and after correction. White asterisks mark the centroids 
of the bins that have been corrected. The correction clearly removes most of the asymmetry, proving the usefulness of the method applied.\\
\indent Figure \ref{correctedregions} presents for each galaxy the map of the RMS deviations of the fitted continuum after removal of a linear slope. We plot in black the bins that do not meet the requirements for the correction, for which we have artificially put the RMS to 0 (for plotting purposes 
only); in these bins the Fe5015 index is calculated on the original emission-cleaned spectra. The maps are oriented with the 
horizontal and vertical axis aligned, respectively, to the long- and to the short- axis of the {\tt SAURON} field. The extension of the corrected area is mainly related to the signal-to-noise of the data: galaxies with poor signal-to-noise will 
require higher binning and the big bins will undergo the correction.\\
\indent As for the uncertainties on the measured indices, the most worrying factor is the separation of 
emission and absorption lines. Simulations presented in Appendix A of Paper V show that the accuracy in recovering the emission-line 
fluxes does not depend on the A/N ratio between the line
amplitude and the noise level in the stellar continuum, but only on the signal-to-noise level 
in the stellar continuum itself. These simulations were meant to investigate the 
accuracy of emission-line fluxes measurements for ellipticals and lenticulars. In the case of signal-to-noise = 60, 
the typical uncertainties in the line fluxes obtained in Appendix A of Paper V translates in errors in the equivalent width of the 
emission line of $\approx$ 0.08 \AA, and in similar errors in the absorption line indices.
\citet{reynier} presents absorption line-strengths for 
early-type spirals observed with {\tt SAURON}; in some of the galaxies studied there, the emission 
lines are stronger by a factor up to 100, relative to the absorption lines, than in the ellipticals or lenticulars of Paper V with the strongest
emission-lines.  Therefore, for the objects in Paper XI the situation is much closer to 
what happens in our spirals. There, the authors report results of similar simulations of the reliability of the gas cleaning, 
exploring a range in A/N ranging up to 100. The results are such that the errors in the absorption line indices do not vary appreciably with 
increasing A/N. When coming to the data, one has to consider that the spectra are packed close together, so that the absorption lines 
are affected by neighbouring spectra; despite the reduction tries to minimize this contamination, there might be some residual effects, particularly in 
presence of strong emission lines. Additional errors can derive from template mismatch in the fitting procedure and from the continuum correction described above.
 Therefore, throughout the rest of the Paper we adopt a conservative value of 0.2 \AA\, for the uncertainties 
in H$\beta$ and Mg{\textit{b}} and 0.3 \AA\ in Fe5015. The same figures were used in Paper XI.

\section{line-strength maps of 18 late-type galaxies}
Figures \ref{maps1}-\ref{maps9} present the absorption line-strength maps of the 18 galaxies, ordered by increasing NGC number. For each galaxy, in the first row we show 
the total intensity obtained by integrating the full {\tt SAURON} spectra in the wavelength direction, the galaxy NGC number (and the UGC 
also, when available), and the two-dimensional map of the characteristic time-scale of star formation $\tau$, in Gyr (see Section \ref{continuum}). 
Overlapped in pink on the intensity map are the contours delimiting the outer limits of the so-called `bulge' and `disc regions', defined in Section \ref{correlationsec}. In the second row we present the two-dimensional maps 
of H$\beta$, Fe5015 and Mg{\textit{b}}, where the measured indices have been calibrated to the Lick/IDS system and are expressed in equivalent widths and 
measured in \AA. The maps of Fe5015 have been corrected as explained in Section \ref{correctionsec}; the corrected bins are marked 
with a white dot. In the third row, we present the maps of age (expressed in units of decimal logarithm and measured in Gyr), metallicity (in units of decimal logarithm, 
with the solar metallicity as zero point) and abundance ratios obtained with the one-SSP approach (see Section \ref{singlessp}). 
Overplotted on each map are the isophotal contours. The maps are all plotted with the same spatial scale, and 
oriented with the horizontal and vertical axis aligned, respectively, to the long- and to the short- axis of the {\tt SAURON} field for presentation purposes; the orientation is the same as in \citet{ganda} and as in Figures
\ref{correctionexample} and \ref{correctedregions} in this Paper. The relative directions of North and East are indicated by the arrow 
above the galaxy's name. The maximum and minimum of the plotting ranges and the colour table are given in the tab attached 
to each map. To make the comparison among different objects more immediate, we adopted the same plotting ranges for all of the 
galaxies, but in Appendix \ref{individuals} we will enlighten the spatial structures within the single maps, when noticeable, by inspecting the maps 
plotted on a `galaxy-by-galaxy range'.

\renewcommand{\thefigure}{\arabic{figure}\alph{subfigure}}
\setcounter{figure}{5}
\setcounter{subfigure}{1}
\begin{figure*}
\begin{center}
{\includegraphics{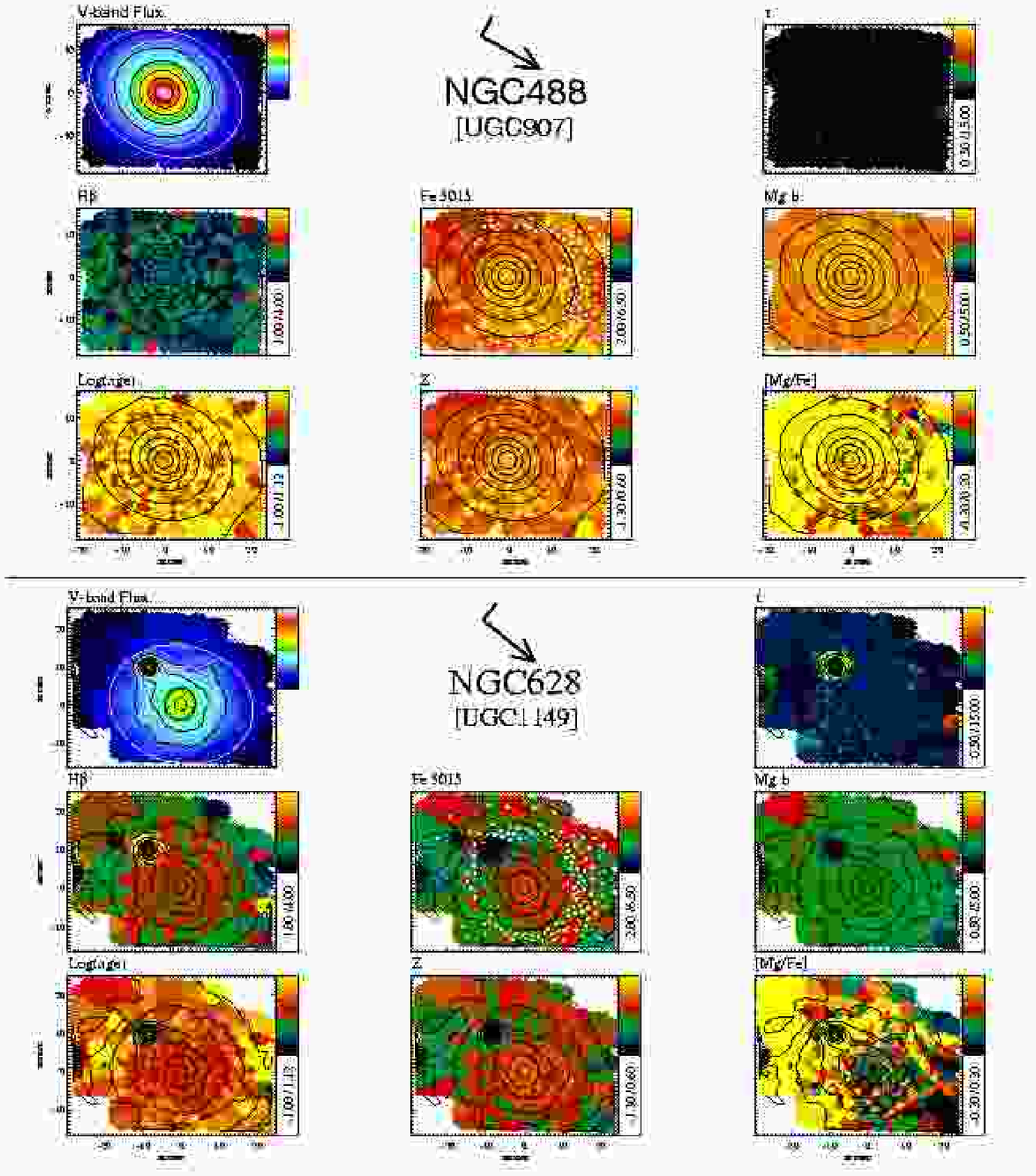}}
\end{center}
\caption{Maps for NGC\,488 and NGC\,628. For each galaxy: \textit{first row}: reconstructed image; 
NGC and UGC numbers and time-scale $\tau$ for star formation in Gyr; \textit{second row}: maps of H$\beta$, Fe5015 and Mg{\textit{b}} equivalent widths, in
\AA; the white dots on the Fe5015 map mark the bins where the correction described in Section \ref{correctionsec} has been applied; 
\textit{third row}: age (in units of decimal logarithm and measured in Gyr), metallicity and [Mg/Fe] abundance ratio from the one-SSP analysis. 
For a description of the methods, see Section \ref{agesection}. Overplotted
on each map are the isophotal contours. On the intensity map also the contours defining the outer limits of the so-called `bulge' and `disc' regions (see Section 
\ref{correlationsec}) are overplotted, in pink. All the maps are presented with the same spatial scale; the arrow above the galaxy's name
gives the direction of North and East. The plotting ranges are indicated together with the colour bar at the right side of each map.} 
\label{maps1}
\end{figure*}
\addtocounter{figure}{-1}
\addtocounter{subfigure}{1}

\clearpage
\begin{figure*}
\begin{center}
{\includegraphics{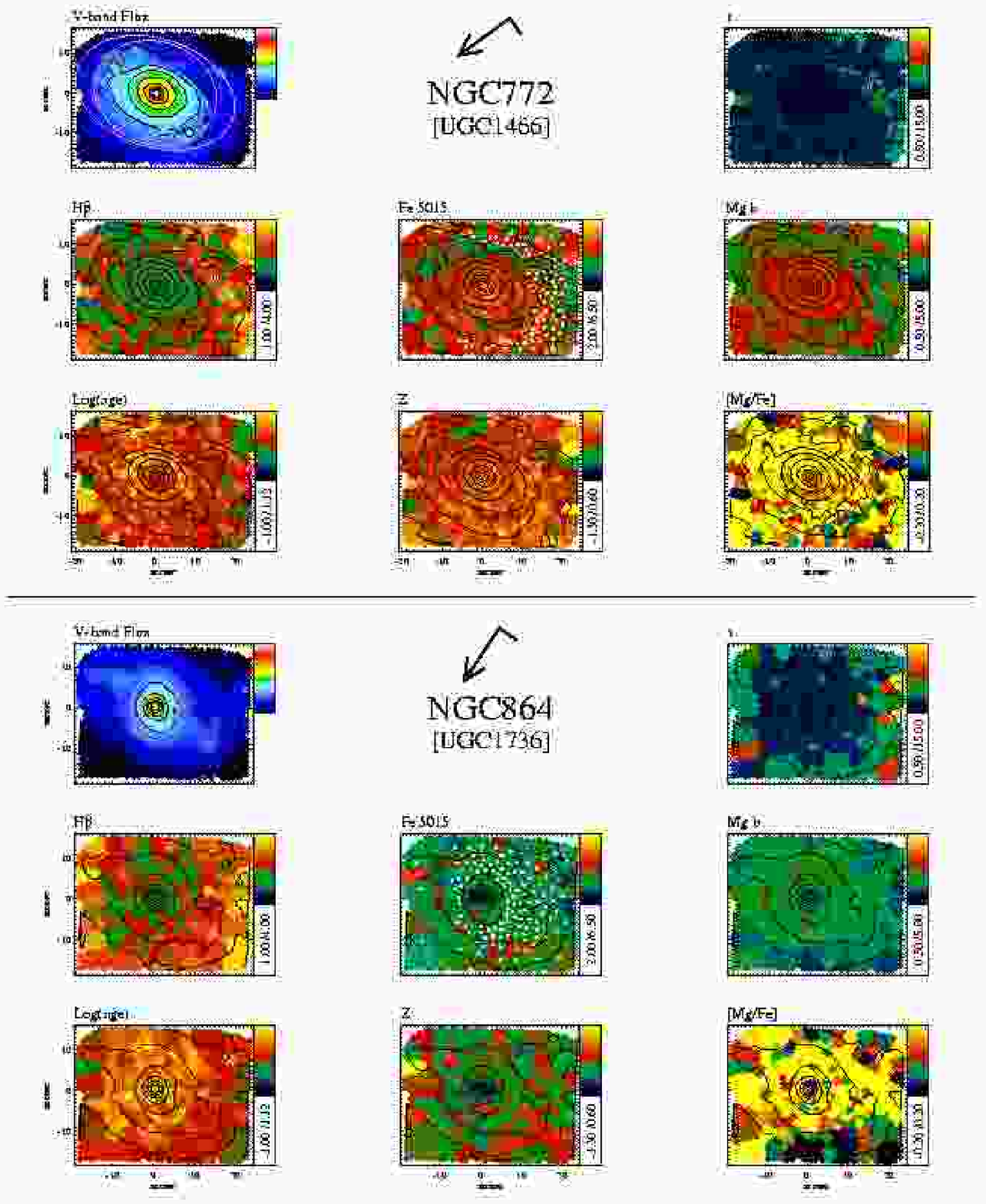}}
\end{center}
\caption{As in Figure \ref{maps1} for NGC\,772 and NGC\,864.}\label{maps2}
\end{figure*}
\addtocounter{figure}{-1}
\addtocounter{subfigure}{1}

\clearpage
\begin{figure*}
\begin{center}
{\includegraphics{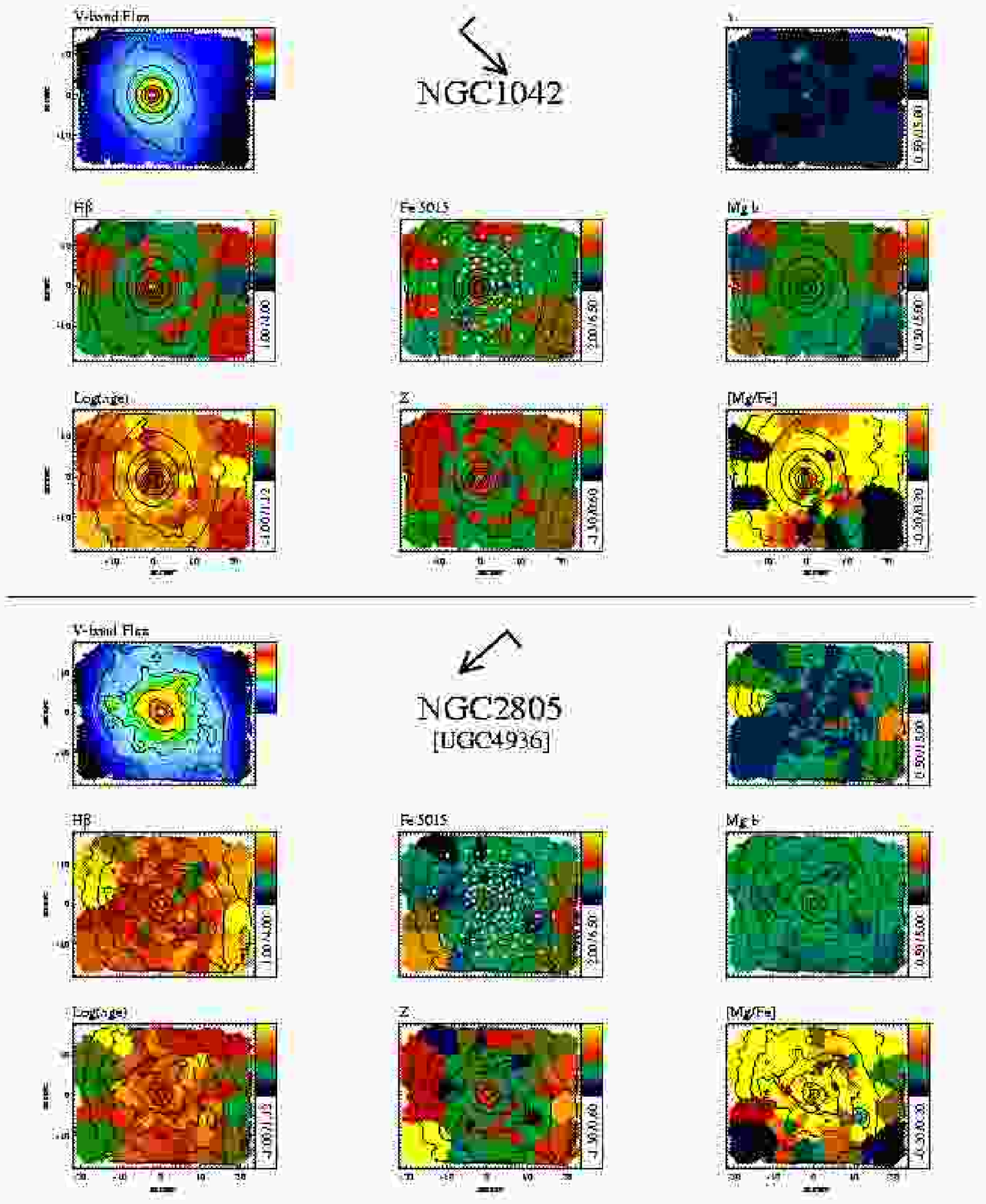}}
\end{center}
\caption{As in Figure \ref{maps1} for NGC\,1042 and NGC\,2805.}\label{maps3}
\end{figure*}
\addtocounter{figure}{-1}
\addtocounter{subfigure}{1}

\clearpage
\begin{figure*}
\begin{center}
{\includegraphics{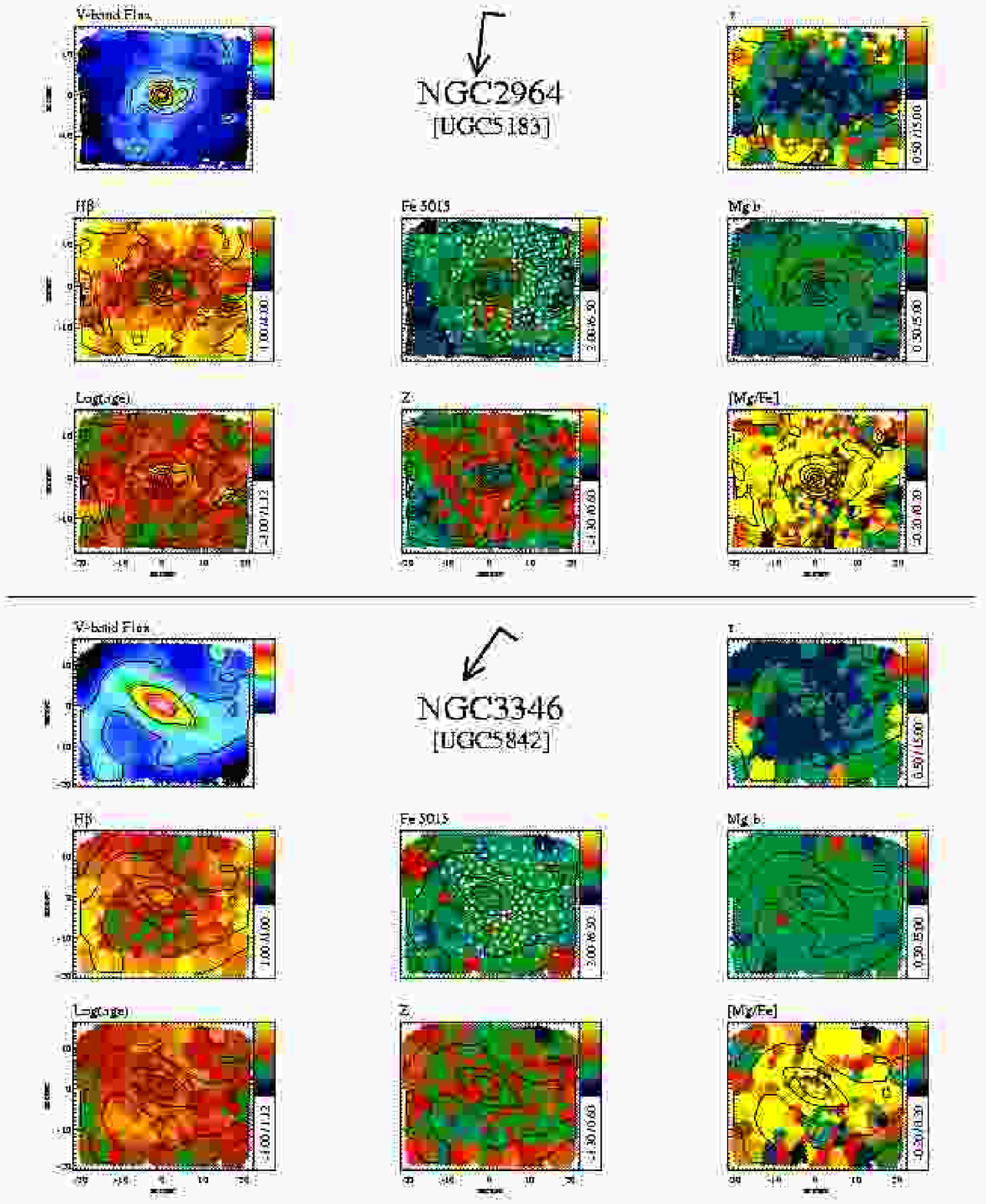}}
\end{center}
\caption{As in Figure \ref{maps1} for NGC\,2964 and NGC\,3346.}\label{maps4}
\end{figure*}
\addtocounter{figure}{-1}
\addtocounter{subfigure}{1}

\clearpage
\begin{figure*}
\begin{center}
{\includegraphics{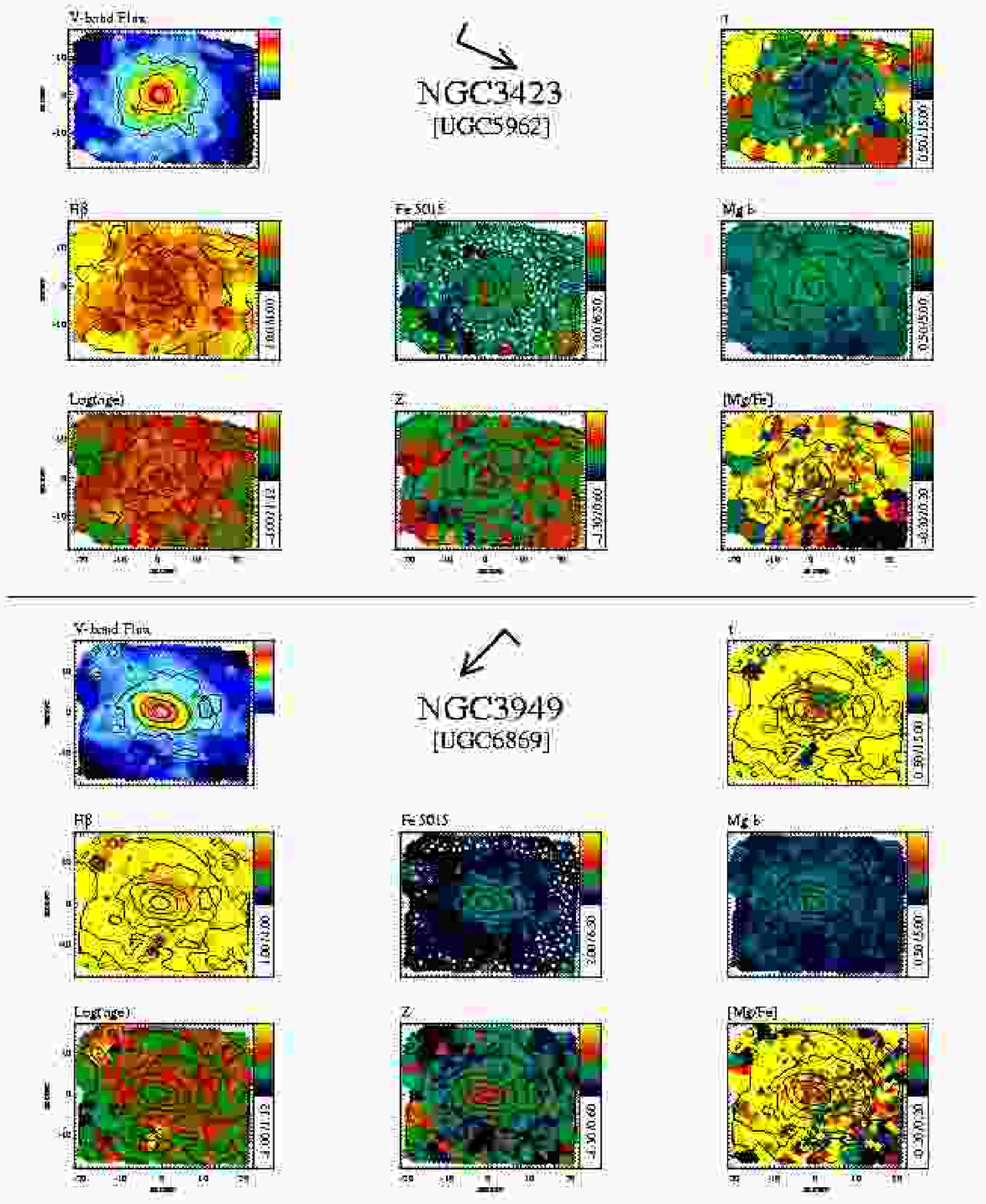}}
\end{center}
\caption{As in Figure \ref{maps1} for NGC\,3423 and NGC\,3949.}\label{maps5}
\end{figure*}
\addtocounter{figure}{-1}
\addtocounter{subfigure}{1}

\clearpage
\begin{figure*}
\begin{center}
{\includegraphics{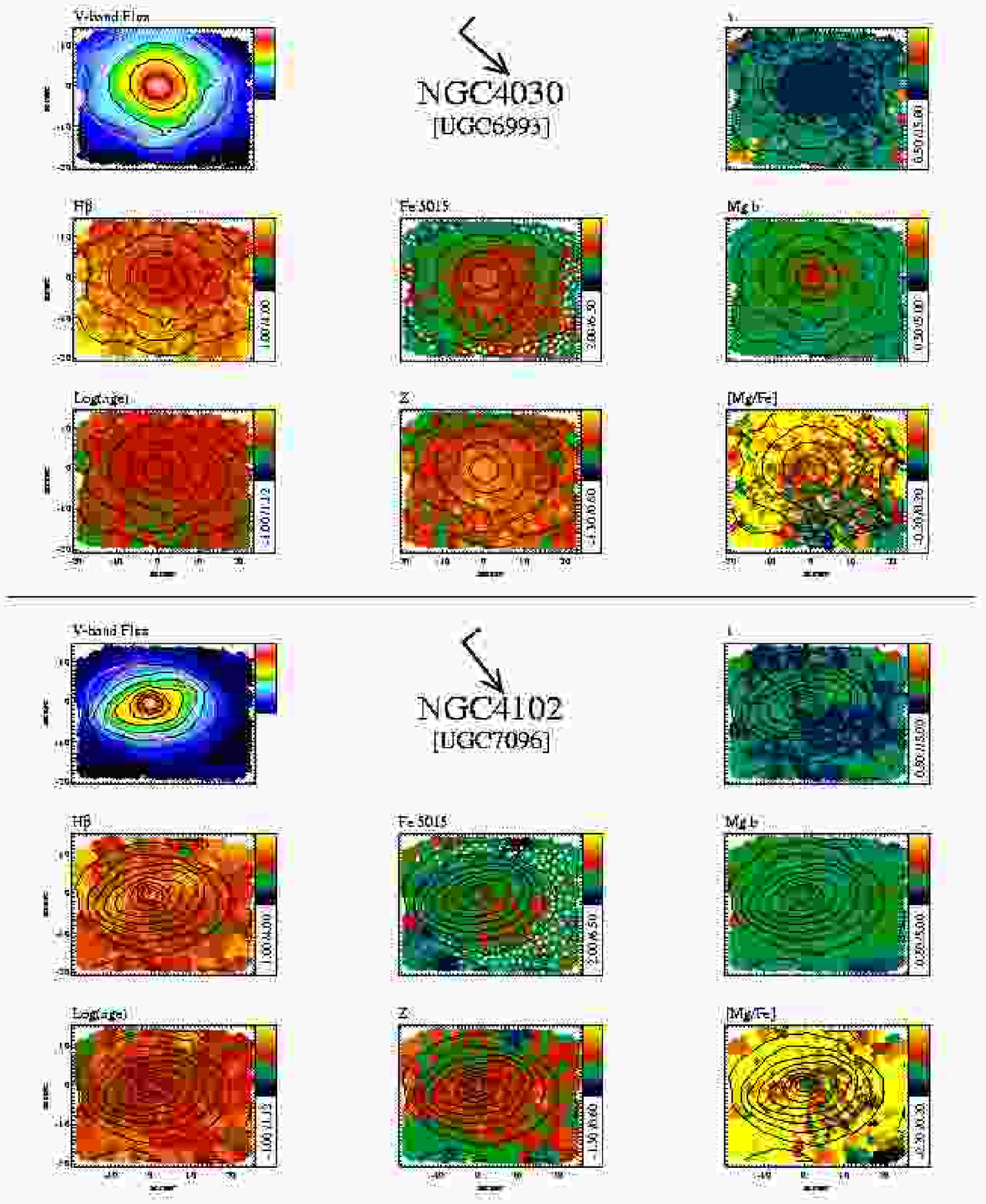}}
\end{center}
\caption{As in Figure \ref{maps1} for NGC\,4030 and NGC\,4102.}\label{maps6}
\end{figure*}
\addtocounter{figure}{-1}
\addtocounter{subfigure}{1}

\clearpage
\begin{figure*}
\begin{center}
{\includegraphics{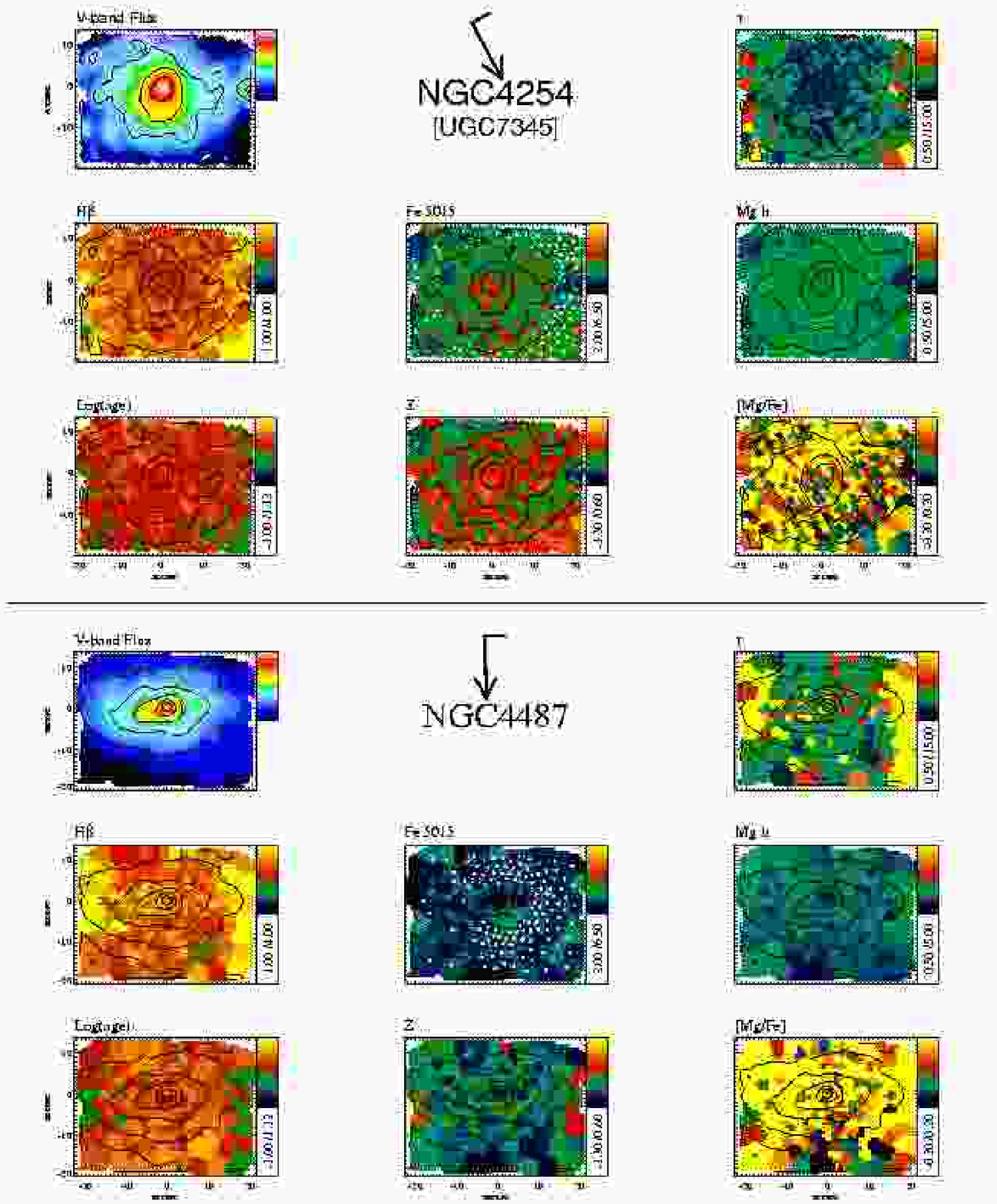}}
\end{center}
\caption{As in Figure \ref{maps1} for NGC\,4254 and NGC\,4487.}\label{maps7}
\end{figure*}
\addtocounter{figure}{-1}
\addtocounter{subfigure}{1}

\clearpage
\begin{figure*}
\begin{center}
{\includegraphics{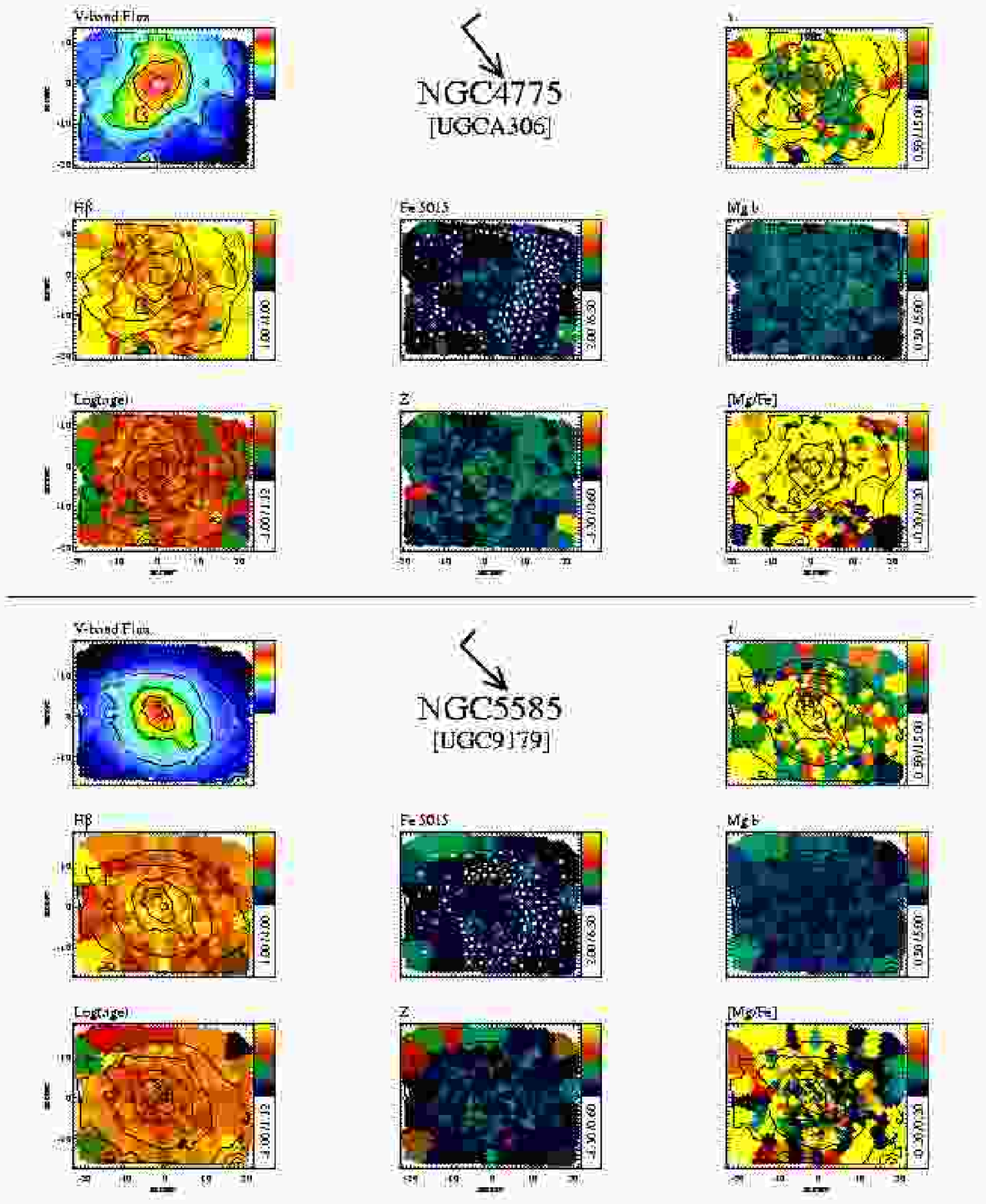}}
\end{center}
\caption{As in Figure \ref{maps1} for NGC\,4775 and NGC\,5585.}\label{maps8}
\end{figure*}
\addtocounter{figure}{-1}
\addtocounter{subfigure}{1}

\clearpage
\begin{figure*}
\begin{center}
{\includegraphics{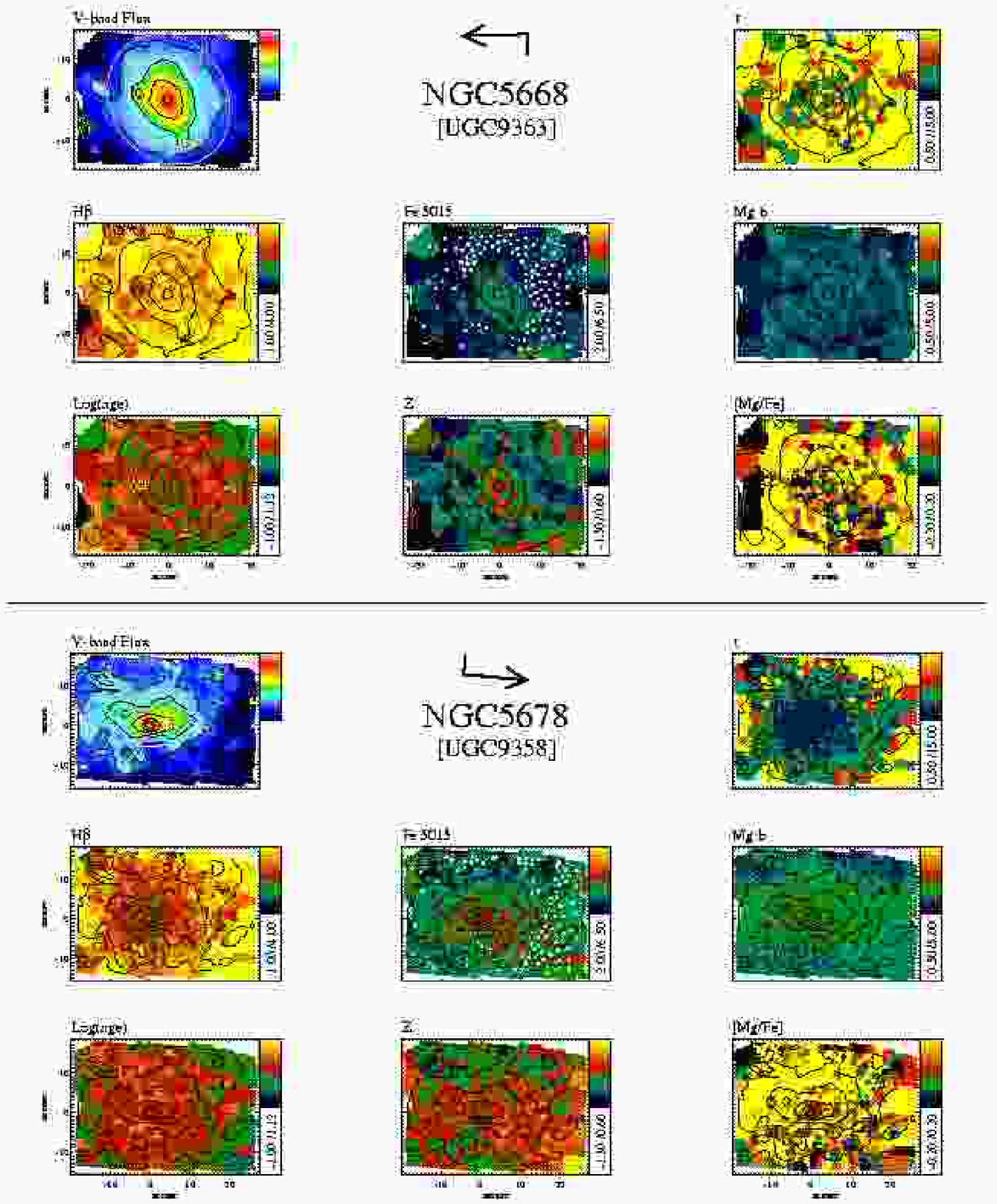}}
\end{center}
\caption{As in Figure \ref{maps1} for NGC\,5668 and NGC\,5678.}\label{maps9}
\end{figure*}
\setcounter{subfigure}{0}
\clearpage

\section{Correlations between line-strength indices and other galactic parameters}\label{correlationsec}
In this Section we analyse the dependence between the line-strength indices and other galactic parameters; we mainly focus on central properties. 
For each galaxy we defined three regions: the central circular aperture of 1\farcs5 radius (corresponding to 
the innermost nine single-lens spectra); a so-called `bulge region', defined as the region within which the light due to 
the fitted exponential disc falls below 50\% of the total light, with the exclusion of the central aperture; and a so-called `disc region', a 3\arcsec\/-thick elliptical annulus external to 
the `bulge region'\footnote{The inner semimajor axis of the `disc region' corresponds to the outer semimajor axis of the `bulge region' plus 2\arcsec\/. A slightly different definition of the disc region has been adopted in the case 
of NGC\,628, in order to avoid contamination from a foreground star situated $\approx$ 13\arcsec\/ south of the centre.}. We averaged the spectra in these regions 
before spatial binning, and on the resulting spectrum calculated the line-strength 
indices, after emission removal\footnote{For the aperture measurements, we never applied the continuum correction to the Fe5015 index 
described in Section
\ref{correctionsec}: in several cases, even the disc region is affected only partially by the correction; in addition to this, 
the bulge and disc region  
measurements are obtained averaging spectra from spatially separated areas, which helps in minimising the impact 
of the continuum problem in the
Fe5015 spectral region.}. The actual numbers for the definition of the bulge and disc regions (semi-major axis of the ellipses, position angles and
ellipticities) were taken from our own photometric analysis and
bulge-disc decomposition based on archival NIR images (Ganda
et al., in preparation). There we model the galaxies as an exponential disc and a S\'ersic (1968) bulge (which often turns out to be very tiny, with effective 
radius smaller than $\approx$ 5\arcsec\/) and apply a 
two-dimensional decomposition from which we can estimate the bulge extension (cfr. \citealt{edo} for the method and Ganda et al., in preparation, for the actual analysis of the
profiles). For clarity, in the top left map in Figures \ref{maps1}-\ref{maps9} we overplot in pink the elliptical contours defining the outer 
limits of the `bulge region' and of the `disc region'. In any case, we want to warn the reader that the `bulge' and `disc' regions here
introduced are mainly a tool for us to investigate radial 
variations, and do not refer to distinct components, since the so-called `bulge' is affected by a huge contamination from the disc, and an accurate decomposition 
is beyond the scopes of the present Paper, and will be addressed in detail in the mentioned paper in preparation.\\
\indent In Table \ref{tableproperties} we list the morphological type (from the RC3) and the stellar velocity dispersion, with its associated error, measured on the central aperture 
spectra, since we will make use of these quantities in the following Sections and Figures; the errors on the velocity dispersion are based on the scatter 
of the velocity dispersion values that we measure for the single-lens spectra within the central aperture, and do not take into account template mismatch. Table \ref{indicesvalues} lists the index values calculated, 
galaxy by galaxy, on the central aperture and on the bulge and disc regions. 
\begin{center}
\begin{table}
\begin{center}
 \begin{tabular}{c l c c c}
\hline \hline
NGC & Type &T &$\sigma$&$\pm$$\Delta\sigma$\\
\hline
\,\,488& SA(r)b & 3.0 & 197&4\\
\,\,628& SA(s)c & 5.0 & 52&6\\
\,\,772& SA(s)b & 3.0 & 120&4\\
\,\,864& SAB(rs)c & 5.0  & 65&20\\
1042& SAB(rs)cd & 6.0& 55&13\\
2805& SAB(rs)d & 7.0 & 46&18\\
2964& SAB(r)bc & 4.0 & 102&11\\
3346& SB(rs)cd & 6.0 & 48&23\\
3423& SA(s)cd & 6.0 &48&20\\
3949& SA(s)bc & 4.0 &60&10\\
4030& SA(s)bc & 4.0 &100&2\\
4102& SAB(s)b? & 3.0 &153&8\\
4254& SA(s)c & 5.0  &72&13\\
4487& SAB(rs)cd & 6.0  &51&21\\
4775& SA(s)d & 7.0 & 41&21\\
5585& SAB(s)d & 7.0 &37&23\\
5668& SA(s)d & 7.0&52&22\\
5678& SAB(rs)b & 3.0 &102&9\\
\hline
\end{tabular}
\end{center}
\caption{Hubble type (RC3 through NED); numerical morphological type (from RC3, indicated as `T-type' in some of the following
pictures); stellar velocity dispersion (in km s$^{-1}$) and its associated error, also in km s$^{-1}$, measured on the central aperture spectrum, for all galaxies.}
\label{tableproperties}
\end{table}
\end{center}
\begin{center}
\begin{table*}
\begin{center}
 \begin{tabular}{@{}c|c c c c c|c c c c c |c c c c c}
\hline \hline
NGC & &&H$\beta_{\mathrm{centre}}$ &H$\beta_{\mathrm{bulge}}$ &H$\beta_{\mathrm{disc}}$ &&& Fe5015$_{\mathrm{centre}}$ &   Fe5015$_{\mathrm{bulge}}$ &  Fe5015$_{\mathrm{disc}}$&&&Mg{\textit{b}}$_{\mathrm{centre}}$ & Mg{\textit{b}}$_{\mathrm{bulge}}$& Mg{\textit{b}}$_{\mathrm{disc}}$\\
\hline
\,\,488&&&1.961&1.966&1.961&&&6.122 & 5.635&  5.615&&&4.387& 4.112&4.042\\
\,\,628&&&2.713&2.604&2.486&&&5.247&   4.556&  3.936&&&2.561& 2.424&2.274\\
\,\,772&&&2.316&2.355&2.765&&&5.495&   4.746&  4.506&&&3.285& 3.028&2.558\\
\,\,864&&&2.167&2.380&2.404&&&2.697&   3.115&  3.686&&&1.564& 1.859&2.378\\
1042&&&3.327&2.533&2.172&&&4.570&   4.046&  3.767&&&2.200& 2.498&2.356\\
2805&&&3.324&2.878&3.013&&&3.975&   3.400&  3.303&&&2.343& 2.121&2.058\\
2964&&&3.033&2.957&2.808&&&3.313&   3.476&  4.025&&&1.736& 2.049&2.264\\
3346&&&3.488&3.289&2.838&&&3.732&  3.643&  3.915&&&1.939& 2.002&2.247\\
3423&&&2.846&2.935&3.299&&&4.334&   3.528&  3.092&&&2.137& 1.869&1.692\\
3949&&&3.944&3.846&3.726&&&3.760&   3.623&  3.175&&&1.744& 1.772&1.505\\
4030&&&2.769&3.086&3.343&&&5.287& 4.512&   3.886&&&2.996&2.524& 2.144\\
4102&&&3.283&3.257&3.173&&&4.083&   3.950&  3.736&&&1.904& 2.143&2.121\\
4254&&&2.676&2.955&3.177&&&4.121&   3.954&  3.699&&&2.470&2.149&1.864\\
4487&&&3.167&3.452&3.275&&&2.410&  2.692&  3.175&&&1.319&1.487&1.666\\
4775&&&3.560&3.482&3.465&&&3.071&   2.980& 2.376&&&1.609&1.564&1.412\\
5585&&&3.816&3.490&3.236&&& 2.643&   2.453&  2.354&&&1.198&1.201&1.351\\
5668&&&3.661&3.584&3.825&&&3.756&   3.037&  2.546&&&1.717&1.464&1.265\\
5678&&&3.015&2.617&2.881&&&4.566&   4.328&  4.112&&&2.367&2.484&2.380\\
\hline
\end{tabular}
\end{center}
\caption{Line-strength indices calculated on the central apertures and bulge and disc regions, defined as explained in Section
\ref{correlationsec} for the 18 galaxies. The indices are intended as equivalent widths and expressed in \AA.}
\label{indicesvalues}
\end{table*}
\end{center}
\subsection{Index - index relations}\label{indexvsindexsec}
\renewcommand{\thefigure}{\arabic{figure}}
\begin{figure}
{\includegraphics[width=0.99 \linewidth]{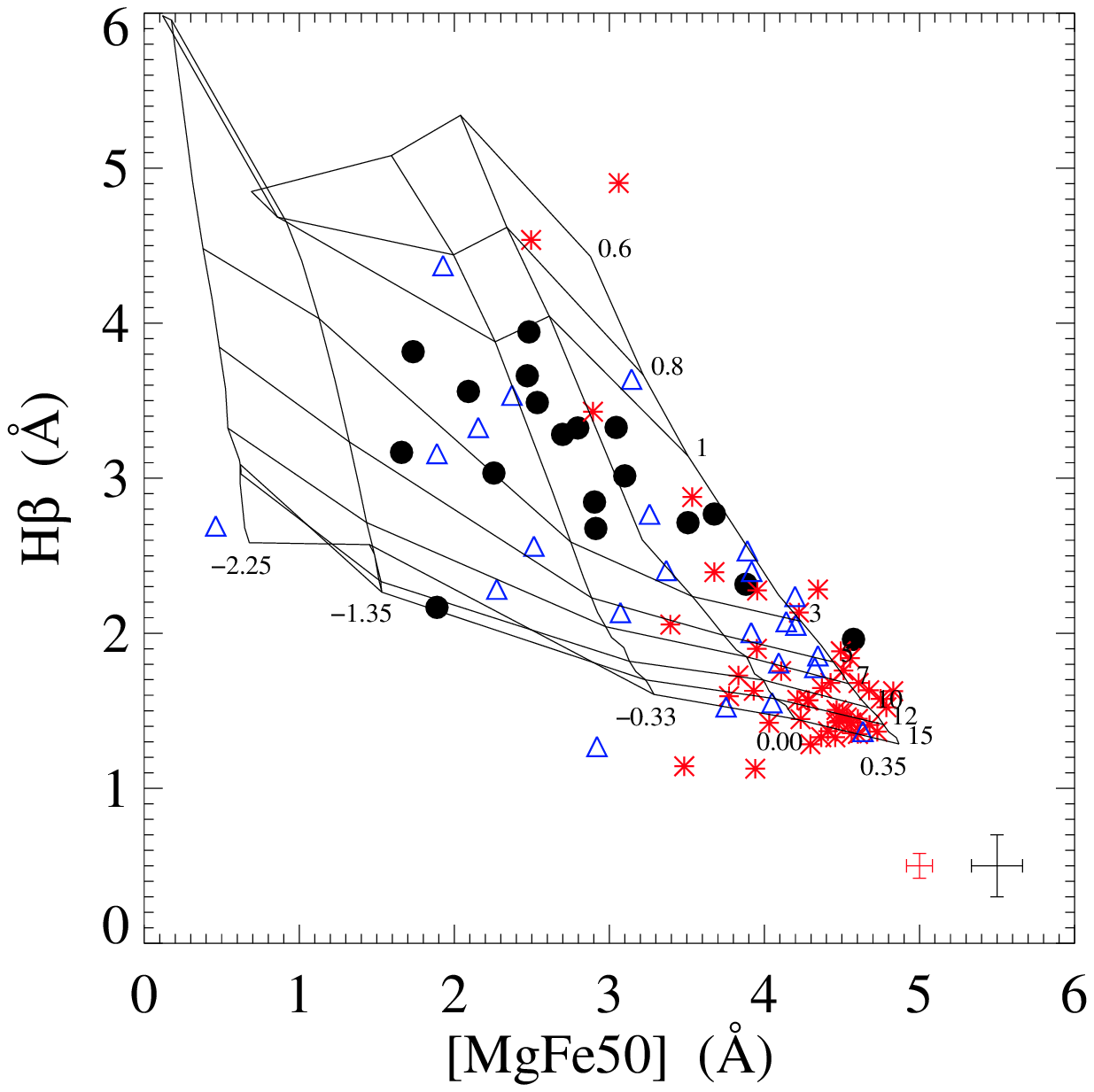}}
\caption{H$\beta$ index against [MgFe50] (in \AA), for three samples of galaxies observed with {\tt{SAURON}}: the black symbols represent the central apertures of our 
late-type spirals, the red symbols are the E and S0 (central aperture of 1\farcs335 radius) and the blue ones the Sa galaxies (central aperture). The solid lines represent
SSP models from \citet{thomas}, limited to models with solar abundance ratios; ages of the models vary from 0.6 to
15 Gyr (top to bottom); metallicity (in decimal logarithm) from $-$2.25 to +0.35 (left to right). Representative error bars are added at the bottom right of the 
panel; the black one refers to both the early- and late-type spirals samples, while the red one, smaller, refers to the E/S0 galaxies.}
\label{corr_mgfe50}
\end{figure}
In order to derive `zero-order' estimates on the luminosity weighted ages and metallicities of our galaxies, we show in Fig. \ref{corr_mgfe50} an 
age/metallicity diagnostic diagram. Here we use the abundance ratio insensitive index
\begin{equation}
\mathrm{[MgFe50]}= \frac{0.69 \times \mathrm{Mg{\textit{b}}} + \mathrm{Fe5015}}{2}
\end{equation} 
(for details on the index definition see Kuntschner et al., in preparation) as metallicity indicator and H$\beta$ as age indicator. 
In  Fig. \ref{corr_mgfe50} we plot the indices calculated on the central 1\farcs5 aperture of our galaxies (black symbols). 
We overplot with red symbols the elliptical and
lenticular galaxies from the {\tt SAURON} survey (based on the data presented in \citet{paper6}), with reference to measurements within apertures of
radius 1\farcs335, that the authors 
of that paper extracted in order to match the Lick aperture (see \citet{paper6} for details); the blue symbols represent central apertures of 1\farcs2 radius for the early-type spirals from the same survey \citep{reynier}. 
Overplotted in the Figure are also the models of Thomas, Maraston \& 
Bender (2003), limited
to the models with [Mg/Fe]=0; they are labelled with the corresponding age and metallicity. As noted in \citet{reynier}, there is a smooth transition between 
the ellipticals/lenticulars and the early-type spirals, 
with most E/S0 populating the region of the old, metal-rich models, and the Sa spanning a larger range in age and reaching lower metallicities. Our 18 
later-type spirals occupy a region of the diagram that has little intersection with the E/S0s, having generally younger ages and being clearly 
more metal poor. We also notice that the parameter region occupied by our galaxies is completely included in the one spanned by the Sa galaxies. We warn the reader that the 
interpretation in terms of ages and metallicities is here done on the basis of SSP models, which might not provide a good description 
of spiral galaxies, for which the star formation history could be characterised by several bursts spread over time or by a constant star formation rate. A more detailed analysis of the age, 
metallicities and the star formation history is presented in Section \ref{agesection}.\\
\indent In Fig. \ref{corr_index} the left panel shows the line index Fe5015 against Mg{\textit{b}} for the central aperture of our 18 galaxies (black symbols); 
as in Fig. \ref{corr_mgfe50}, we show also the points corresponding to the E/S0 (red symbols) and Sa (blue) of the {\tt SAURON} survey. 
The solid lines represent the 
SSP model grid from \citet{thomas} with [Mg/Fe]=0; the dashed lines correspond instead to the Thomas models with [Mg/Fe]=0.5. As also shown in \citet{reynier} with 
a similar figure, it is clear that most of the ellipticals can be fitted better by models over-enhanced in $\alpha$-elements, while the 
Sa bulges present a range in
enhancement. Our later-type spirals instead are in all cases compatible with models with solar-abundance ratios. In the middle and left panels we show, only for our
sample of late-type spirals, the measurements for the previously defined bulge and disc regions respectively in the (Mg{\textit{b}}, Fe5015) plane, along with the 
models from \citet{thomas}, drawn only in the case of solar abundance ratios. We see that our spirals are compatible with solar abundance ratios, 
within the uncertainties; we do not detect relevant 
global variations in the 
appearance of the diagram when considering the different regions (central aperture, bulge and disc region): the single galaxies move in the diagram, but they
remain in the region of the models with solar abundance ratios. This differs from what seen in the early-type spirals of the {\tt SAURON} survey: there, away from the centre
($\approx$ 10\arcsec\/ from the centre) the galaxies have
generally higher [Mg/Fe] than in the centre. Super-solar abundance ratios are possibly caused by the presence of populations with several ages. An increase of [Mg/Fe] when moving outwards is detected also by \citet{jablonka07} and by \citet{moorthy} 
for samples of galaxies of type varying between S0 and Sc. This might be due to a change from SSP-like stellar populations in the 
centre to populations with a large range in age outside. In particular, \citet{jablonka07} measures small increments in abundance ratio from the central region 
to the bulge effective radius; these gradients stay rather constant among bulges; \citet{moorthy} find that most of their galaxies have positive or zero 
gradient in [Mg/Fe] within the bulge-dominated region; as mentioned in the Introduction, they divide their sample in red
(B$-$K $>$ 4) and blue bulges, and, according to their analysis, 
the red bulges are overabundant in the centre, while the blue bulges have
solar abundance ratios in the centre, and the disc-dominated regions also have approximately solar abundance ratios. Therefore, our results are not necessarily in
contradiction with theirs, as we will also see in Appendix \ref{gradientssec}, where we briefly discuss the radial profiles of the indices and the population
parameters.
\begin{figure*}
\begin{center}
{\includegraphics[width=0.99 \linewidth]{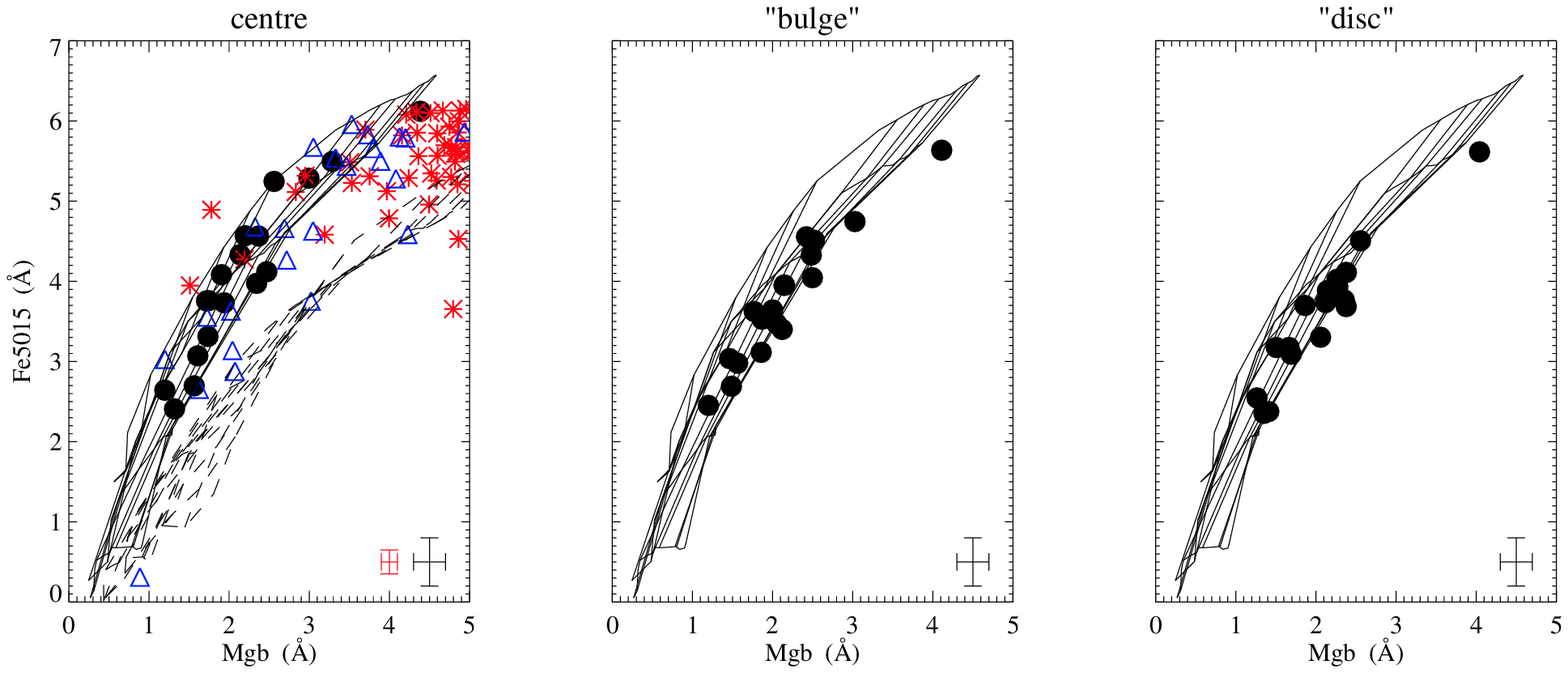}}
\caption{Left panel: index - index diagram showing Fe5015 against Mg{\textit{b}} for the central aperture of our late-type spirals (black symbols), while the red symbols are the E and S0 
(within a 1\farcs335 central aperture) and the blue ones the Sa (central aperture), observed with {\tt SAURON}. The solid lines represent the SSP models from 
\citet{thomas}, limited to models with solar abundance ratios; the dashed lines correspond instead to the models with
[Mg/Fe]=0.5. Middle panel: same as the previous, 
for the `bulge region' apertures of the 18 late-type spirals only. 
Right panel: same as the previous, for the `disc region' apertures of the 18 late-type spirals only; in the middle and right panels we only plot the model 
grid for solar abundance ratios. All the indices are expressed as equivalent width and measured in \AA ; the three panels share the same vertical axis. Representative error 
bars are added at the bottom right of each panel; the black one refers to both the early- and late-type spirals samples, while the red one refers to the E/S0 galaxies.}
\label{corr_index}
\end{center}
\end{figure*}

\subsection{Index - Hubble type relations}
\begin{figure*}
{\includegraphics[width=0.99 \linewidth]{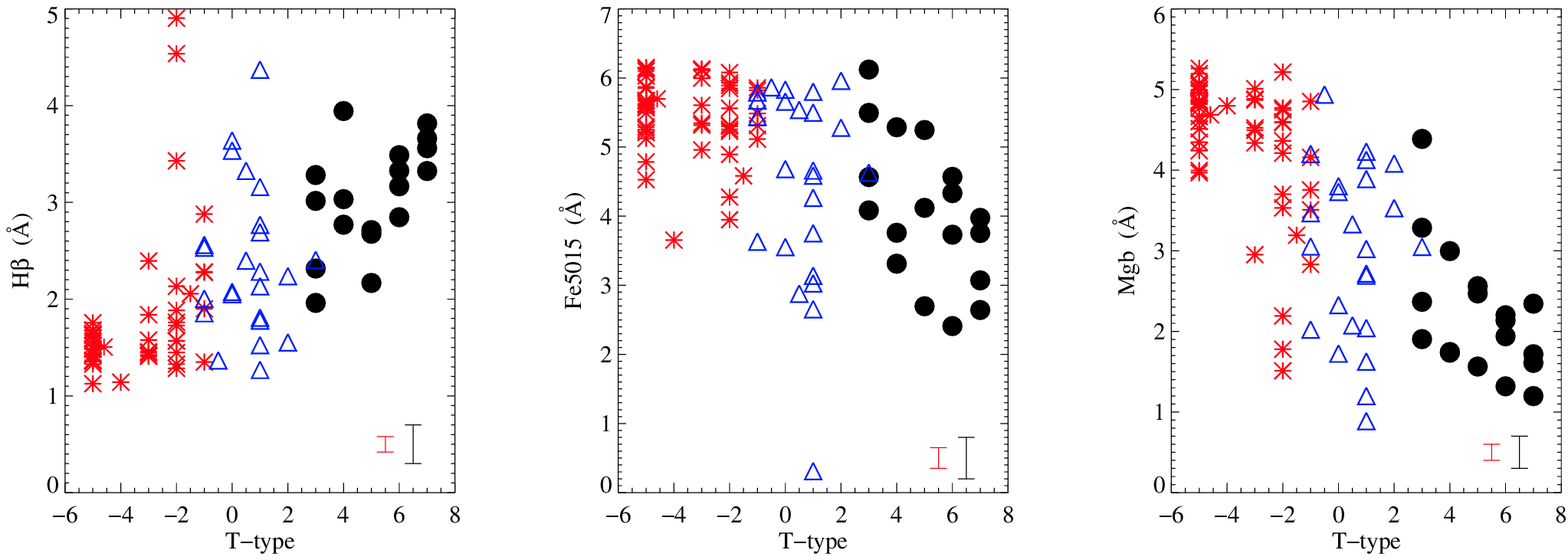}}
\caption{Line-strength indices (in \AA) as function of morphological type. Black symbols are our sample galaxies (central aperture),
red are the E and S0 (central 1\farcs335 aperture) and blue the Sa galaxies (central aperture) of the {\tt SAURON} survey. The left panel refers to H$\beta$, the middle one to 
Fe5015 and the right one to Mg{\textit{b}}. Representative error 
bars are added at the bottom right of each panel; the black one refers to both the early- and late-type spirals samples, while the red one refers to the E/S0 galaxies.}
\label{corr_type}
\end{figure*}
The three panels in Fig. \ref{corr_type} present the relation between the central line indices (respectively, H$\beta$, Fe5015 and Mg{\textit{b}}) and the 
morphological type (see Table \ref{tableproperties}). Again, we use black symbols for the 18 late-type spirals under investigation, red ones for the ellipticals and lenticulars and 
blue for the early-type spirals of the {\tt SAURON} survey. We see an overall increase of H$\beta$ and a decrease of Fe5015 and Mg{\textit{b}} going towards later types. 
We also see that for ellipticals the range spanned in equivalent width is in general quite small, it becomes larger for lenticulars and Sa galaxies and smaller again in
the later-type spirals. This result is illustrated quantitatively in Table \ref{indicesranges}, where we list for each index the range in equivalent width 
spanned in the three samples, together with the minimum and maximum values assumed by the indices: as one can see, the range in H$\beta$ is larger for the E/S0 than
for the Sa sample, while for the other indices the range is larger for the early-type spirals; in all three indices, the Sb-Sd galaxies span 
a narrower range than the Sa. In the same table, we also list the average value of the indices in the three samples, together with their standard deviation 
and the standard deviation after subtraction of the observational errors, which indicates the amount of scatter that cannot be explained by the errors. 
\begin{center}
\begin{table*}
\begin{center}
 \begin{tabular}{@{}l c c c}
\hline \hline
index & E/S0 & Sa & Sb-Sd\\
\hline
H$\beta$$_{range}$ & 3.78 [1.13-4.90]& 3.10 [1.27-4.37] & 1.98 [1.96-3.94]\\
Fe5015$_{range}$ & 2.50 [3.66-6.15]& 5.65 [0.31-5.96]& 3.71 [2.41-6.12]\\ 
Mg{\textit{b}}$_{range}$ & 3.75 [1.51-5.26]& 4.05 [0.89-4.94]& 3.19 [1.20-4.39]\\
\hline
$\langle$H$\beta$$\rangle$ $\pm$ $\sigma$& 1.79 $\pm$ 0.75 (0.67)& 2.39 $\pm$ 0.77 (0.57)& 3.06 $\pm$ 0.56 (0.36)\\
$\langle$Fe5015$\rangle$ $\pm$ $\sigma$&  5.45 $\pm$ 0.56 (0.41)& 4.51 $\pm$ 1.40 (1.10)& 4.06 $\pm$ 1.04 (0.74)\\
$\langle$Mg{\textit{b}}$\rangle$ $\pm$ $\sigma$& 4.33 $\pm$ 0.87 (0.77)& 2.99 $\pm$ 1.07 (0.87)& 2.19 $\pm$ 0.77 (0.57)\\
\hline
\end{tabular}
\end{center}
\caption{Upper part: ranges spanned in equivalent width (\AA) by the three indices, for the three samples represented in Fig. \ref{corr_type}; between brackets are 
indicated the minimum and maximum values. Lower part: average values of the indices in the three samples, 
with the relative standard deviations; the numbers in
parenthesis indicate the standard deviation after subtraction of the observational error.}
\label{indicesranges}
\end{table*}
\end{center}
\begin{figure*}
{\includegraphics[width=0.99 \linewidth]{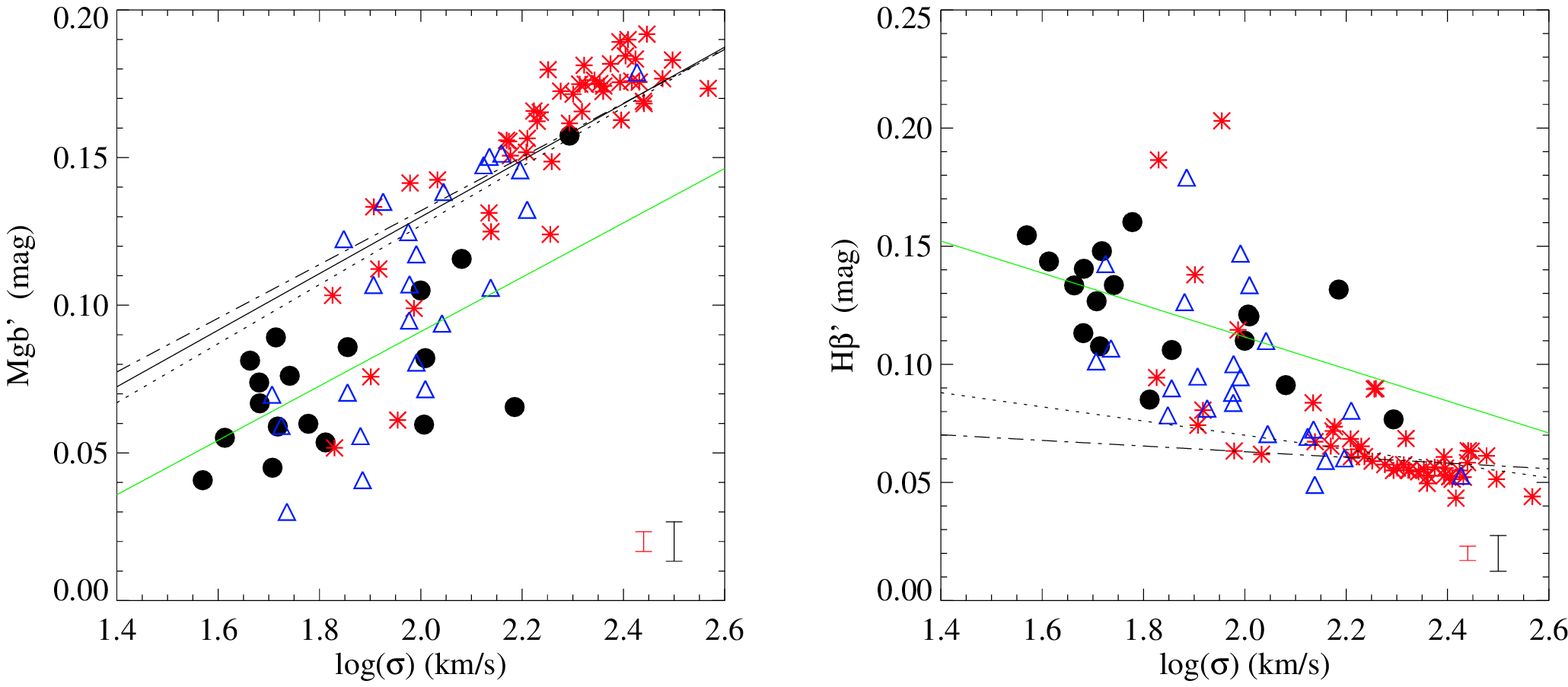}}
\caption{Left panel: central Mg{\textit{b}}$^{'}$, expressed in magnitudes, against central velocity dispersion, in 
units of decimal logarithm. The black symbols represent our own sample (central aperture),
the red the 
E and S0 and the blue the Sa galaxies of the {\tt SAURON} survey. Right panel: central H$\beta ^{'}$, expressed in magnitudes, 
against central velocity dispersion, in units of decimal logarithm; the colour code for the symbols is the same as in the previous case. 
The black solid line overplotted in the left panel represents the relation found by \citet{Jorgensen} for
early-type galaxies in 10 cluster. The dotted and dash-dotted black lines overplotted in both panels are the relations obtained by \citet{patricia06} 
for low- and high-density environments, respectively, and the green solid lines are the relations determined using our own sample; see text in Section \ref{indexsigma} for details. Representative error 
bars are added at the bottom right of each panel; the black one refers to both the early- and late-type spirals samples, while the red one refers to the E/S0 galaxies.}
\label{corr_sig}
\end{figure*}

\subsection{Index - velocity dispersion relation}\label{indexsigma}
Figure \ref{corr_sig} shows the central Mg{\textit{b}} and H$\beta$ indices against the central velocity dispersion of the stars (see Table
\ref{tableproperties}), 
in the left and right panels respectively. 
As in the previous Figures, the black symbols represent the late-type galaxies, the red the ellipticals and lenticulars and the blue the Sa observed with 
{\tt SAURON} (central apertures). 
In these plots we express the indices in magnitudes, as it is often done in the literature, using the definition:
\begin{equation}
\mathrm{index'} = \mathrm{-2.5 \times \log \left(1 - \frac{index}{\Delta\lambda}\right)},
\end{equation}
where index and index$^{'}$ are measured respectively in \AA\, and magnitudes and $\Delta\lambda$ is the width of the index bandpass (cfr. \citealt{paper6} for
the full index definition).\\
\indent A tight and well known relation between magnesium indices (Mg{\textit{b}}$^{'}$ or Mg$_2$) and $\sigma$ holds for ellipticals (see for 
example \citealt{terlevich}, \citealt{burstein}, \citealt{bender}, \citealt{Jorgensen}, \citealt{colless}, 
Falc\'on-Barroso, Peletier \& Balcells 2002, \citealt{bernardi}, \citealt{worthey}, 
\citealt{denicolo}).
In Fig. \ref{corr_sig} (left panel) we overplot with a black solid line the relation obtained by \citet{Jorgensen} for a sample of 207 E and S0 
galaxies in 10 clusters:
\begin{equation}
\mathrm{Mg_{2}} = 0.196 \times \log(\sigma) - 0.155, 
\label{jorgensenequation}
\end{equation}
after having converted the Mg$_2$ index to Mg{\textit{b}}$^{'}$ using a least-squares fit to all the models from \citet{vazdekis96} with 
 Mg$_2$ $>$ 0.10, holding the relation:
\begin{equation}
\mathrm{Mg{\textit{b}}^{'}} = \mathrm{0.489\times Mg_{2} +0.014} .
 \end{equation}
One could think that this transformation, using models that do not take into account differences in abundance ratios, could be misleading. But in practice the 
Mg{\textit{b}}$^{'}$ - Mg$_2$ relation has very little scatter, even for different [Mg/Fe]. In any case, we also compare the position of our spirals in the 
($\log\left(\sigma\right)$, Mg{\textit{b}}$^{'}$) plane with the relations determined by \citet{patricia06} in a fully empirical way for a sample of 98 early-type galaxies 
drawn from different environments (the field, poor groups, Virgo, Coma and some Abell clusters). The black dotted and dash-dotted lines overplotted in Fig. \ref{corr_sig} represent their fitted Mg{\textit{b}}$^{'}$ - 
$\sigma$ relation for low- and high-density environments, respectively, and lie both very close to the relation of \citet{Jorgensen}, especially for high velocity
dispersions; similarly, in the right panel we overplot their H$\beta$ - $\sigma$ relations. The relations of \citet{Jorgensen} and
\citet{patricia06} are both determined performing a fit down to velocity dispersions below 100 km s$^{-1}$ ($\approx$ 40 km s$^{-1}$ in the case of \citealt{patricia06}), 
therefore exploring the range of velocity dispersions that we test with our late-type spirals. The small differences between these relations might be due 
to the different statistics and to the different environments: \citet{Jorgensen} themselves noticed that the relation differs slightly from cluster to cluster, in the sense that galaxies in clusters with lower $\sigma$ 
have systematically lower Mg$_2$. We also performed a linear fit to our own data: the green solid line overplotted in both panels represents the corresponding relation.\\
\indent In Table \ref{relationscoeff} we list the coefficients defining the relations taken from the literature, together 
with the coefficients determined for our own sample, in the form:
\begin{equation}
\mathrm{index'} = \mathrm{ a_{index'} + b_{index'} \times \log (\sigma)}.
\end{equation}
\indent As already pointed out by \citet{reynier} for the Sa galaxies of the {\tt SAURON} survey (blue symbols in the Figure), it is clear from Fig. \ref{corr_sig} and from Table 
\ref{relationscoeff} that the Mg{\textit{b}}$^{'}$ - $\sigma$ relation for ellipticals represents an upper envelope for the 
spirals; some of the spirals follow the relation (particularly if we consider the Sa), but the majority of them 
lie below it; in particular, the linear fit to our own data (green line) runs almost parallel to the line from \citet{Jorgensen}, but significantly below it.
Similarly, in the case of the H$\beta^{'}$ - $\sigma$ relation, the ellipticals represent a lower envelope \citep{reynier} and many spirals lie above the 
relation. Deviations from the relations for early-type galaxies may be driven by stochastic processes like star formation: as argumented by \citet{schweizer}, the line defined by the 
galaxies of \citet{Jorgensen} corresponds to old stellar populations, while the deviations would be due to younger stars. For our objects (black dots) the effect of young populations as deviation from the line defined by E/S0s galaxies is evident.\\
\begin{center}
\begin{table}
\begin{center}
 \begin{tabular}{@{}l c c c c}
\hline \hline
REF &  a$_{Mg{\textit{b}}^{'}}$&  b$_{Mg{\textit{b}}^{'}}$ & a$_{H\beta^{'}}$ & b$_{H\beta^{'}}$\\
\hline
Jor & $-$0.062 & 0.096 & $-$ & $-$ \\
San$_{LD}$ & $-$0.073 & 0.1 & 0.130 & $-$0.030 \\ 
San$_{HD}$ & $-$0.050 & 0.091 & 0.087 & $-$0.012\\
KG & $-$0.093 & 0.092 & 0.247 & $-$0.068\\
\hline
\end{tabular}
\end{center}
\caption{Coefficients of the index$^{'}$ - $\sigma$ relation, in the form: index$^{'}$ = a$_{index'}$ + b$_{index'} \times \log(\sigma)$; the first column lists the 
references (Jor=\citealt{Jorgensen}; San$_{LD}$=\citealt{patricia06}, low-density environment; San$_{HD}$=\citealt{patricia06}, high-density environment; KG= this work), the second and third the coefficients for the relation with Mg{\textit{b}}$^{'}$, the fourth and fifth the coefficients for the 
relation with H$\beta^{'}$.}
\label{relationscoeff}
\end{table}
\end{center}

\section{Age, metallicity, star formation history}\label{agesection}
The ultimate aim of measuring line-strengths is to understand the distribution of 
stellar ages, metallicities and abundance ratios, and to get hints on the star formation histories. Using a combination of line-strength indices, 
one can constrain stellar population models, 
which allow to translate the line-strengths values into measurements of age, 
metallicity and abundance ratio.\\
\indent Trying to address this issue, we investigated two main 
approaches: from one side, one can assume that a galaxy can be viewed as a Single Stellar Populations (SSPs), 
and compare observations with theoretical SSPs in order to determine the best-fitting population; this is a `classical' approach 
to stellar population studies. From the other side, one can 
view a galaxy's population as the time evolution of an initial SSP; making some assumptions on the initial metallicity and 
the metal enrichment history, by comparison of observations and models one can retrieve information on the star formation history. 
Nowadays, this is a topic of greatest interest: recently, several groups have been working on it and developed tools. Here we mention 
noticeable examples such as {\tt STECMAP} and its extension {\tt STECKMAP}\,\footnote{http://astro.u-strasbg.fr/Obs/GALAXIES/stecmap\_eng.html} 
and {\tt MOPED}. {\tt STECMAP} and {\tt STECKMAP} have been 
developed by Ocvirk and collaborators (for a description, see \citealt{pierre1}, \citealt{pierre2}), and try to recover the star formation
history via spectral fitting, without any \textit{a priori} assumption on the star formation history; the only condition imposed is that 
the star formation history is a smooth function of time; 
{\tt MOPED}, developed by Panter and collaborators, is a method that allows to analyse huge quantities of spectra using the entire
spectral range, and assumes that a galaxy can be viewed as superposition of single bursts (for details see Panter, Heavens \& Jimenez 2003).\\
\indent In this Section we will follow these two roads (SSP and recovery of star formation history), on the basis of our 
measured Lick indices.

\subsection{Single stellar population analysis}\label{singlessp}
\begin{figure*}
{\includegraphics[width=0.99\linewidth]{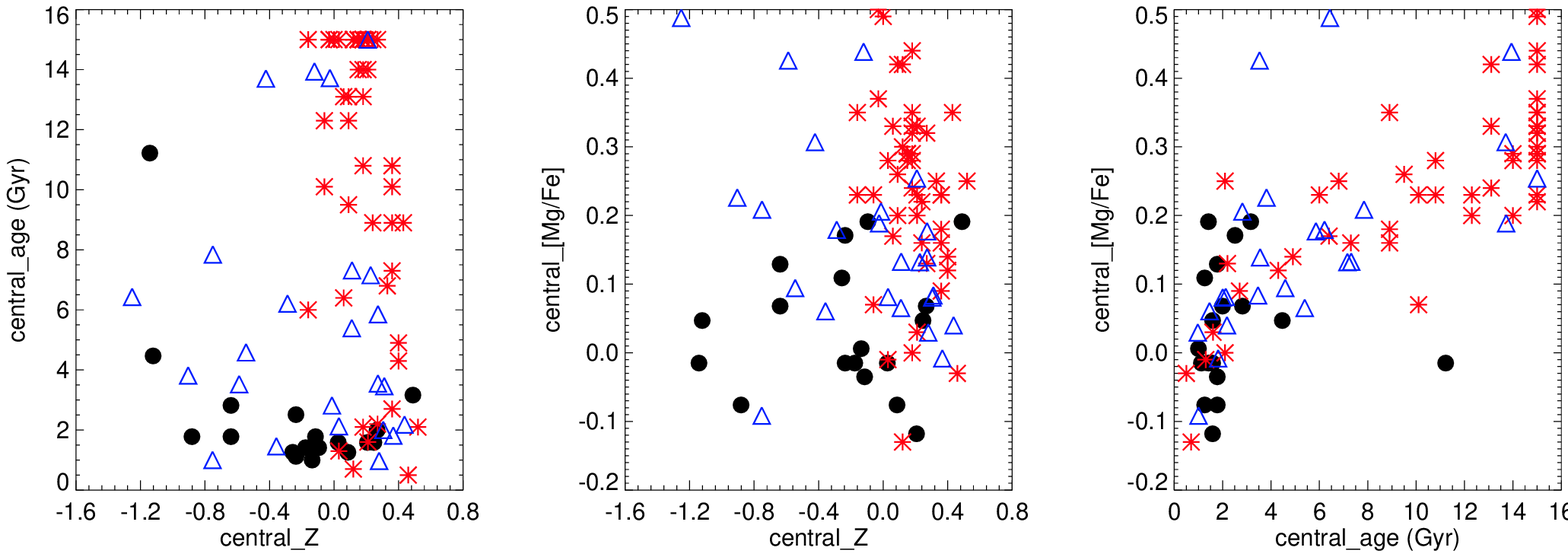}}
\caption{Central aperture values of the stellar population parameters from our one-SSP analysis. Left panel: central age (in Gyr) against central metallicity (in decimal
logarithm, with the solar metallicity as zero point); middle panel: central abundance ratio against central metallicity; right panel: central abundance ratio against central age. In black we represent our sample
galaxies, while in red the E and S0 and in blue the Sa galaxies observed with {\tt SAURON}.}
\label{1ssp_fullrange}
\end{figure*}

\begin{figure*}
\begin{center}
{\includegraphics[width=0.79 \linewidth]{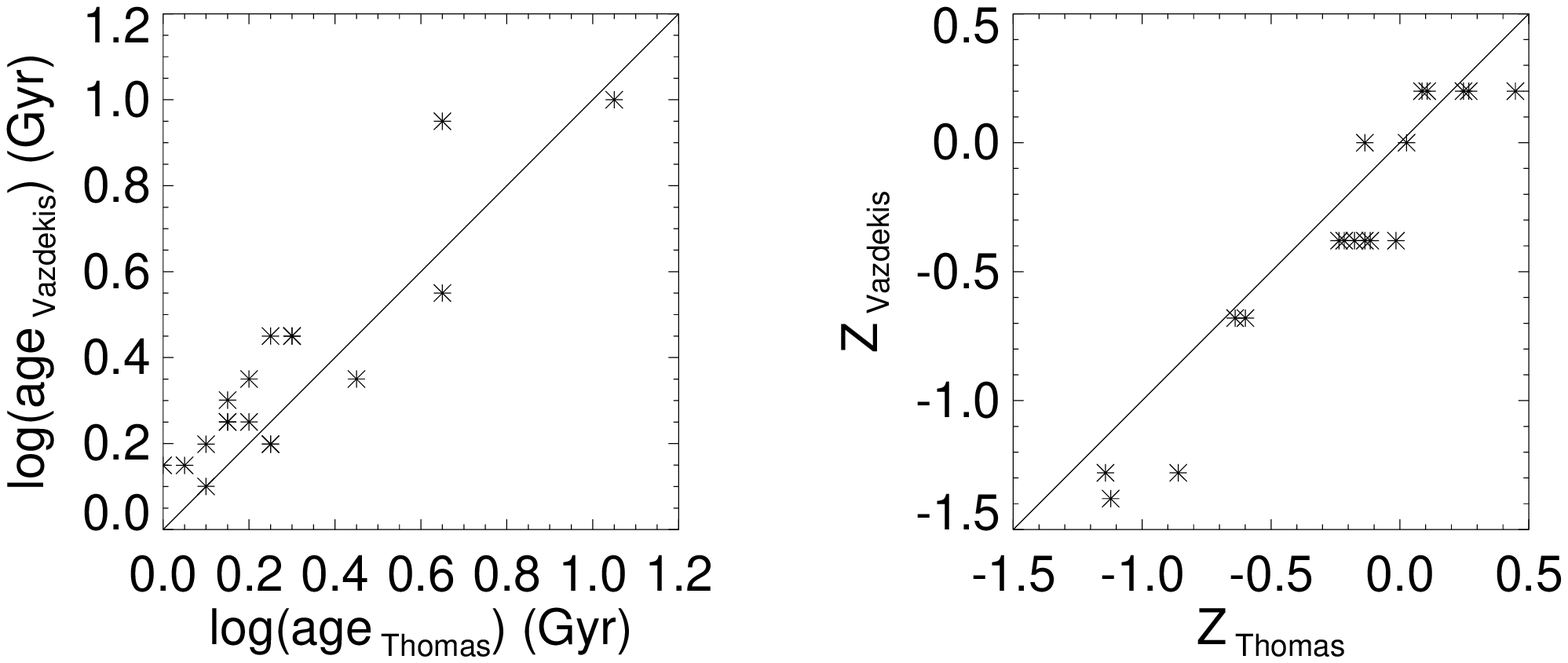}}
\caption{Comparison between the ages (left panel, in Gyr) and metallicities (right panel) obtained from our one-SSP approach using the models of \citealt{thomas} (horizontal 
axis, limitately to nearly-solar abundance ratios) and \citealt{vazdekis} (vertical axis). The solid lines overplotted represent the 1:1 relation.}
\label{compare_vaz_thomas}
\end{center}
\end{figure*}

Following the examples of \citet{richard} and \citet{reynier}, we used the single-burst stellar population models 
of \citet{thomas} to compare with our observations and estimate the ages, metallicities and abundance ratios of our galaxies. 
To avoid strong discretisation 
effects on the derived parameters, we interpolated the original grid of model indices, obtaining a cube of $\approx$ 225000
individual models, with  $-$2.25 $\leq$ Z $\leq$ +0.67, 0.1 $\leq$ age $\leq$ 15 Gyr and $-$0.2 $\leq$ [Mg/Fe] $\leq$ 0.5. 
We then determined the model closest to our observations (the set of three indices) for each Voronoi bin in our galaxies, via a $\chi^2$ minimisation technique, 
consisting in minimising 
contemporaneously the distance between all of the observed and the model indices, weighted with the observational error on the line-strengths 
(see Section \ref{correctionsec} for the actual figures). We attributed to the galaxy bin 
the age, metallicity and abundance ratio of the selected model. The third row in Figures \ref{maps1}-\ref{maps9} shows the age, 
metallicity and abundance ratio maps obtained in this way. 
We are aware that it is an over-simplification to represent a galaxy's population using a SSP, however, we want to
apply this method as a `zero-order' estimate, before moving to somewhat more 
sophisticated ones. Fig. \ref{1ssp_fullrange} plots against each other the age, metallicity and abundance ratio values calculated over 
the central aperture
spectrum. Our sample galaxies are represented with black symbols. Overplotted with red symbols are the ellipticals and lenticular galaxies 
of the {\tt SAURON} survey (Kuntschner et al., in preparation), while the blue symbols represent the early-type spirals of 
the {\tt SAURON} survey \citep{reynier}. 
Despite the problems intrinsic to this approach, from this plot we can see that, as expected, 
our galaxies are on average younger and more metal poor than early-type galaxies, and have on average a value of 
[Mg/Fe] closer to zero, ranging from slightly sub-solar to slightly over-abundant. This confirms what we saw in Fig. \ref{corr_index}, 
where we noticed that 
the late-type spirals were compatible with models with [Mg/Fe]=0.\\
\indent In order to assess the dependence of our results on the set of models used, we performed the following exercise. We took the model spectra from 
the SSP library of \citet{vazdekis} and calculated the line-strength indices on them, obtaining a grid of 322 model indices. Then we determined, galaxy by galaxy, 
the model that best matched our observations, deriving age and metallicity estimates. Abundance ratios are fixed to solar values in 
the library of \citet{vazdekis}. In Fig. \ref{compare_vaz_thomas} we plot against each other the ages and metallicities obtained for the central 
1\farcs5 apertures of our galaxies using the Thomas (horizontal axis) and the Vazdekis (vertical axis) models. Overplotted with solid lines are the 1:1 relations. 
We see that there are no dramatic differences between the two sets of measurements, though the Vazdekis models tend in several cases 
to predict slightly older ages and 
lower metallicities with respect to the Thomas ones. The small residual differences do not depend on the fixed abundance ratios of the Vazdekis models: the ages and
metallicities on the horizontal axis in Fig. \ref{compare_vaz_thomas} are obtained for this test considering only the Thomas models closest to solar 
abundance ratios. The differences in the metallicity could be instead at least partially due to the coarseness in Z of the model grid obtained from the Vazdekis
models.\\
\indent As also noticed in \citet{reynier}, in some cases an unconstrained SSP fit cannot provide a suitable representation of a galaxy. 
In the case of a complex star formation history, not approximable with a single, instantaneous burst, the SSP-equivalent parameters must be interpreted 
with caution, as a `zero-order' estimate, as we already stressed. As an example, we remind the reader the case of the star 
formation ring in NGC\,4321 discussed by \citet{emma}. They show that in the ring the line-strength indices suggest the superposition of two 
components: 
a young stellar population and an old and metal-rich one. When they force a single SSP, they obtain low metallicity and old age. But it is known
\citep{zaritsky} that the HII regions of this galaxy have super-solar metallicity: the one-SSP solution obtained by \citet{emma} represents 
therefore an inconsistent description. As clearly visible from the first panel in Fig. \ref{1ssp_fullrange}, some of our galaxies also populate the region of the 
(metallicity, age) plane characterised by low metallicities (below $-$1.0) and old ages (above 10 Gyr). This is the case for the 
central aperture of NGC\,864, for which our 
$\chi^2$ technique selects a model with Z $\approx$ $-$1.14 and age $\approx$ 11 Gyr. The line-strength maps for this object 
(see Fig. \ref{maps2}) show that 
in the very center the Fe5015 and Mg{\textit{b}} indices assume low values, and so does H$\beta$, on a large inner region. The low indices 
drive the metal poor old population, but a SSP description might be too unrealistic for this object.\\
\indent More generally, if for example a galaxy has undergone two separate bursts of star formation, the SSP-equivalent 
age will be biased towards the age of the youngest stars, and the SSP-equivalent metallicity will be biased towards the metallicity of the old population, 
as studied and described in detail by \citet{paolo}.\\
\indent In order to put some physically motivated constraints to our one-SSP analysis, we decided to constrain the metallicity of our objects 
 in a narrow range around the value 
in the model grid 
closest to the one determined from a relation between central stellar velocity dispersion and metallicity for early-type galaxies. 
In practice, we aim to apply a scaling relation obtained for spheroids (ellipticals and lenticulars). Our late-type galaxies do not host relevant 
bulge components 
(Ganda et al., in preparation), so the regions that we probe with our data do not fall 
under this category. 
Also, unlike for spheroids, $\sigma$ is not a good mass indicator in disc galaxies. But in any case, 
using a `$\sigma$ relation' for spheroids is a `zero-order' approach to extrapolate the metal content of the old stars in
low-mass galaxies. We investigated several possibilities for a metallicity-sigma relation for early-type galaxies. The resulting relations can differ from each other significantly, 
particularly at the low-$\sigma$ end, because of differences in the methods, in the models used, and in the underlying assumptions.\\
\indent A possible and common way is to go through stellar population 
models, estimate the ages and metallicities by comparison of observations and SSP models, and then fit a relation with $\sigma$. This approach has been extensively investigated 
in the literature; we will refer to it as ` the model approach'. As a prototype, we refer to the work of \citet{thomas05}, who studied 124 early-type galaxies in high- and low-density environments and 
derived their ages, metallicities and element ratios. For the galaxies in high-density environments, they retrieved the following relation:
\begin{equation}
Z = 0.55 \times \log(\sigma) - 1.06.
\label{zsigmaeq_thomas}
\end{equation}
See also \citet{harald2000} for a similar result.\\
\indent Another possibility, to which we will refer as `the index - $\sigma$ approach', is to assume a set of index - $\sigma$ relations for early-type galaxies, extrapolate them to the range of $\sigma$ under investigation, 
from the measured $\sigma$ infer the indices and from these, via a comparison with models, an estimate for the metallicity. In practice, we used the 
H$\beta^{'}$ - $\sigma$, Fe5015$^{'}$ - $\sigma$ , Mg{\textit{b}}$^{'}$ - $\sigma$ relations published in \citet{paper6} for the 48 ellipticals and lenticulars of the main {\tt SAURON}
survey. Using these relations and the measured values for the central aperture velocity dispersions of our galaxies, we obtained a set of index values. We then
compared them with the models of \citet{thomas} via the previously described $\chi^{2}$ minimisation, but choosing only among models with solar abundance
ratios and age
$\approx$ 12.6 Gyr. In this way we come to the estimate of the metallicity of an old spheroid at the observed velocity dispersion. We will adopt the estimates obtained
in this way to put a constraint on the metallicity.\\
\indent Another approach, which does not involve any stellar population modeling, relies on empirical calibrations and tight observed scaling relations 
between metal indices and velocity dispersion. Consider the 
Mg$_2$ - $\sigma$ relation for elliptical and lenticular galaxies given by \citealt{Jorgensen} (Equation \ref{jorgensenequation}). 
If we combine this relation with the Mg$_{2}$ - metallicity calibration from Buzzoni, Gariboldi \& Mantegazza (1992)\footnote{These authors calibrate the 
dependence of Mg$_{2}$ from stellar parameters using an observed stellar library and find a general relation, that reduces to the one reported here (Equation 
\ref{calibeq}) in the parameter range spanned by elliptical galaxies.}:
\begin{equation}
\mathrm{Mg}_{2} = 0.135 \times Z + 0.28,
\label{calibeq}
\end{equation}
we obtain a relation between metallicity and velocity dispersion, holding for early-type galaxies:\footnote{We warn the reader about the fact that the Mg$_2$
index has a slightly different definition in the papers of \citet{Jorgensen} and
\citet{buzzoni}, and that we have not taken this difference into account in deriving Equation \ref{zsigmaeq}.}. 

\begin{equation}
Z = 1.452 \times \log(\sigma) - 3.222.
\label{zsigmaeq}
\end{equation}
We call this last approach `the empirical approach'. \\
\begin{figure}
\begin{center}
{\includegraphics[width=0.99 \linewidth]{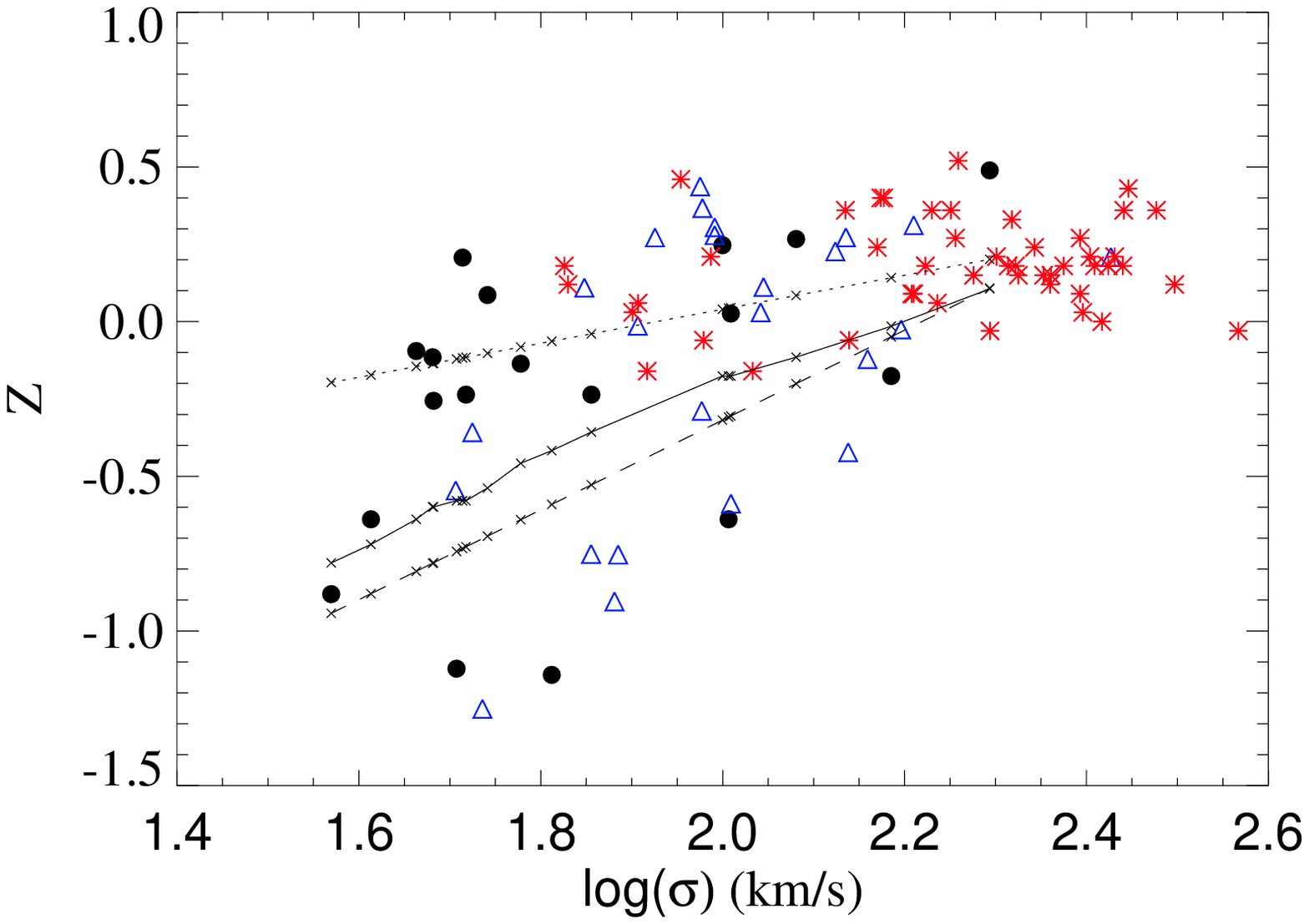}}
\end{center}
\caption{Relation between central metallicity and central velocity dispersion (in decimal logarithm and units of km s$^{-1}$) for the E and S0
galaxies (Kuntschner et al., in preparation, red symbols) and the Sa (\citealt{reynier}, 
blue symbols) of the {\tt{SAURON}} survey and for our late-type spirals (black dots), from our one-SSP analysis. The overplotted lines represent the relation metallicity - velocity 
dispersion obtained in different ways, drawn on the $\sigma$ range of our data only: the dotted line reproduces Equation
\ref{zsigmaeq_thomas} (`model approach'), the dashed line Equation \ref{zsigmaeq} (`empirical approach') and 
the solid lines represents a relation based on the assumption of a set of index - $\sigma$ relations. See text for a more complete description.}
\label{zsigma}
\end{figure}
\indent In Fig. \ref{zsigma} we show in the ($\log\left(\sigma\right)$, Z) plane the distribution of the elliptical and lenticular galaxies in the 
{\tt SAURON} survey (red symbols, from Kuntschner et al., in preparation), the Sa galaxies from the same survey (blue symbols, from \citealt{reynier}), 
together with our own
galaxies (black dots). All the values refer to the central apertures, and the metallicities are obtained from the one-SSP analysis. The lines 
overplotted in the Figure represent the various metallicity-velocity dispersion relations for E/S0 galaxies described above. The dotted line 
reproduces Equation \ref{zsigmaeq_thomas} (from `the model approach'), the dashed line Equation \ref{zsigmaeq} (from `the empirical approach') and the solid line connects 
the points representing our galaxies in the ($\log\left(\sigma\right)$, Z) plane according to 
`the index - $\sigma$' approach, the second one we presented.\\
\begin{figure*}
{\includegraphics[width=0.99 \linewidth]{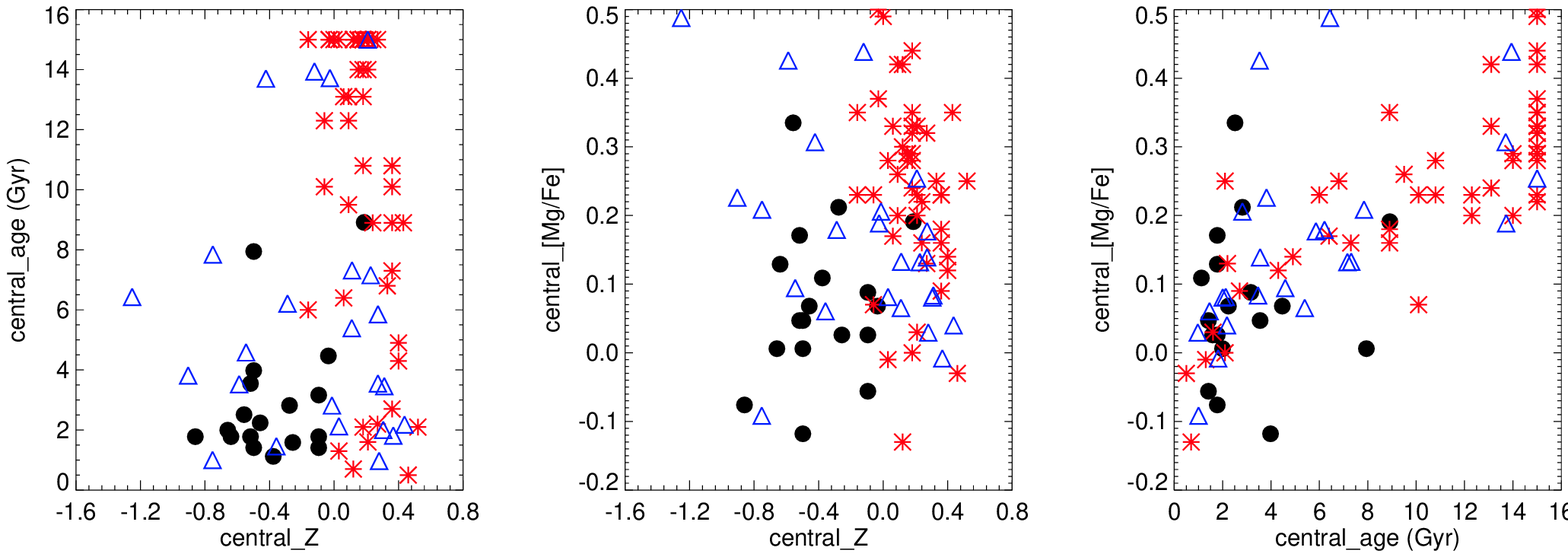}}
\caption{Same plots as in Figure \ref{1ssp_fullrange}, but referring to the case with constrained metallicity (see text in Section \ref{singlessp} for more details), for 
our late-type spirals (black symbols); for the E/S0 (red symbols) and the Sa (blue symbols) galaxies no constraint on metallicity is applied.}
\label{1ssp_zconst}
\end{figure*}
\begin{figure}
\begin{center}
{\includegraphics[width=0.99 \linewidth]{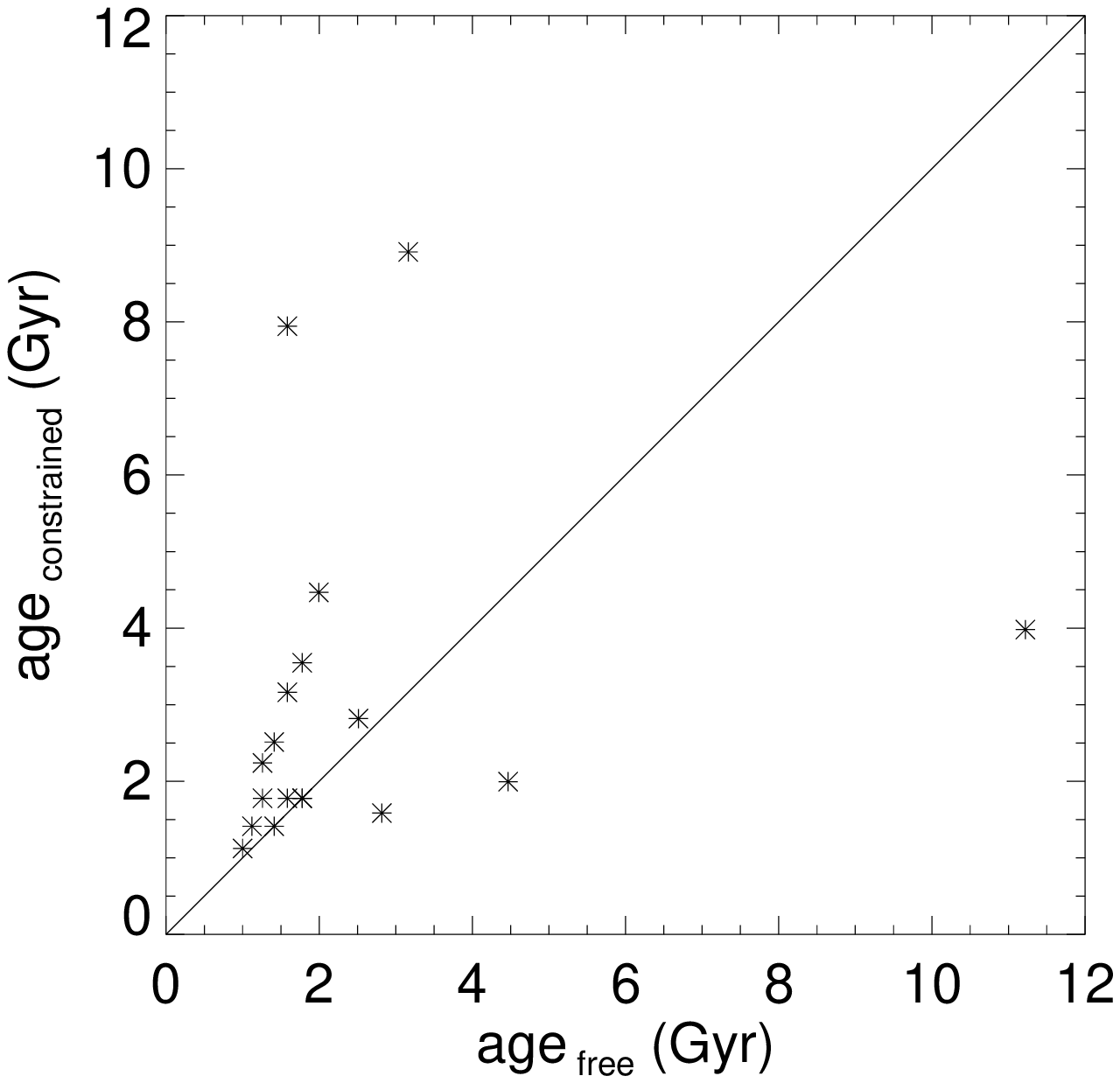}}
\caption{Constrained (y-axis) vs unconstrained (x-axis) age (in Gyr), measured on the central apertures of our galaxies with the one-SSP approach.}
\label{freevsconst}
\end{center}
\end{figure}
\indent We re-evaluated our age estimates by imposing a constraint on the metallicity, allowing it to vary only in a narrow range ($\pm$ 0.1 dex)
 around the metallicity given, galaxy by
galaxy, by the solid black line in Fig. \ref{zsigma} (`the index - $\sigma$ approach' discussed above). We
preferred this approach to the other two because it is based on empirical relations, and does not rely
exclusively on models, unlike `the model approach', and because it uses more than just magnesium to
establish the metallicity of the old population, contrary to 'the empirical approach'. Fig. \ref{1ssp_zconst} is equivalent to Fig. \ref{1ssp_fullrange}, but with the constraints on metallicity.
In Fig. \ref{freevsconst} we plot the constrained ages against the unconstrained ones: we can notice that for very young ages the values are close, and differ more and more
going to older ages. The fact that in some cases the constrained age differs from the unconstrained might indicate either that the imposed metallicity is 
inappropriate or that the one-SSP assumption is wrong, and the galaxy's star formation history is not approximable with an instantaneous burst.\\
\indent In Table \ref{popsvalues} we list the population parameters measured, both in the unconstrained and in the constrained case, 
on the central aperture of our galaxies.
\begin{center}
\begin{table}
\begin{center}
 \begin{tabular}{@{}c c c c c c c }
\hline \hline
NGC & age & age$_{c}$ & Z & Z$_{c}$ & [Mg/Fe] & [Mg/Fe]$_{c}$\\
\hline
\,\,488 &  3.162& 8.913& \,\,\,\,0.489 &  \,\,\,\,0.187 & \,\,\,\,0.191 &\,\,\,\,0.191\\
\,\,628 &  1.585&  7.943&\,\,\,0.207& $-$0.498 &  $-$0.118 &\,\,\,\,0.006\\
\,\,772 &  1.995&  4.467&\,\,\,0.267&  $-$0.035 &\,\,\,\,0.068 &  \,\,\,\,0.068\\
\,\,864 &  11.220&  3.981&  $-$1.142& $-$0.498 &  $-$0.015 & $-$0.118\\
1042 &  1.259&  2.239&\,\,\,0.086 & $-$0.458 &  $-$0.076 & \,\,\,\,0.068\\
2805 &  1.413&  2.512& $-$0.095& $-$0.558 & \,\,\,\,0.191 & \,\,\,\,0.335\\
2964 &  2.818&  1.585&  $-$0.639& $-$0.256 &\,\,\,\,0.068 &\,\,\,\,0.026\\
3346 &  1.259&  1.778&  $-$0.256& $-$0.518 &\,\,\,\,0.109 & \,\,\,\,0.171\\
3423 &  1.778&  3.548&  $-$0.115& $-$0.518 &  $-$0.035 & \,\,\,\,0.047\\
3949 &  1.000& 1.122& $-$0.136& $-$0.377 & \,\,\,\,0.006 & \,\,\,\,0.109\\
4030 &  1.585&  3.162& \,\,\,0.247& $-$0.095 &\,\,\,\,0.047 & \,\,\,\,0.088\\
4102 &  1.413&  1.413&$-$ 0.176& $-$0.095 &  $-$0.015 &  $-$0.056\\
4254 &  2.512&  2.818&  $-$0.236& $-$0.276 &\,\,\,\,0.171 & \,\,\,\,0.212\\
4487 &  4.467&  1.995&  $-$1.122& $-$0.659 & \,\,\,\,0.047 &\,\,\,\,0.006\\
4775 &  1.778&  1.778&  $-$0.639& $-$0.639 &\,\,\,\,0.129 & \,\,\,\,0.129\\
5585 &  1.778&  1.778&  $-$0.881& $-$0.860 & $-$ 0.076 & $-$0.076\\
5668 &  1.122&  1.413& $-$0.236& $-$0.498 &  $-$0.015 & \,\,\,\,0.047\\
5678 &  1.585&  1.778&\,\,\,\,0.026& $-$0.095 &  $-$0.015 & \,\,\,\,0.026\\
\hline
\end{tabular}
\end{center}
\caption{Population parameters from our one-SSP analysis. The subscript `\textit{c}' refers to the quantities obtained 
in the case where we constrain the metallicity within a narrow range ($\pm$ 0.1 dex) around the metallicity given by the 
Z - $\sigma$ relation. Age is in Gyr and metallicity in decimal logarithm, with solar metallicity as zero point.}
\label{popsvalues}
\end{table}
\end{center}
\subsection{Two-population analysis}\label{2ssp_analysis}
\begin{figure*}
\begin{center}
{\includegraphics[width=0.99 \linewidth]{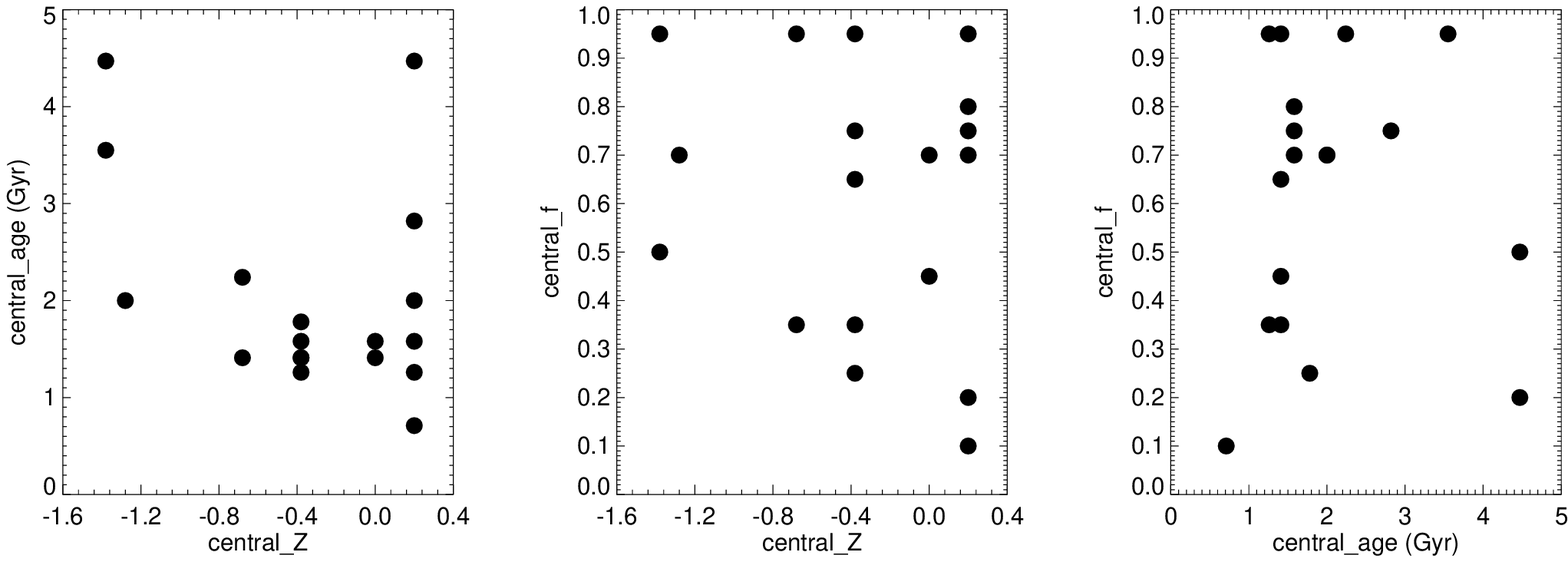}}
\end{center}
\caption{Fitted parameters from our two-SSP analysis, measured on the central apertures. In the left panel we have the age (in Gyr) against
the metallicity (in decimal
logarithm, with the solar metallicity as zero point)
of the young component $pop\_young$; in the middle its mass fraction against its metallicity; in the right panel, the
mass fraction of the young component against its age.}
\label{2ssp_zfree}
\end{figure*}

\begin{center}
\begin{figure*}
{\includegraphics[width=0.99 \linewidth]{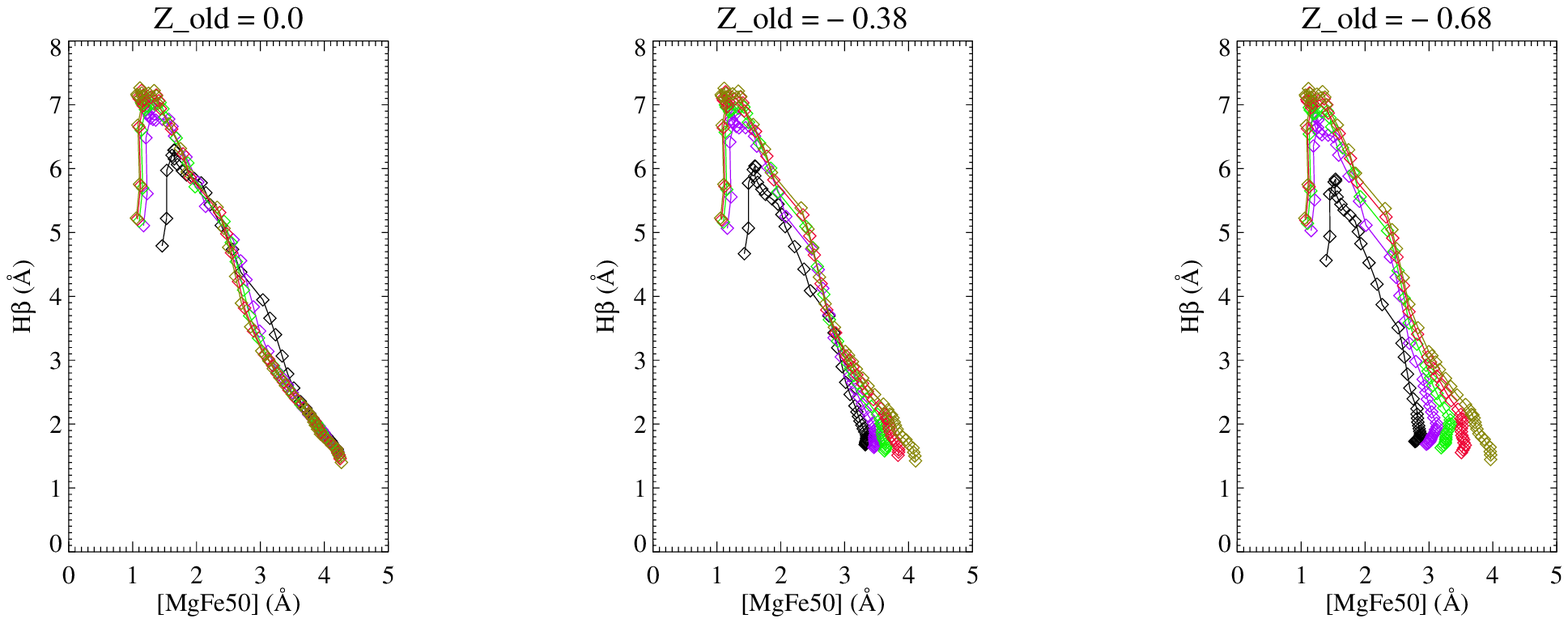}}
\caption{Composite models for different choices of the metallicity of the old component (indicated at the top of each panel) and solar metallicity of 
the young one represented by points in the 
([MgFe50], H$\beta$) plane. Different colours refer to different values of the mass fraction of the young component in the composite model. The effects of age and fraction of the young component on the line-strength indices of the composite models are degenerate.}
\label{degeneracy_2ssp}
\end{figure*}
\end{center}

\begin{figure*}
{\includegraphics[width=0.99 \linewidth]{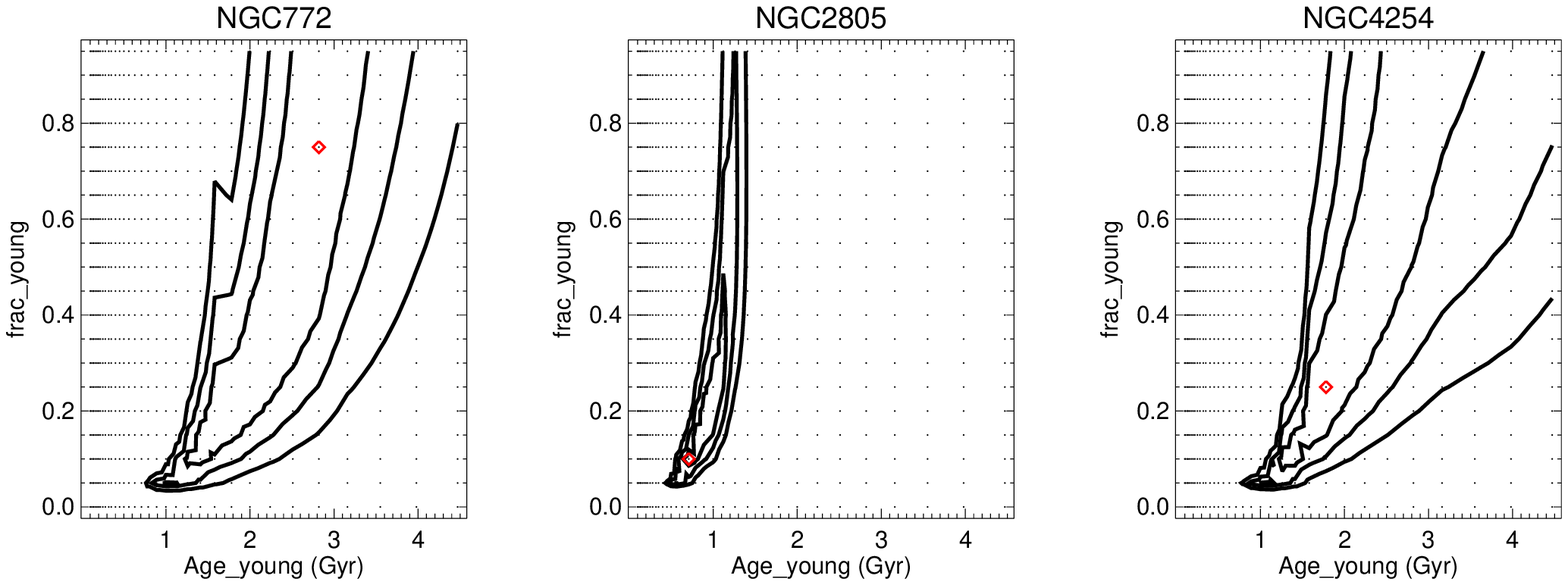}}
\caption{Contours of $\Delta\chi^2$ projected on the plane (age, fraction) of the young component at the 1,2,3-$\sigma$ confidence level, 
for the central apertures of NGC\,772, NGC\,2805, NGC\,4254; the red symbol marks the best-fitting model. These 
plots clearly show that the two-SSP approach is unable to constrain the fitted parameters, and that age and mass fraction of the young component are degenerate.}
\label{contours_2ssp}
\end{figure*}

A galaxy is most likely characterised by stellar populations that are more complex than SSPs; therefore, in order to 
try to build a more realistic picture of our galaxies, we investigated more sophisticated approaches. As an attempt, we explored 
a two-SSP scenario, describing the galaxy 
as the superposition of an old population and a younger one. In practice, we want to describe the galaxy population as:
\begin{equation}
pop\_tot = (1-f) \times  pop\_old + f \times pop\_young,
\end{equation}
where $pop\_old$ is a $\approx$ 12.6 Gyr old SSP with metallicity fixed to the value close to the one obtained from the Z - $\sigma$ relation 
(on the basis of what we called `the index - $\sigma$ approach') and $pop\_young$ is a 
SSP with age below 5 Gyr and free metallicity; the limit of 5 Gyr was chosen in order to have a population younger than the old component, but at the same 
time not limited to very young ages; $f$ is the mass-fraction of the young component and varies between 0 and 1. To do this, we 
take the model spectra from \citet{vazdekis} and select, galaxy by galaxy, the 12.6 Gyr old model with the metallicity closest to the one given by the solid line in
Fig. \ref{zsigma}, among those available in the model grid, and combine it with all the other SSP models in the library, with varying mass fractions. In this way we obtain for each galaxy a 
set of 6762 `composite model spectra', with varying age and metallicity of the young component and relative mass fraction. On these spectra we compute the line-strength indices, obtaining for each galaxy a 
grid of `composite model indices'. Since we use the model library of \citet{vazdekis}, the abundance ratio is fixed to solar values, which 
for our objects is a quite fair approximation, as we saw in the previous Section. We then use the same $\chi^2$ technique as for the one-SSP analysis 
and determine, spectrum by spectrum, the composite model closest to our observed line-strengths. This gives us an estimate for the age and metallicity of the young 
component and its mass fraction.\\
\indent Figure \ref{2ssp_zfree} plots against each other the parameters fitted on the spectrum averaged in a central aperture of 1\farcs5 radius. We see that the young
component spans a range in age from below 1 to $\approx$ 4.5 Gyr, a range in Z from $-$1.38 to +0.2 and that its mass contribution to the 
global galaxy population $f$ varies from $\approx$ 10\% (NGC\,2805) to $\approx$ 95\% (NGC\,1042,\,2964,\,3949,\,4487).\\
\indent The main problem of this approach is the degeneracy of the parameter space, which must act as a warning for the reader against the robustness of the 
ages and mass fraction estimates for the young component presented in Fig. \ref{2ssp_zfree}. This degeneracy is illustrated in Figures 
\ref{degeneracy_2ssp} and \ref{contours_2ssp}. In Fig. \ref{degeneracy_2ssp} we plot, for different choices of the metallicity of the old component and 
solar metallicity for the young one, the points representing the models in the ([MgFe50], H$\beta$) diagram. Different colours represent different weights 
of the young component in the composite models, from 0.1 (black) to 0.9 (gold). From these plots it is evident that it is not possible to significantly 
discriminate the effects of age and mass fraction of the young component. Fig. \ref{contours_2ssp} shows the projection of the $\Delta\chi^2$ 
space (metallicity, age and mass fraction of the young component) on the (age, fraction) plane and the contours representing the 1,2,3-$\sigma$ confidence level, 
based on the $\Delta\chi^2$ for two degree of freedom, for the central aperture of some of the galaxies. We see that very large regions of the parameter space 
are equivalent in terms of $\chi^2$, and that even in cases where the age is rather well constrained (see NGC\,2805), the mass fraction is badly determined. Therefore, we 
see that we cannot extract solid conclusions from this method, despite its interesting principle, and we decided not to push it further and not to present the two-dimensional
maps of the inferred parameters.

\subsection{Continuous star formation}\label{continuum}
\begin{figure*}
\begin{center}
{\includegraphics[width=0.99\linewidth]{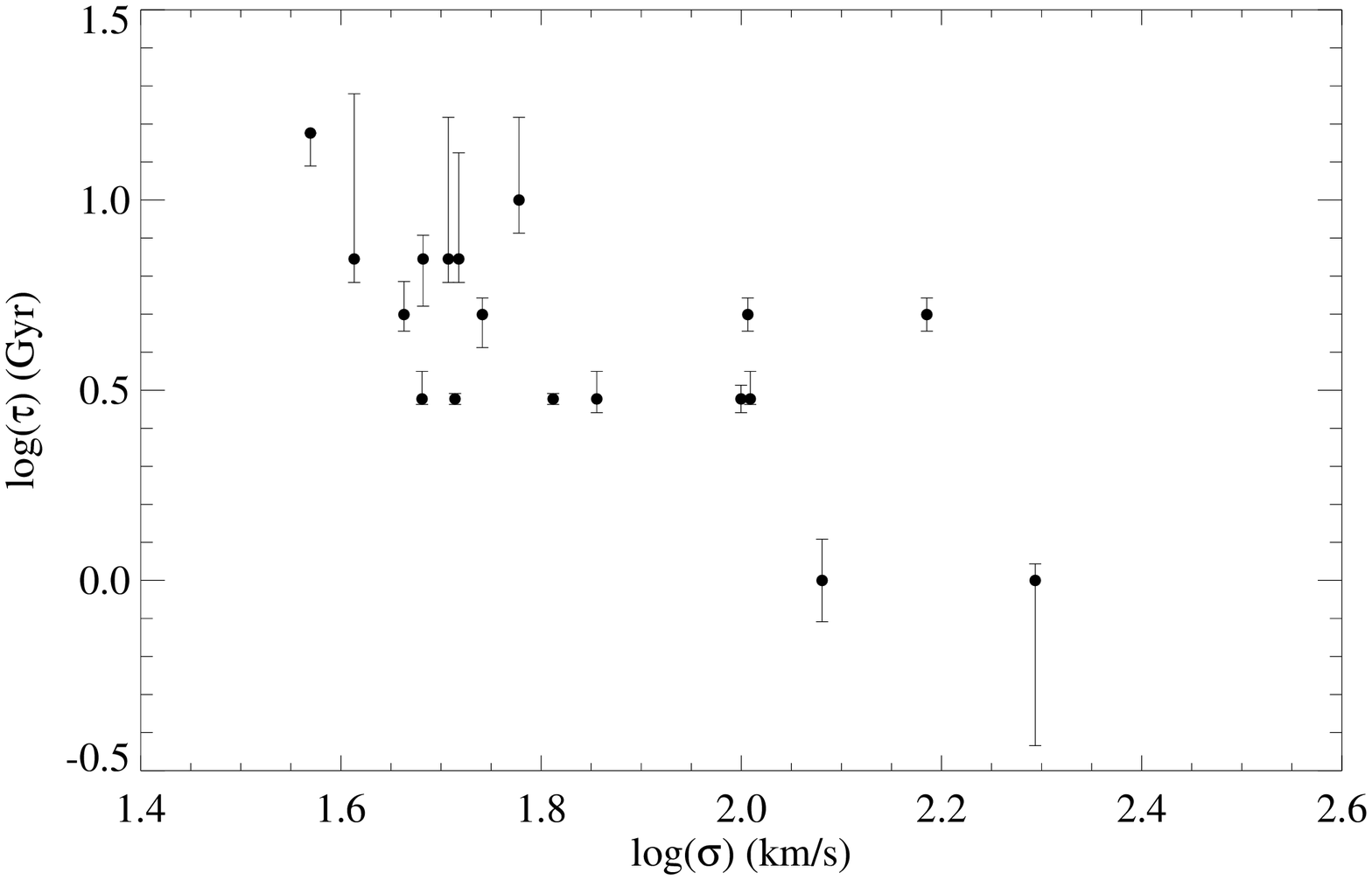}}
\caption{Central aperture values for the $e$-folding time-scale $\tau$ against central velocity dispersion $\sigma$ (both in units of decimal logarithm), 
in an exponentially declining star formation scenario; 
the $\tau$ values are obtained selecting among models with age = 10 Gyr and $\epsilon$ = 1; the error bars come from a rough estimate on the basis 
of the 1-$\sigma$ contour plots of the projection of $\Delta\chi^2$ on the ($\tau$-metallicity) plane.}
\label{tauvssigma}
\end{center}
\end{figure*}

\begin{figure*}
{\includegraphics[width=0.99\linewidth]{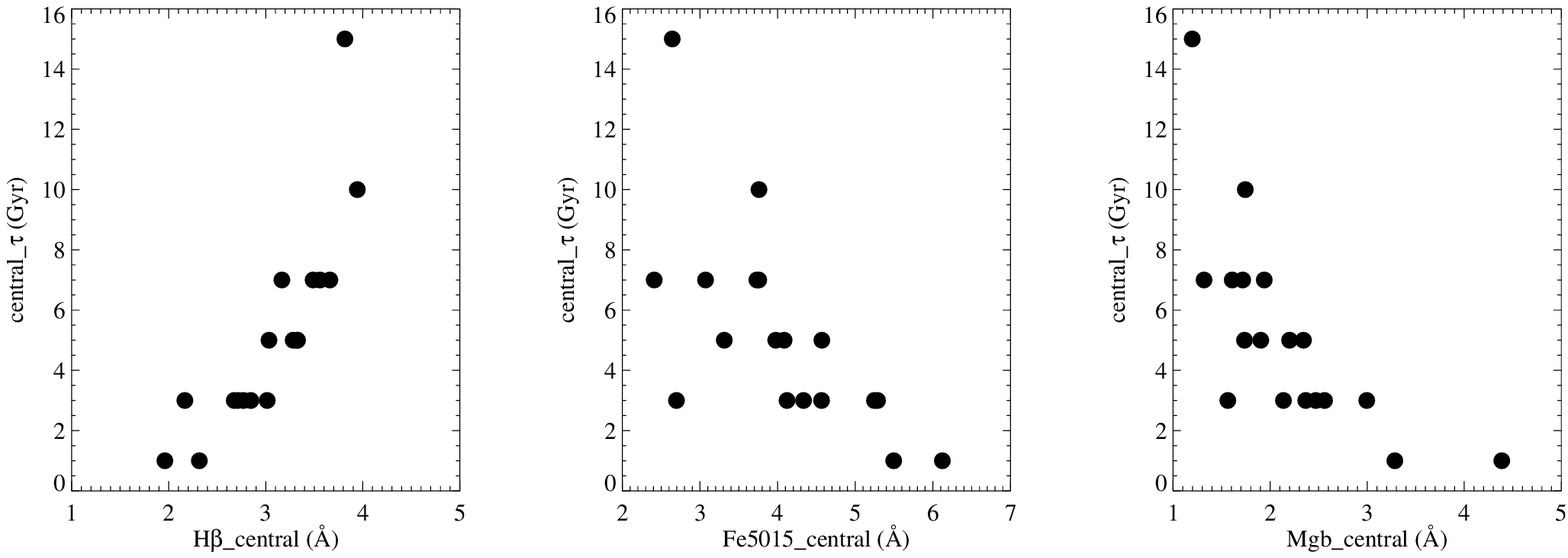}}
\caption{Central aperture values for the $e$-folding time-scale $\tau$ (in Gyr) of an exponentially declining star formation history, against the central line-strength indices (expressed in \AA): H$\beta$ in the left panel, 
Fe5015 in the middle and Mg{\textit{b}} in the right one; the $\tau$ values are obtained selecting, in the fit, among models with age = 10 Gyr and $\epsilon$ = 1 (see text for
details).}
\label{tauvsindices}
\end{figure*}

{\emph {`The star formation history of galaxies is imprinted in their integrated light'}}, state Bruzual \& Charlot in a paper where they present a new 
model for the spectral evolution of stellar populations at ages between $1\times10^4\,$ and 
$2\times10^{10}\,$ yr for a range of metallicities \citep{BC}. In practice, they use the {\em isochrone synthesis} technique to compute the 
spectral evolution of stellar populations. This technique relies on the fact that any stellar population can be expanded in a series of
instantaneous starbursts or SSPs. So, the spectral energy distribution at time $t$  of a stellar population can be
written as in Equation 1 in \citet{BC}:
\begin{equation}
F_\lambda(t)=\int_0^t\,\psi(t-t')\times S_\lambda\left[t',\zeta(t-t')\right]\,dt'\,,
\label{convolbc}
\end{equation}
where $\psi(t)$ and $\zeta(t)$ are respectively the star formation rate and the metal-enrichment law, as a function of time, and 
$S_{\lambda}\left[t',\zeta(t-t')\right]$ is the power radiated by an SSP of age $t'$ and metallicity $\zeta(t-t')$ per unit wavelength
and unit initial mass. \citet{BC} compute their models with various choices for the initial mass function IMF, the metallicity of the
initial SSP and the stellar evolution prescription. They also provide as online material some tools that allow the user to compute 
the spectral evolution -in time- of a SSP for different star formation history
scenarios\footnote{The program we used are available from the web page http://www.cida.ve/$\sim$bruzual/bc2003 .}. We chose to
use these tools in order to apply an evolutionary analysis to our data. Starting from the SSP models from \citet{BC} obtained using the Chabrier (2003) IMF, 
for different choices of the initial metallicity (namely, Z = $-$1.69, $-$0.69, $-$0.39, 0.0, 0.39), we computed the
time evolution with constant and with exponentially declining star formation rate: respectively, $\psi(t)=const$ and 
$\psi(t)=[1 M_{\odot} + \epsilon M_{PG}(t)]\,\tau^{-1} exp(-t/\tau)$, where $\tau$ is the $e$-folding time-scale and $M_{PG}(t)$ is the 
mass of gas processed into stars and re-ejected into the interstellar medium at the time $t$; the parameter $\epsilon$ sets the fraction of the 
ejected gas that is re-used in new star formation episodes; we only explored the cases $\epsilon=0$ (no recycling) and $\epsilon=1$
(all of the ejecta go into new stars). In any case, the actual value of
$\epsilon$ does not change our results significantly. These are both plausible scenarios: the constant star 
formation history has long been
considered as a suitable scenario for spirals: {\emph {`star formation has proceeded at relatively constant rate over the lifetimes 
of most late-type spiral galaxies'}} \citep{kennicut83}, while the exponentially declining star formation describes the situation of
a starburst of duration $\tau$; we notice that the case $\tau \rightarrow \infty$ approximates the constant star formation scenario, while the case
$\tau \rightarrow 0$ reproduces an instantaneous burst or SSP. We 
computed the spectral evolution for the chosen metallicities, for both cases $\epsilon=0$ and $\epsilon=1$ and for several choices
of $\tau$: $\tau$ = 0.5, 1.0, 3.0, 5.0, 7.0, 10.0, 12.0 and 15.0 Gyr. The tools provided by Bruzual \& Charlot also include an option for computing the effect of
attenuation by dust, but we decided not to use this possibility, in order not to further increase the number of parameters at play.\\
\indent For all of the selected cases we calculated the time evolution of the Lick indices using the programs from Bruzual \& Charlot 
and focussed on the situation at $t=$ 10 Gyr, 
building a model grid of Lick indices for different metallicities, $\tau$ and $\epsilon$ (0 or 1) and age fixed at 10 Gyr 
and then compared them with the line-strength indices 
of the {\tt SAURON} spectra, via the $\chi^2$ minimisation technique used also in the previous Sections. In this way, we come to an estimate for
the time-scale $\tau$, $\epsilon$ and the metallicity. Fixing the age to 10 Gyr is equivalent to assume that the galaxy existed for 10 Gyr, and to try to describe how stars were 
formed over that time. We tested that the fitted parameters ($\tau$ and metallicity) had negligible variations when 
fixing the value of $\epsilon$, therefore we repeated our estimates allowing only models with $\epsilon = 1$ and will refer to this case from now on.\\
\indent In Figures \ref{maps1}-\ref{maps9} the top right panel shows, galaxy by galaxy, the two-dimensional maps of $\tau$, obtained for the exponentially declining
SFR, for age fixed to 10 Gyr and $\epsilon = 1$. Interesting features of the individual $\tau$ maps, whenever present, are highlighted in Appendix \ref{individuals}. As for the 
metallicity maps, obtained in the $\chi^{2}$ minimisation and not shown in this Paper, we can state that in the majority of cases they 
resemble in their structure the metallicity maps obtained in the one-SSP approach; the actual numbers are different, since the models used in the 
two approaches are different and also the metallicity ranges and values explored by the model grids differ, but the spatial structure is generally similar.\\
\begin{table}
\begin{center}
 \begin{tabular}{c c c c c c c c}
\hline \hline
NGC & $\tau$ &\,&NGC & $\tau$&\,&NGC & $\tau$\\
\hline
\,488&1&\,&2964&5&\,&4254&3\\
\,628&3&\,&3346&7&\,&4487&7\\
\,772&1&\,&3423&3&\,&4775&7\\
\,864&3&\,&3949&10&\,&5585&15\\
1042&5&\,&4030&3&\,&5668&7\\
2805&5&\,&4102&5&\,&5678&3\\
\hline
\end{tabular}
\end{center}
\caption{Best-fitting values for the star formation time-scale $\tau$ (in Gyr) measured on the central apertures of our 
galaxies, obtained for the choices age = 10 Gyr and $\epsilon$ =1, in an exponentially declining star formation rate scenario.}
\label{tauvalues}
\end{table}
\indent If we focus on the values obtained for the central apertures, we find a correlation between 
the best-fitting $\tau$ and the
central velocity dispersion, as shown in Fig. \ref{tauvssigma} where we plot log($\tau$) against log($\sigma$). We see that 
higher $\sigma$ galaxies, which are larger and more massive, tend to have shorter star formation time-scales, more consistent with an instantaneous burst scenario, while 
at the low-$\sigma$ end the smaller galaxies tend to have larger $\tau$, indicating a star formation history spread over time.
Therefore, the time-scale $\tau$ represents a physically meaningful parameter, parametrizing a smooth transition from SSP-like to
constant SFR.\\
\indent Figure \ref{tauvsindices} shows that rather tight correlations exist also between the central $\tau$ and the central line-strength indices. These plots tell 
us that old, metal-rich galaxies (low H$\beta$, high Fe5015 and Mg{\textit{b}} values) have shorter star formation time-scales, while young and more metal poor 
galaxies (higher H$\beta$, lower Fe5015 and Mg{\textit{b}}) have a star formation history more extended 
in time: we find again a smooth transition between a SSP-like and constant star formation. The central aperture values of $\tau$ used in Figures 
\ref{tauvssigma} and \ref{tauvsindices} are listed in Table \ref{tauvalues}.\\
\indent We also investigated the constant star formation history scenario, by selecting the corresponding option when using the programs from 
Bruzual \& Charlot, building a grid with the model indices for the exponentially declining star formation rate and adding those for constant star 
formation rate, for the various metallicity values selected and for $\epsilon$ = 1 and age fixed to 10 Gyr. When applying our $\chi^2$ minimisation procedure, 
we found that the constant star formation was preferred
in the regions where the fit with the exponential star formation only had selected the highest values allowed for the time-scale
$\tau$. This confirms our 
interpretation of $\tau$ as a parameter describing a continuous transition from constant SFR to SSP.\\
\indent We then repeated the analysis for different values of the fixed age of the evolved models, spanning the range 1-18 Gyr and found that the 
$\tau$ values depend on the fixed age, given the existence of a strong degeneracy between age and $\tau$, while the metallicities 
are more robust.\\
\indent The main interpretative difficulty of this method is indeed due to the 
degeneracies in the parameter space, which are difficult to break with the few observed Lick indices that we measure. Nevertheless, 
by inspecting the contours of $\Delta\chi^2$ in the ($\tau$-metallicity) plane, for several choices of the fixed age of the evolved models, 
we can in some cases exclude certain star formation scenarios in favour of
others, even though we cannot give a precise, quantitative estimate of the
fitted parameters. For example, in the case of NGC\,488, if we look at the two-dimensional $\Delta\chi^2$ space (metallicity, $\tau$), 
for a fixed age of the evolved models), and look at the contours in the ($\tau$-metallicity) plane representing the 3-$\sigma$ confidence level, 
based on the $\Delta\chi^2$ for one degree of freedom, we notice that it lies in a region far from the high-$\tau$ end, and this happens 
independently on the chosen values of the fixed age. Even if we perform the minimisation by allowing free age, $\tau$ and metallicity, and examine the 
projection of the $\Delta\chi^2$ contours in the ($\tau$-age) plane, we see that for all ages we can exclude large $\tau$ values at a high confidence 
level. This must be regarded as an exercise, since we are not interested in fitting simultaneously age and $\tau$, but it gives us 
insight on the impact of the degeneracies and of our assumptions on the robustness of our conclusions. Thus, for NGC\,488 we can safely state that we can exclude a constant star formation scenario in favour of an exponentially 
declining star formation rate, with a short time-scale $\tau$ (closer to an instantaneous burst than to a constant SFR). This is illustrated in the first column 
in Fig. \ref{chi2contours1}. Similar 
hints for the star formation history of other galaxies will be given, whenever possible, in Appendix \ref{individuals}, dedicated to the description of the single galaxies
individually. For all galaxies, the $\Delta\chi^2$ contours for the fit at different fixed ages and with free age are presented in Figures \ref{chi2contours1}-\ref{chi2contours3}.
\renewcommand{\thefigure}{\arabic{figure}\alph{subfigure}}
\setcounter{figure}{20}
\setcounter{subfigure}{1}
\begin{center}
\begin{figure*}
{\includegraphics[width=0.89 \linewidth]{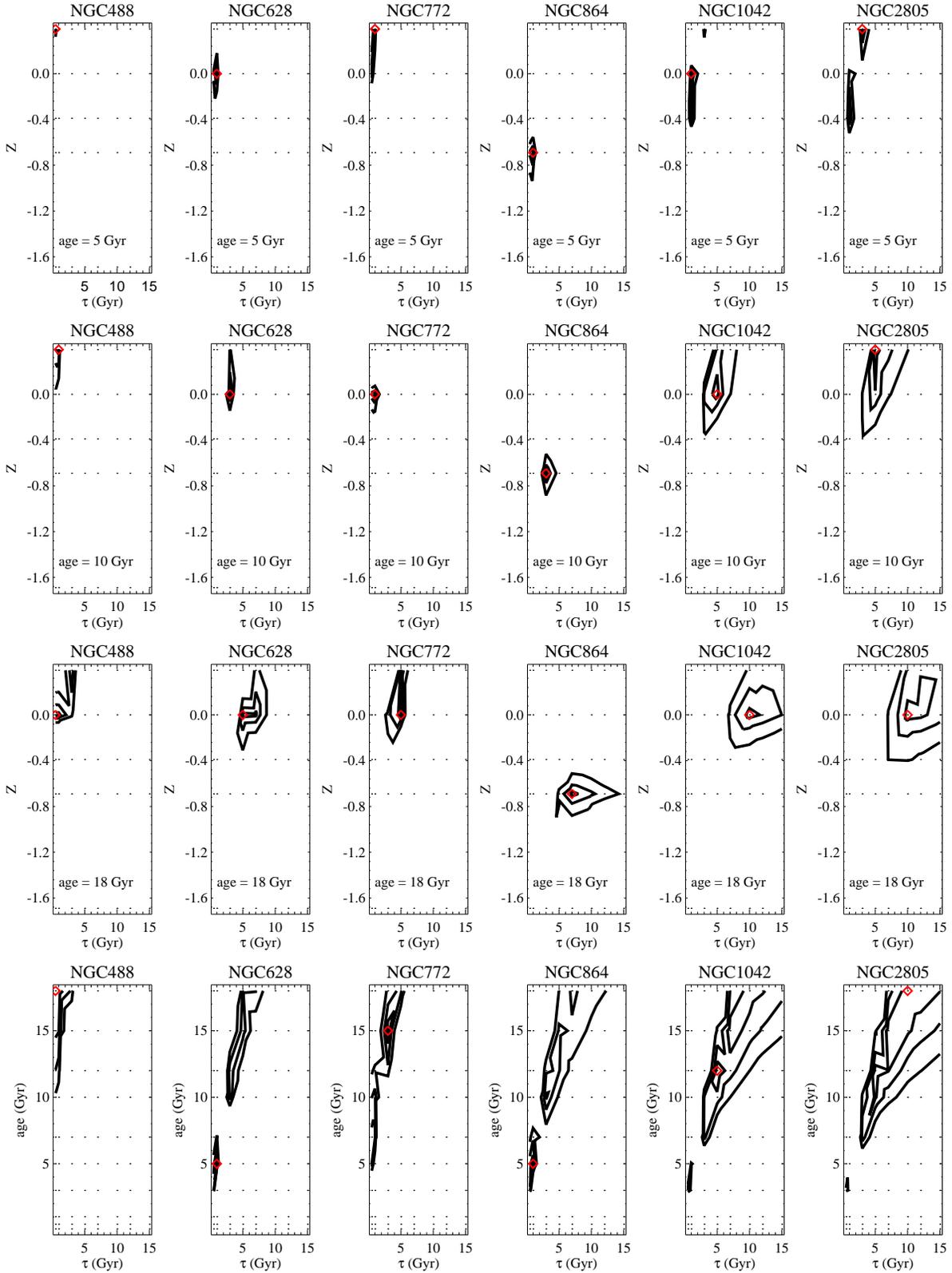}}
\caption{\textit{First column}: projections of the 1-, 2- and 3-$\sigma$ $\Delta\chi^2$ contours on the ($\tau$-metallicity) plane for the central aperture of NGC\,488, 
for different fixed ages in the fit (5 Gyr, first row, 10 Gyr, second row, 18 Gyr, third row); projections of the 1-, 2- and 3-$\sigma$
$\Delta\chi^2$ contours on the ($\tau$-age) plane for the central aperture of NGC\,488, for the fit performed with free age, $\tau$ and metallicity, fourth row; \textit{second to fifth column}: as in the first 
column, for NGC\,628, 772, 864, 1042 and 2805; in all plots, the red symbol indicates the best-fitting model. These plots demonstrate that for NGC\,488, 628 and
772 we can exclude at a high confidence level a constant star formation scenario, while, for example, for NGC\,2805 high values of $\tau$ are not highly unlikely at all
ages. }
\label{chi2contours1}
\end{figure*}
\addtocounter{figure}{-1}
\addtocounter{subfigure}{1}
\begin{figure*}
{\includegraphics[width=0.89 \linewidth ]{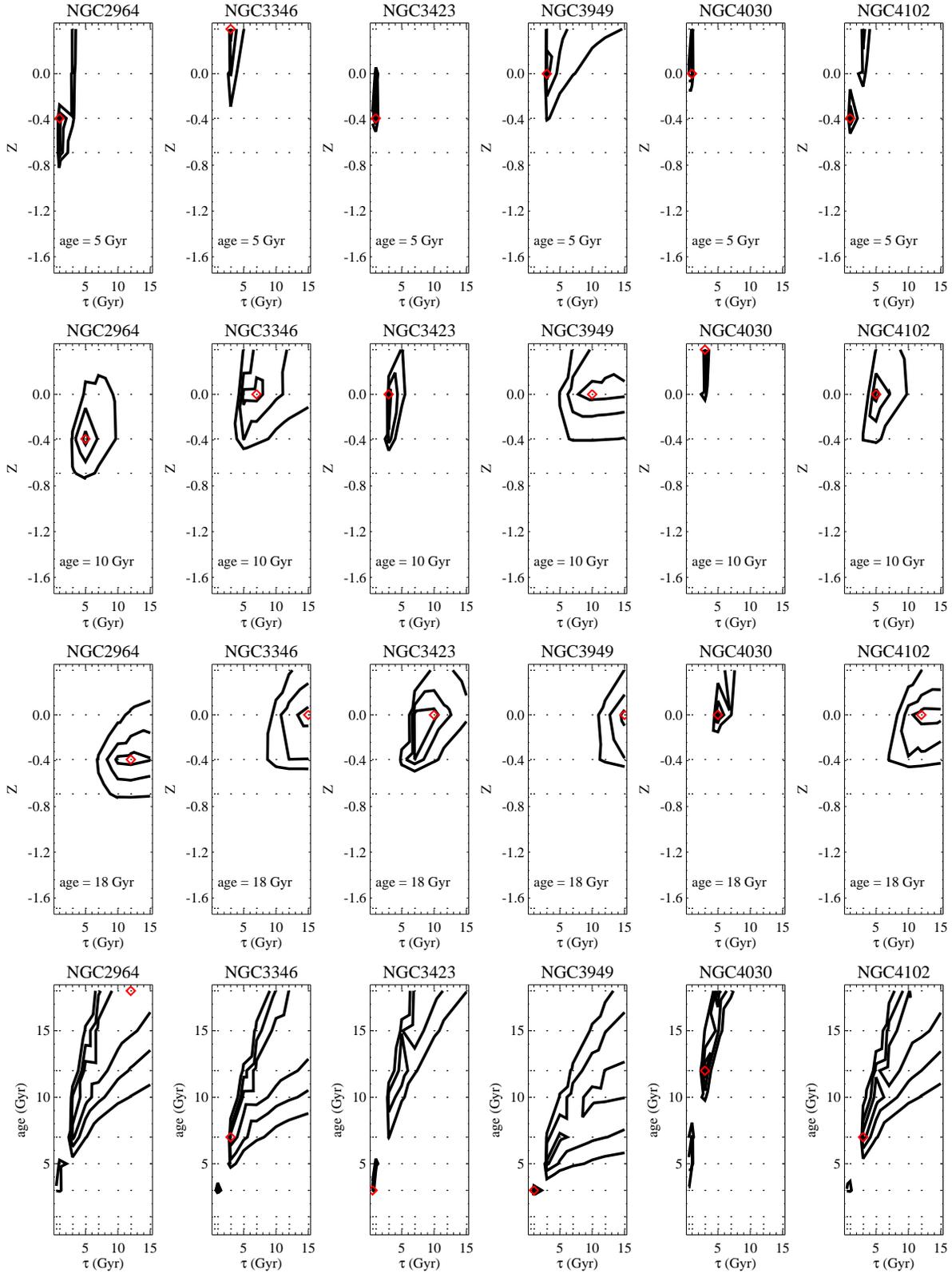}}
\caption{As in Figure \ref{chi2contours1}, for NGC\,2964, 3346, 3423, 3949, 4030 and 4102.}
\label{chi2contours2}
\end{figure*}
\addtocounter{figure}{-1}
\addtocounter{subfigure}{1}
\begin{figure*}
{\includegraphics[width=0.89 \linewidth ]{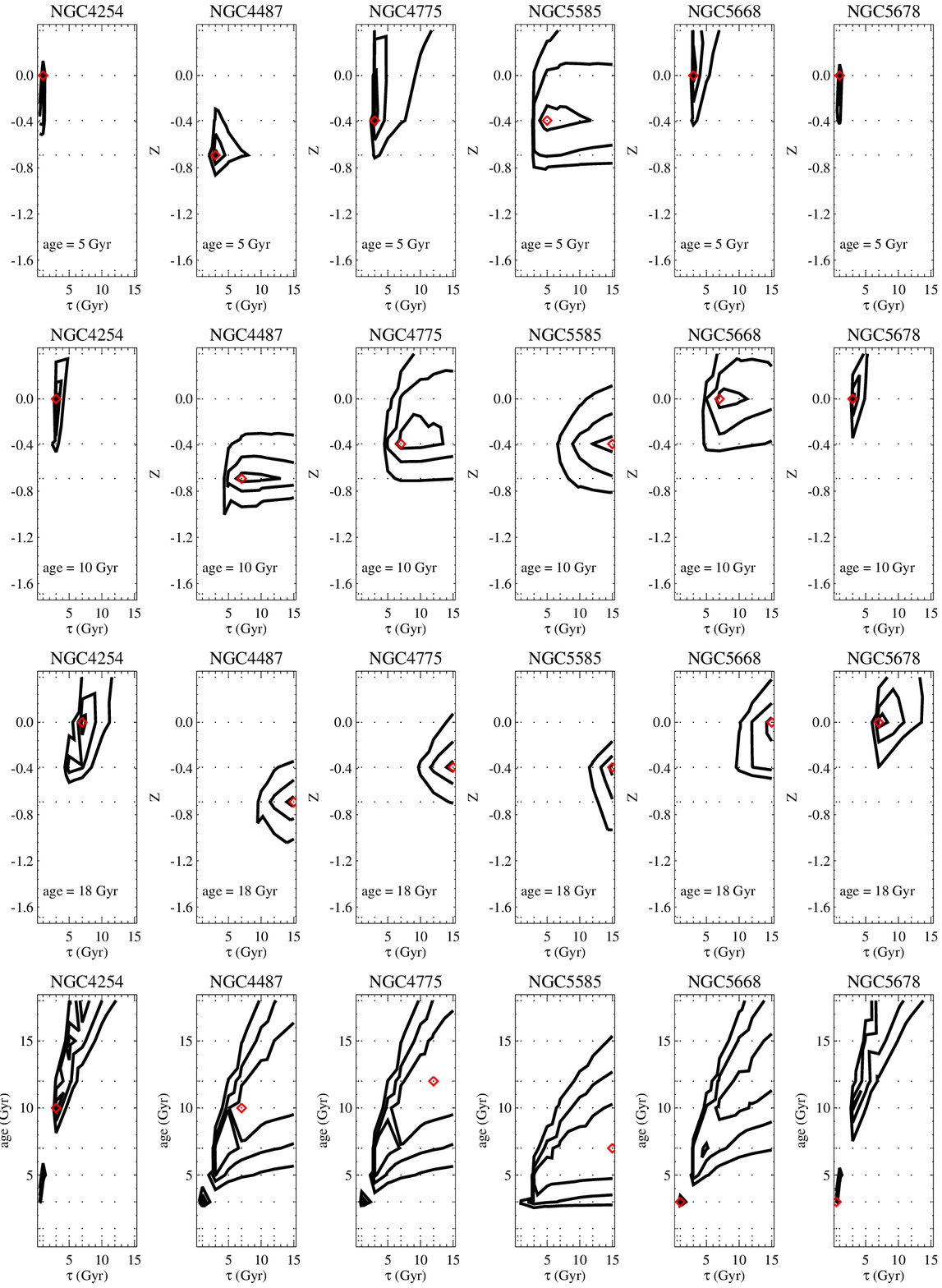}}
\caption{As in Figure \ref{chi2contours1}, for NGC\,4254, 4487, 4775, 5585, 5668 and 5678.}
\label{chi2contours3}
\end{figure*}
\end{center}
\setcounter{subfigure}{0}

\section{SUMMARY}
In this Paper we presented the first integral-field measurements of line-strength indices and stellar population 
parameters for a sample of late-type spiral galaxies. This work is based on data 
of 18 spirals spanning a range in type between Sb and Sd, observed 
with {\tt SAURON}, which provides measurements of the Lick indices H$\beta$, Fe5015 and Mg{\textit{b}} over a field of 
view covering the nuclear region of our objects.\\
\indent Here we summarise the main achievements of the Paper:
\begin{itemize}
\item We presented the two-dimensional maps of the line-strength indices and analysed the values extracted on a central aperture of 1\farcs5 radius.
\item We looked at the relations with morphological type: we found that late-type spirals do differ from earlier-type galaxies in 
terms of their populations, having globally higher H$\beta$ values and lower metal indices than ellipticals, lenticulars and early-type 
spirals. This suggests younger ages and lower metallicities.
\item We investigated the relations between indices and central velocity dispersion: late-type spirals do not 
obey the index - $\sigma$ relations found for early-type galaxies: our galaxies lie above the H$\beta$ - $\sigma$ and 
below the Mg{\textit{b}} - $\sigma$ relations determined for 
the elliptical galaxies, that correspond to old populations.
\item We determined SSP-equivalent ages, metallicities and abundance ratios by comparison of our observed indices with models: 
ages are mainly young, metallicities low and abundance ratios close to solar; the range spanned in the parameters is narrower than for early-type spirals.
\item We investigated the possibility of recovering the star formation history: we assumed that the galaxies formed 10 Gyr ago, and that they experienced an 
exponentially declining or constant star formation rate since then, and compared our indices with the evolved models of \citet{BC}; 
the fit returned an estimate for the $e$-folding time-scale for star formation $\tau$, which parametrizes a smooth transition from a 
SSP-like star formation history to a constant SFR: small $\tau$ approximate an instantaneous burst and very large 
$\tau$ reproduce a constant SFR; for the central apertures of our galaxies, we found a large range in $\tau$ values, from 1 to 15 Gyr, 
covering basically the whole parameter interval that we explored.
\item Interestingly, we found a trend between the fitted $\tau$ and the central velocity dispersion, in the sense that galaxies 
with large $\sigma$ tend to have small $\tau$ and galaxies with low $\sigma$ tend to have large $\tau$.
\end{itemize}
\indent Our observations and results nicely fit in the scenario summarised by \citet{kennicutt98}: there are two modes of star formation; one takes place 
in the extended discs of spiral and irregular galaxies, the other in compact gas discs in the nuclear regions of galaxies. The first has 
been detected mainly via H$\alpha$ surveys and it is strongly dependent on the morphological type: the star formation rate 
increases by a factor of $\approx$ 20 going from Sa to Sc galaxies. Similar information is derived from observations of the UV continuum and 
broadband visible colours. The second mode of star formation 
can be found in the circumnuclear regions; the physical conditions in these regions are in many cases distinct from the more extended star forming discs. 
The circumnuclear star formation is characterised by the absolute range in SFR, much higher spatial concentration of gas and stars and its burst-like nature \citep{kennicutt98}.
Circumnuclear star formation is largely decoupled from Hubble type. Studies on the dependence of nuclear H$\alpha$ emission in star
forming nuclei as a function of galaxy type (\citealt{stauffer82}, \citealt{keel83}, Ho, Filippenko, \&  Sargent 1997) showed that the 
detection frequency of HII region nuclei is a monotonic function of type, increasing from 0\% in elliptical galaxies to 8\% in S0, 22\%
in Sa, 51\% in Sb, and 80\% in Sc - Im galaxies. Among the galaxies with nuclear star formation, the
H$\alpha$ luminosities show the opposite trend; the average extinction-corrected luminosity of
HII region nuclei in S0 - Sbc galaxies is nine times higher than in Sc galaxies.
The conclusion of these studies, as summarised by \citet{kennicutt98}, is that the bulk of the total nuclear star formation 
in galaxies is weighted towards the earlier Hubble types, even though the frequency of occurrence is higher in the late types. 
According to \citet{kormendy04}, the central star formation  accounts for 10 - 100\% of the total SFR
of spiral galaxies. The highest fractions occur in early-type spiral  galaxies, which typically
have low SFRs in their outer discs \citep{kennicuttkent}.\\
\indent The work presented in this Paper supports this picture: if we put together the information gathered from the sample of E/S0s of Paper VI, 
the sample of Sa galaxies of Paper XI and our own sample, and look at the behaviour along the Hubble sequence, we find that early-type spirals 
show a larger range in age than both E/S0s and late-type spirals; this is consistent with a scenario where the star formation in elliptical 
and lenticular galaxies is not very important, in early-type spirals it is dominated by short bursts and in late-type galaxies it is more 
quiescent. This is also supported by our finding that big spirals tend to have an SSP-like star formation history, while 
for smaller ones the star formation history is better approximated by a constant over time.\\
\indent The following steps in the analysis of the {\tt SAURON} data here presented and discussed will be a full investigation of the 
spatial information, with a more detailed study of the radial profiles (here shown in Appendix \ref{gradientssec}) and the gradients, 
and an interpretation of the connection between features in the line-strengths maps and in the morphological appearance. 
The interdependencies between the populations and the kinematical structures \citep{ganda} will also be examined.\\
\indent The main limitation of our work is the degeneracy in the parameter space in the different approaches we investigated. Our analysis 
relies on three data-points only (the line-strength indices within the {\tt SAURON} spectral range), which are too few to break the degeneracies and 
constrain the fitted parameters. Therefore, it is urgent to complement the data presented here with spectra with a broader spectral coverage. 
Observations at other wavelengths -both shorter and longer- would help building a complete picture: both the UV and the NIR 
spectral regions can provide population indicators much less affected by the model degeneracies found in the optical. In addition, different 
sub-populations contribute to the integrated light in different spectral ranges, therefore a multi-wavelength study would 
be a step forward towards a full understanding of the formation and evolution of our objects.

\section{Acknowledgments}
We kindly acknowledge Isabel P\'erez for very useful discussion and comment. We thank the referee for interesting comments that 
improved the paper. The {\tt SAURON} - related projects are made possible through grants 614.13.003 and 781.74.203 from
ASTRON/NWO and financial contributions from the Institut National des Sciences
de l'Univers, the Universit\'e Claude Bernard Lyon~I, the universities of
Durham, Leiden, Groningen and Oxford, the British Council, PPARC grant
`Extragalactic Astronomy \& Cosmology at Durham 1998--2002', and the
Netherlands Research School for Astronomy NOVA. KG acknowledges support for the Ubbo Emmius PhD 
 program of the University of Groningen. MC acknowledges support from a PPARC Advanced Fellowship (PP/D005574/1). 
 GvdV acknowledges support provided by NASA through Hubble Fellowship
grant HST-HF-01202.01-A awarded by the Space Telescope Science
Institute, which is operated by the Association of Universities for
Research in Astronomy, Inc., for NASA, under contract NAS
5-26555.
 This project made use of the HyperLeda and NED databases.


\appendix
\renewcommand{\thefigure}{A\arabic{figure}\alph{subfigure}}
\setcounter{subfigure}{1}
\clearpage
\begin{figure*}
{\includegraphics{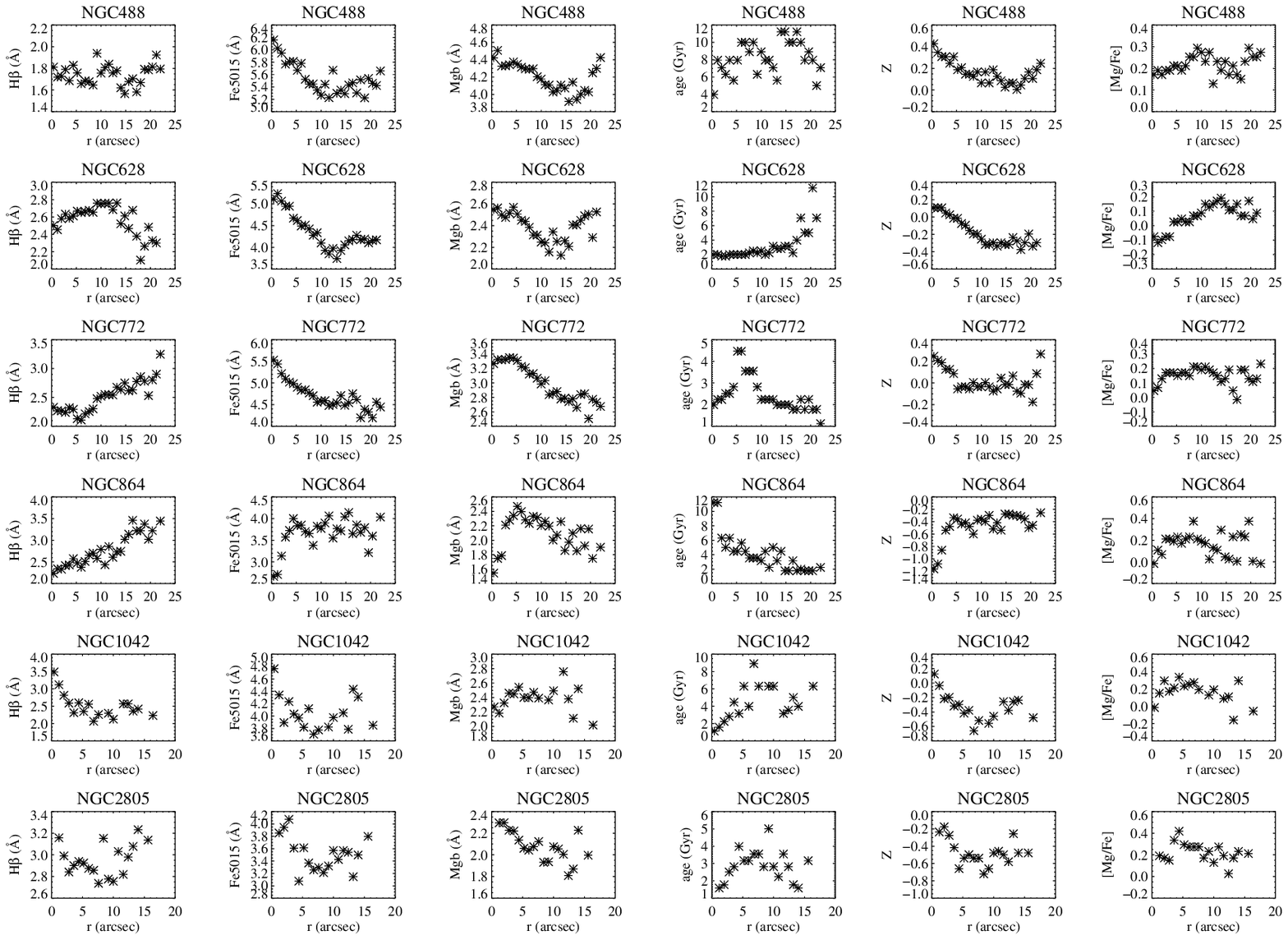}}
\caption{Azimuthally averaged radial profiles of the line-strength indices H$\beta$, Fe5015, Mg{\textit{b}}, (expressed as equivalent widths and measured in
\AA) and of the one-SSP analysis parameters age (in Gyr), metallicity and
abundance ratio against galactocentric distance (expressed in arcsec); every row 
presents a different galaxy. See text in Appendix \ref{gradientssec} for a more detailed description. Here we show NGC\,488, NGC\,628, NGC\,772, NGC\,864, NGC\,1042 and NGC\,2805.}
\label{profiles1}
\end{figure*}
\addtocounter{figure}{-1}
\addtocounter{subfigure}{1}
\begin{figure*}
{\includegraphics{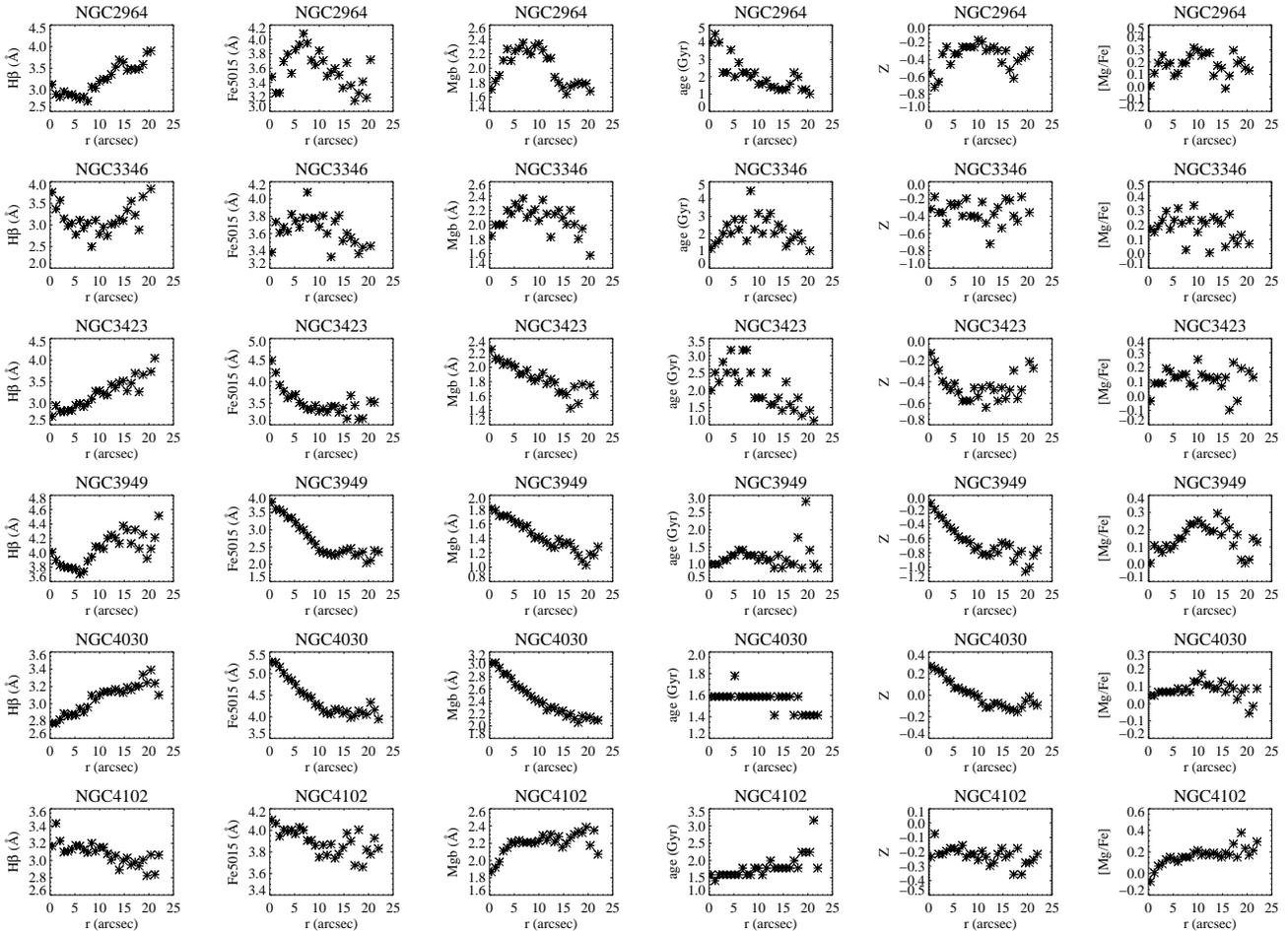}}
\caption{Same as Figure \ref{profiles1} for NGC\,2964, NGC\,3346, NGC\,3423, NGC\,3949, NGC\,4030 and NGC\,4102.}\label{profiles2}
\end{figure*}
\addtocounter{figure}{-1}
\addtocounter{subfigure}{1}
\begin{figure*}
{\includegraphics{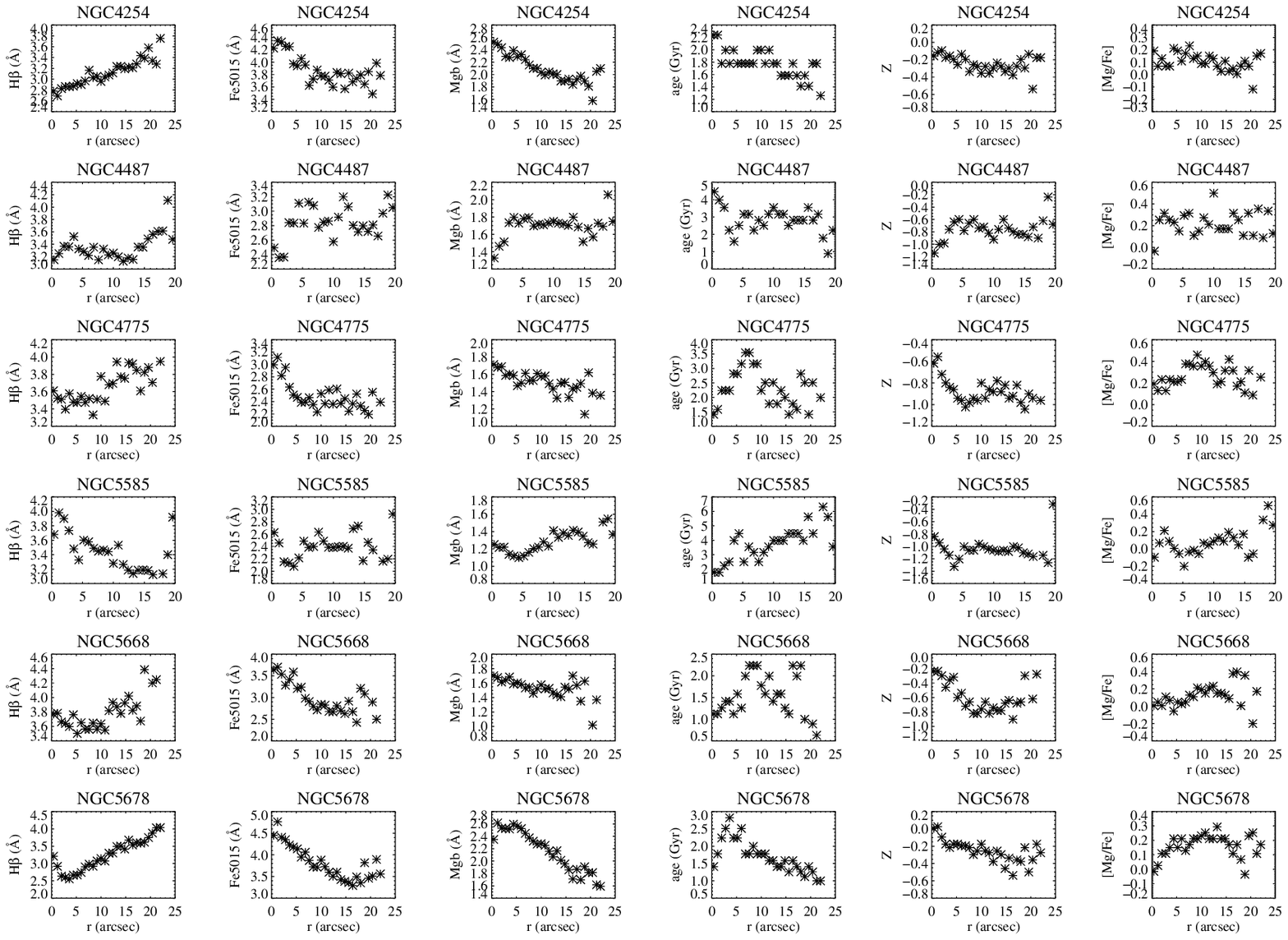}}
\caption{Same as Figure \ref{profiles1} for NGC\,4254, NGC\,4487, NGC\,4775, NGC\,5585, NGC\,5668 and NGC\,5678.}\label{profiles3}
\end{figure*}
\setcounter{subfigure}{0}
\section{Radial behaviour}\label{gradientssec}
We briefly investigated the radial variations of the line-strength indices and age, metallicity and abundance ratios. 
From the corresponding maps, we calculated 
radial profiles of H$\beta$, Fe5015, Mg{\textit{b}} and age, Z and [Mg/Fe] from the one-SSP approach, by averaging the maps 
on concentric annuli of 0\farcs8 
width. We present these profiles in Figures \ref{profiles1}-\ref{profiles3}. 
These plots show that there is a wide variety of radial behaviours among late-type 
spirals: the line-strength indices can be roughly constant in radius (H$\beta$ profile for NGC\,488), rising outwards 
(even very steeply, see for example 
the H$\beta$ profile for NGC\,5678), rising inwards (many Fe5015 and Mg{\textit{b}} profiles). 
In some cases there are also complicated structures, with 
`oscillating' profiles: see for example the Fe5015 profile for NGC\,864 or NGC\,2805 and several others. 
Some of the `oscillating' profiles might just reflect the noisy nature of the data for some low-surface brightness objects 
(for example: the Fe5015 profile 
for NGC\,2805), but others are probably related to real variations of the indices within the galaxy. 
Despite all this variety, we conclude that in most cases the H$\beta$ absorption strength increases with radius, 
while Fe5015 and Mg{\textit{b}} decrease. The fact that the metal lines show most 
often negative gradients was already known for early-type galaxies, while in the case of H$\beta$ for elliptical galaxies the gradients are consistent with a flat 
relation with radius, or present a mild positive outwards radial gradient (see \citealt{paper6} and references therein).\\
\indent The stellar population parameters show the same heterogeneous behaviours. Roughly, the age shows most often a complex behaviour with radius, reflecting features 
recognizable in the maps in Figures \ref{maps1}-\ref{maps9},
while the metallicity in many cases is decreasing, in agreement with \citet{moorthy}, who measure mostly negative metallicity gradients for a
sample of S0-Sc galaxies;   
in the majority of the galaxies, the metallicity profile resembles closely the aspect of the Fe5015 profile. 
Looking at the abundance ratio
profiles in more detail, we see that the central values are usually very close to solar; many of the profiles stay 
approximately constant with radius, after an initial increase in the
bulge-dominated region, which can be very tiny (for a
quantitative description of the bulges in these galaxies, we
refer the reader to Ganda et al., in preparation). This is also consistent with
the findings of \citet{moorthy}, that we briefly summarised
in Section \ref{indexvsindexsec}: they suggest that the bulges have positive or null gradients in abundance ratio
with increasing radius, while the discs have solar abundance ratios. In early-type galaxies, instead, the radial profiles of 
abundance ratios are rather flat, as found by Kuntschner et al. (paper in preparation) for the {\tt SAURON} E/S0 galaxies and by \citet{patricia} for a sample 
of 11 E/S0 galaxies; the latter authors do find radial abundance ratio gradients, both positive and negative, but rather shallow.

\section{Individual galaxies}\label{individuals}
\subsection*{NGC\,488}
NGC\,488 is an unbarred Sb galaxy and it is the most elliptical-like (featureless) object within our sample. 
NGC\,488 has regular stellar and gaseous kinematics, with regular rotation and
velocity dispersion smoothly decreasing outwards \citep{ganda}. The line-strength maps
present high values for Fe5015 and Mg{\textit{b}} and low values for H$\beta$ 
over the central regions. As one can notice from the values reported in Table \ref{indicesvalues}, 
this galaxy has the highest Fe5015 and Mg{\textit{b}} in the sample, and the lowest H$\beta$. The ages are mainly old and the
metallicities high. This is indeed the most metal-rich galaxy in the sample (the central
metallicity from the one-SSP approach is $\approx$ 0.49, then it decreases outwards). This galaxy is well described by a one-SSP: 
as shown in Fig. \ref{chi2contours1} (first column) on the basis of our 
continuous star formation analysis, NGC\,488 is very unlikely 
to have experienced a constant star formation over time; the observed indices are instead reproduced by a short 
starburst, approximating a SSP. The time-scale of star formation $\tau$ is short over the whole field. From the profile in Fig. \ref{profiles1} 
we can see that 
the [Mg/Fe] profile is approximatively constant around the value $\approx$ 0.2: abundance ratios are slightly super-solar, confirming the fact that this galaxy resembles an
elliptical more than the later objects in our sample. 
Other suggestions in this sense come from the position it occupies in the diagrams in 
Figures \ref{corr_mgfe50},
\ref{corr_index}, \ref{1ssp_fullrange}, where it lies in the region
occupied by the ellipticals.\looseness=-2

\subsection*{NGC\,628}
NGC\,628, known also as M74, is a grand-design Sc galaxy. \citet{cornett} concluded that the star formation history of NGC\,628 
varies with galactocentric 
distance; Natali, Pedichini \& Righini (1992) suggested that the galaxy could be seen as an inner and an outer disc 
characterised by different stellar
populations; according to them, the transition between the two regions is located at $\approx$ 8-10 kpc from the centre. 
Our observations of NGC\,628 are disturbed by the presence of a foreground star which falls close to the centre
 ($\approx$ 13\arcsec\/ southern). The stellar kinematics is characterised by slow projected rotation, and velocity dispersion 
 decreasing in the central zones, indicating a  cold central region, maybe an inner disc \citep{ganda}.
This galaxy is part of the `control sample' of non-active objects studied by \citet{cid1}, \citet{cid2}, \citet{cid3}, 
who analyse the nuclear spectrum, describing it as a mixture of $10^8$-$10^9$ yr-old and older stars.\\
\indent As can be seen from the profiles reported in Fig. \ref{profiles1}, NGC\,628 has ages (one-SSP approach) below 3 Gyr out to $\approx$ 15\arcsec\/, 
and becomes older at larger radius; the features are roughly in agreement with the above mentioned findings of Cid-Fernandes and collaborators; 
the same trend in age is observed, on the common radial range, also by \citet{lauren}, on the basis of {\tt GMOS} data independently 
analysed. The metallicity decreases outwards. The `bumps' recognizable in the Fe5015, 
Mg{\textit{b}} and (but less pronounced) metallicity profiles at $\approx$ 13\arcsec\/ from the centre are due to the foreground star. From our continuous star formation analysis, as 
shown in Fig. \ref{chi2contours1} (second column), we can 
state that this galaxy is not described by a constant star formation scenario.\looseness=-2

\subsection*{NGC\,772}
NGC\,772, also called Arp78, is an Sb galaxy characterised by a particularly strong spiral arm; it forms a pair at 3\farcm3 
with the E3 galaxy NGC\,770. 
From the kinematical point of view, NGC\,772 is characterised by a clear drop of the stellar 
velocity dispersion in the central zones, possibly corresponding to an inner
disc, as suggested also by the anticorrelation between velocity and the Gauss-Hermite moment h$3$ \citep{ganda}. In the line-strength maps we can clearly see an extended central region with
high Fe5015 and Mg{\textit{b}} and low H$\beta$; indeed, from the central aperture values reported in Table \ref{indicesvalues} we can see that 
this galaxy is one of those with the highest metal indices, surpassed only by NGC\,488. In the H$\beta$ map it is possible to see, surrounding the central depression, regions with higher values corresponding to 
the spiral arms. A similar behaviour can be seen in the [OIII]/H$\beta$ map presented by \citet{ganda}, which shows a structure
resembling the spiral pattern. Estimating ages and metallicity by comparison with SSP models, we infer that ages are mainly 
young-intermediate and the central metallicity is among the 
highest in our sample. The age and abundance ratio maps display the spiral arms structure as well, 
like the index maps. Fig. \ref{chi2contours1} (third column) demonstrates that for this galaxy constant star formation can 
be ruled out at a high confidence level: high $\tau$ values are never likely. The $\tau$ map shown in Fig. \ref{maps2} (for age fixed to 10 Gyr in the
fitting procedure) presents
a spatial structure that partially resembles the H$\beta$ map and the spiral arms pattern. 
NGC\,772 is part of the sample studied in the series of papers by Cid Fernandes et al. (\citealt{cid1}, \citealt{cid2}, 
\citealt{cid3}), who are engaged in a 
project to examine the stellar populations of low-luminosity active galactic nuclei. They classify this object as `young-TO', i.e. a 
transition object between a LINER and an HII nucleus, with weak [O I] $\lambda 6300$ emission; they find that the nuclear spectrum is dominated by $10^8$-$10^9$ yr-old
 stellar populations, 
with a contribution up to $\approx$ 32\% from populations younger than $10^7$ yr and a similar contribution also from 
older stellar populations ($\geq$ 10$^9$ yr). Their data probe distances up to several hundred parsecs from the nucleus; they 
measure in NGC\,772 significant radial variations of the stellar populations, with absorption line-strengths increasing in the inner 3\arcsec\/
indicating a younger population in the very centre, especially referring to the Ca II K line; the Mg I absorption strength presents instead less prominent variations over 
the spatial range of their data. Our data cover a different spectral range and a different spatial extension, but we also concluded 
that the populations in the inner parts have intermediate ages, and that the age (from the one-SSP analysis) increases in the inner 
$\approx$ 5\arcsec\/ (see Fig. \ref{profiles1}).\looseness=-2

\subsection*{NGC\,864}
This is a barred Sc galaxy where a nuclear radio source with the linear size of 
$\approx$ 300 pc has been detected \citep{ulverstad}; according to \citet{n864asym} its HI profile is symmetric 
in velocity but asymmetric in 
intensity; they also measure large-scale asymmetries in the two-dimensional kinematics, such as a warp. The mentioned 
asymmetries occur in the outer parts, well beyond the extent of
the {\tt SAURON} field. The asymmetries could have been induced by a past 
encounter with a companion, passing outside the optical disc but within the HI disc; the companion would have dispersed 
its stars and dark matter after the merging, consistent with the fact that NGC\,864 appears to be remarkably isolated.\\
\indent Our line-strength maps show a very sharp central dip in both Fe5015 and Mg{\textit{b}}; in the region surrounding the 
dip, the Fe5015 and Mg{\textit{b}} values rise. The H$\beta$ absorption also takes on low values over an extended central 
region; as seen also in Table \ref{indicesvalues}, the indices measured on the central aperture of this galaxy are among the lowest in the sample. 
From the one-SSP analysis, given the structure of the line-strength maps, we derive very low metallicity and 
old ($\approx$ 11.22 Gyr) ages for the central aperture.\\
\indent The analysis of the $\Delta\chi^2$ contours, in the continuous star formation approach, shows 
that this galaxy is unlikely to have experienced 
a constant star formation over time (see Fig. \ref{chi2contours1}, fourth column). \looseness=-2

\subsection*{NGC\,1042}
NGC\,1042 is an Scd galaxy forming a pair with NGC\,1035 at a separation of 22$'$ (corresponding to 177 kpc). It has a bright, small
nucleus and otherwise low surface brightness. Our line-strength maps are quite patchy, due to poor S/N in the data and consequent 
heavy binning of the spectra. The Mg{\textit{b}} index takes on low values in the very
centre, and higher ones around it. H$\beta$ is high in the centre and low outside it, being flat over most of the map; as can be seen also from the line-strength 
profiles in Fig. \ref{profiles1}, the line-strength indices stay approximately constant over most of the field, outside a central
zone of radius $\approx$ 3\arcsec\/. The age of the stellar populations (one-SSP analysis) 
reaches a minimum in the centre. The metallicity is everywhere low, and peaks in the centre.\\
\indent The analysis of the $\Delta\chi^2$ contours for the central aperture, in the continuous star formation approach, shows 
that this galaxy is not well described by an SSP (see Fig. \ref{chi2contours1}, fifth column). The $\tau$ map for age fixed to 10 Gyr (see Fig. \ref{maps3}) shows a peak in the
very centre, corresponding to the peak in the H$\beta$ and to the minimum in the Mg{\textit{b}} maps.\looseness=-2

\subsection*{NGC\,2805}
NGC\,2805 is an Sd galaxy seen nearly face-on and it is the brightest member of a multiple interacting system containing also NGC\,2814 (Sb), NGC\,2820 (Sc)
at 13$'$ and IC2458 (I0). HI has been detected (\citealt{reak}, \citealt{bosma}) and there are claims 
that the outer HI layers are warped (see for example \citealt{bosma}). The galaxy seems to be also optically disturbed, since the spiral arms appear to be 
broken up into straight segments.\\
\indent The line-strength maps are rather patchy, displaying though a central peak in all three lines. 
By comparison of our measured indices with SSP models, we can state that the 
ages are young-intermediate over most of the field; that the metallicity peaks in the centre and that abundance ratios are super-solar.\\
\indent The analysis of the $\Delta\chi^2$ contours, in the continuous star formation approach, shows 
that this galaxy is not very well described by an SSP, and that even a constant 
star formation rate cannot be excluded at all ages (see Fig. \ref{chi2contours1}, sixth column).\looseness=-2

\subsection*{NGC\,2964}
NGC\,2964, classified as Sbc, forms a non-interacting pair with the I0 galaxy NGC\,2968 at 5\farcm8.
As discussed by \citet{ganda}, the {\tt SAURON} data reveal that this barred Sbc galaxy hosts an active galactic nucleus. 
When we calculate the 
line-strengths, we find that the Mg{\textit{b}} index assumes low values in an elongated central region, 
higher values around it, and low values again towards the
edges of the field; the Fe5015 index presents a very similar behaviour; 
the H$\beta$ map has instead an opposite appearance, with an extended inner region with low 
index values and a surrounding area with higher values; in the very centre a shallow peak might 
be recognized. These features are recognizable also in the profiles in Fig. \ref{profiles2}. We notice also that the central region with
low Fe5015 and Mg{\textit{b}} corresponds to the central depression in the emission-line [OIII]/H$\beta$ map 
(see Fig. 5 in \citealt{ganda}), possibly indicating 
ongoing star formation. Converting the line-strength maps to age and metallicity via our comparison with the SSP models, we find young-intermediate ages over most of the field, with a peak in the central region, and low metallicity everywhere, with a minimum in the central region, a positive 
gradient out to $\approx$ 5\arcsec\/, and then a rather constant behaviour out to $\approx$ 15\arcsec\/. From the continuous star formation
approach, we cannot exclude at all ages the case of constant star formation, but the $\tau$ map 
for age fixed to 10 Gyr (in the fitting procedure) presents a large central region with low values, possibly suggesting a $\approx$ 5 Gyr long starburst.\looseness=-2

\subsection*{NGC\,3346}
This Scd galaxy is very poorly studied in the literature. Ages of the stellar populations (one-SSP approach) are
young-intermediate and metallicity low over most of the field. From the continuous star formation
approach, we cannot exclude at all ages the case of constant star formation (see Fig. \ref{chi2contours2}, second column).\looseness=-2

\subsection*{NGC\,3423}
This Sbc galaxy is also poorly known. Both the Fe5015 and Mg{\textit{b}} indices peak in an extended central region and 
decrease towards the edge of the field; the H$\beta$ map
presents opposite behaviour. The $\tau$ map (for age fixed to 10 Gyr in the fitting procedure) displays low values 
over a large central area, possibly indicating a region where the bulk of star formation happened in a quite short starburst. \looseness=-2

\subsection*{NGC\,3949}
NGC\,3949 is an Sc galaxy belonging to the Ursa Major cluster.
NGC\,3949 presents quite regular Fe5015 and Mg{\textit{b}} line-strength maps, peaking in an extended central region 
and decreasing rather smoothly moving outwards. The H$\beta$ 
index is low in the central region and higher in the surroundings; interestingly, it reaches a minimum in a zone 
south of the centre where the emission-line ratio [OIII]/H$\beta$ is particularly depressed (see relative Figure in \citealt{ganda}). The values 
measured for the H$\beta$ index are the highest in the sample, as one can see from Table \ref{indicesvalues} and from Fig. \ref{profiles2}. Ages 
(from the one-SSP approach) are young ($\approx$ 1.0 Gyr) in the central region, somewhat higher around it and decreasing 
again towards the edge of the field. The metallicity is everywhere low, and decreasing outwards.\\
\indent Constant star formation cannot be ruled out; it is likely to characterise the
star formation mode over most of the field (see the $\tau$ map for age fixed to 10 Gyr in Fig. \ref{maps5}: $\tau$ assumes high values). Actually, when
performing the fit allowing the code to choose among exponentially declining and constant star formation, constant is selected over 
most of the field.\looseness=-2

\subsection*{NGC\,4030}
This Sbc galaxy is characterised by very regular kinematics for both the stellar and gaseous components \citep{ganda}. 
The line-strength maps also have a smooth appearance, with 
Fe5015 and Mg{\textit{b}} decreasing at larger radius and H$\beta$ increasing. The H$\beta$ map seems to be asymmetric with 
respect to the centre, since the western side
displays higher values than the eastern one. The stellar populations are young over the whole field and the metallicity, among the
highest in our sample (the 
third after NGC\,488 and 772), decreases 
moving outwards; the abundance ratio is approximatively solar. The $\tau$ map for age fixed to 10 Gyr in Fig. \ref{maps6} 
shows a large central depression. The analysis of the $\Delta\chi^2$ contours, in the continuous star formation approach, shows 
that this galaxy is probably not very well described by an SSP, nor by a constant star formation rate.\looseness=-2

\subsection*{NGC\,4102}
NGC\,4102 is an Sb galaxy, classified as a LINER in the NED database; it is known to be a powerful far-infrared galaxy 
\citep{young} and also to have a strong nuclear radio source
\citep{condon}. \citet{devereux} classified it as one of the most powerful 
nearby starburst galaxies. The signs of activity are detectable also in the {\tt SAURON} data: as noted by \citet{ganda}, the [OIII] 
maps trace the outflowing gas, with a region of very high gas velocity dispersion corresponding also to high values in the 
[OIII]/H$\beta$ line ratio map, as expected
for an active object. In that same region, we observe double-peaked line profiles for the [OIII] emission-lines, pointing towards 
the presence of different components in
the emitting material. As for the line-strength indices, Mg{\textit{b}} is low in an inner region, increases around it, 
and decreases again in the outer parts. The Fe5015 map is instead more centrally concentrated. 
The map of H$\beta$ absorption looks more complex and not symmetric. Ages (from our one-SSP analysis) are young all over the field and metallicity is low. Neither in the line-strength nor in the age and metallicity map is there a clear 
connection with the mentioned region of activity. On the basis of our continuous star formation analysis, this galaxy is probably not very well 
described by an SSP.
\looseness=-2

\subsection*{NGC\,4254}
Also known as M99, NGC\,4254 is a bright Sc galaxy located on the periphery of the Virgo cluster, at a projected distance of $3.7^{\circ}$ ($\approx$ 1 Mpc) from the centre 
of the cluster. Its optical appearance is dominated by a peculiar one-arm structure: 
the arms to the northwest are much less defined than the southern arm. This kind of spiral structure could be 
related to an external driving mechanism, but for NGC\,4254 there are no close companions. Phookun, 
Vogel \& Mundy (1993) carried out 
deep HI observations of NGC\,4254, detecting non-disc HI clouds, with velocities not following the disc velocity pattern and an extended low surface density tail northwest 
of the galaxy; they interpreted these observational results as 
infall of a disintegrating gas cloud. They also noticed that there is an hole in the HI emission at the centre of the galaxy, with a 
diameter of $\approx$ 3 kpc. Vollmer, Huchtmeier \& van Driel (2005) propose instead a scenario where NGC\,4254 had a close and rapid encounter with another 
massive galaxy when entering the Virgo cluster: the tidal interaction caused the one-arm structure and the HI distribution and kinematics is due to ram pressure stripping.
Recently, \citealt{soria} studied the X-ray properties of M99, noticing that a phenomenon often associated with tidal interactions and active star
formation -both documented- is the presence of ultraluminous X-ray sources (ULXs). The X-ray emission appears approximatively 
uniformly diffused across the inner $\approx$ 5 kpc from the nucleus. They do find a ULX, amongst the brightest observed, located at
$\approx$ 8 kpc southeast of the nucleus, close to the position where a large HI
cloud seems to join the gas disc. From observations of the radial velocities, they suggest that the cloud is falling onto the galactic disc and try to 
build a link between this collisional event and the formation of the bright ULX.
Our data cover a much smaller spatial extent than the mentioned observations, therefore a comparison is not possible; within our field, both the Fe5015 and Mg{\textit{b}} maps show a regular structure with a central 
concentration; at the northern edge of the field, both Fe5015 and Mg{\textit{b}} seem to increase again; the H$\beta$ map has opposite behaviour, presenting a central depression. 
The ages (from our one-SSP approach) are generally young and the metallicity, quite low everywhere, decreases slightly moving
outwards and shows the same enhancement at the northern edge of the field that we noticed for the Fe5015 and Mg{\textit{b}} line-strength. 
The $\tau$ map (see Fig. \ref{maps7}) shows a central depression. The analysis of the $\Delta\chi^2$ contours, in the continuous star formation approach, shows 
that NGC\,4254 is probably not very well described by an SSP, nor by a constant star formation rate. \looseness=-2

\subsection*{NGC\,4487}
This Scd galaxy forms a pair with the Scd galaxy NGC\,4504, at a separation of 35$'$ (corresponding to $\approx$ 165 kpc).
The Fe5015 and Mg{\textit{b}} maps seem to be low in the very centre and higher
outside it; from the profiles in Fig. \ref{profiles3} one can see that outside the inner circle of $\approx$ 3\arcsec\/ radius, the line-strengths 
keep an approximately constant value, out to $\approx$ 13\arcsec\/. Ages (one-SSP approach) are intermediate-young and metallicities very low, with respectively a maximum and a minimum in the centre. Constant star formation cannot be ruled out.
\looseness=-2

\subsection*{NGC\,4775}
NGC\,4775 is an Sd galaxy, for which very little is known from previous studies. The metal lines 
(Fe5015 and Mg{\textit{b}}) seem to be centrally concentrated, while H$\beta$ is low
over an extended central area, and higher in the surrounding region. The maps are not symmetric with respect to 
the centre. The galaxy is young in the nucleus and has older ages in an annular (extended) region around the 
centre; the metal content is everywhere low, and peaks in the centre. Constant star formation cannot be ruled out, and is
selected over a large part of the field when allowing to choose among constant and exponentially declining star formation.\looseness=-2

\subsection*{NGC\,5585}
NGC\,5585 is a barred Sd galaxy; together with NGC\,5204 (Sm), NGC\,5474 (Scd), NGC\,5477 (Sm) , HoIV (Im) and M101 (Scd) it forms the M101 group. 
The Fe5015 and Mg\textit{b} indices assume values 
among the lowest in the sample; H$\beta$ has instead a value among the highest ones (see Table \ref{indicesvalues}). The 
metallicity retrieved from the one-SSP analysis is everywhere very low. In the continuous star formation approach, 
constant star formation cannot be ruled out; indeed, when allowing to choose among exponentially declining and constant star formation,
in the $\chi^2$ minimisation, for age fixed to 10 Gyr the constant star formation is selected (for the central aperture).\looseness=-2
 
\subsection*{NGC\,5668}
NGC\,5668 is another Sd galaxy, which has a high rate of star formation, as indicated by its large far-infrared and 
H$\alpha$ luminosities (\citealt{schulman}, and references therein).
The metal lines (Fe5015 and Mg{\textit{b}}) appear to be centrally concentrated, while H$\beta$ has a more complex appearance. The stellar 
populations are young and metal poor all over the field; the age rises out to $\approx$ 8\arcsec\/ and then decreases (see 
Fig. \ref{maps9} for a full view of the spatial structure of the age map and Fig. \ref{profiles3} for the azimuthally averaged
profiles). Constant star formation cannot be ruled out at all ages, and is
selected over a large part of the field when allowing to choose among constant and exponentially declining star formation.\looseness=-2
 
\subsection*{NGC\,5678}
This is a barred and very dusty Sb galaxy. 
Fe5015 and Mg{\textit{b}} have a quite regular and centrally concentrated appearance; 
H$\beta$ is depressed in the inner region and increases in a annulus-like structure 
around it. In the very centre H$\beta$ has a local maximum, as shown also in the azimuthally-averaged profile in Fig. \ref{profiles3}. 
Ages (one-SSP approach) are young over most of the field and the metallicity, rather low everywhere, decreases moving outwards. 
The $\tau$ map at age fixed to 10 Gyr (see Fig. \ref{maps9}) shows a large central depression. The analysis of the $\Delta\chi^2$ contours shows 
that NGC\,5678 is probably not very well described by an SSP, nor by a constant star formation rate. 
Like NGC\,628 and 772, this galaxy is part of the sample studied in the Cid Fernandes et al papers (\citealt{cid1}, \citealt{cid2}, \citealt{cid3}); following their classification, 
it is a `young-TO' galaxy, with weak emission in the [O I] $\lambda 6300$ line; according to those authors, the nuclear spectrum is heavily dominated by 
$10^8$-$10^9$ yr old stellar populations, with negligible contribution from stars younger than 
$10^7$ yr. They detect radial variations of the absorption indices, particularly in the Ca II K line, but less clearly evident than in the case of NGC\,772.\\

\makeatletter
\def\thebiblio#1{%
 \list{}{\usecounter{dummy}%
         \labelwidth\z@
         \leftmargin 1.5em
         \itemsep \z@
         \itemindent-\leftmargin}
 \reset@font\small
 \parindent\z@
 \parskip\z@ plus .1pt\relax
 \def\newblock{\hskip .11em plus .33em minus .07em}
 \sloppy\clubpenalty4000\widowpenalty4000
 \sfcode`\.=1000\relax
}
\let\endthebiblio=\endlist
\makeatother


\label{lastpage}

\end{document}